\documentclass[a4paper,12pt]{article}
\pdfoutput=1
\usepackage{feynmp-auto,expdlist}
\usepackage{amsmath, amsfonts, amssymb}
\usepackage{graphicx}
\usepackage{enumerate}
\usepackage{latexsym}
\usepackage{dsfont}
\usepackage{hepnicenames}
\usepackage{enumerate}
\usepackage{soul}
\usepackage[normalem]{ulem}
\usepackage{bbold}
\usepackage{wasysym} 
\usepackage{makecell}
\RequirePackage[ colorlinks=true,urlcolor=blue,anchorcolor=blue,citecolor=blue,filecolor=blue,
               linkcolor=blue,menucolor=blue,linktocpage=true,p dfproducer=medialab,pagebackref=true]{hyperref}
 \hypersetup{
 colorlinks   = true, 
 urlcolor     = ForestGreen, 
 linkcolor    = blue, 
  citecolor   = blue 
}

\newcommand{\blue}[1]{\color{blue}#1\color{black}}

\font\tenrsfs=rsfs10 at 12pt
\font\sevenrsfs=rsfs7
\font\fiversfs=rsfs5
\newfam\rsfsfam
\textfont\rsfsfam=\tenrsfs
\scriptfont\rsfsfam=\sevenrsfs
\scriptscriptfont\rsfsfam=\fiversfs

\numberwithin{equation}{section}

\usepackage{mathrsfs}
\usepackage{braket}
\usepackage{titling}
\usepackage{amsmath}
\usepackage{slashed}
\usepackage{amssymb}
\usepackage{epsfig}
\usepackage{graphicx}
\usepackage{color}
\usepackage{rotating}
\usepackage[margin=1.in]{geometry}
\usepackage[table,xcdraw,dvipsnames]{xcolor}
\usepackage[compress,numbers,sort]{natbib}
\usepackage{colortbl}
\usepackage{pdflscape}
\usepackage{color}
\usepackage{mathtools}
\usepackage{colortbl}
\definecolor{Gray}{gray}{0.95}
\definecolor{RGray}{gray}{0.85}
\definecolor{CGray}{gray}{0.92}

\newcommand{\I}{\mathcal{I}}
\newcommand{\J}{\mathcal{J}}
\newcommand{\F}{\mathcal{F}}

\newcommand{\ybreak}{y_{\slashed{G}}}
\newcommand{\lbreak}{\lambda_{\slashed{G}}}
\newcommand{\yG}{y_{{G}}}

\definecolor{nicered}{rgb}{0.7,0.1,0.1}
\definecolor{nicegreen}{rgb}{0.1,0.5,0.1}
\definecolor{red}{rgb}{1.0, 0, 0}
\definecolor{niceblue}{rgb}{0,0,0.8}
\definecolor{red}{rgb}{1.0, 0, 0}
\allowdisplaybreaks

\definecolor{rosso}{cmyk}{0,1,1,0.4}
\definecolor{rossos}{cmyk}{0,1,1,0.55}
\definecolor{rossoc}{cmyk}{0,1,1,0.2}
\definecolor{blu}{cmyk}{1,1,0,0.3}
\definecolor{blus}{cmyk}{1,1,0,0.6}
\definecolor{bluc}{cmyk}{1,1,0,0.1}
\definecolor{verde}{cmyk}{0.92,0,0.59,0.25}
\definecolor{verdec}{cmyk}{0.92,0,0.59,0.15}
\definecolor{verdes}{cmyk}{0.92,0,0.59,0.4}

\def\det{\mbox{det}\,}

\renewcommand{\bar}{\overline}

\newcommand{\beq}{\begin{equation}}
\newcommand{\eeq}{\end{equation}}
\newcommand{\bea}{\begin{eqnarray}}
\newcommand{\eea}{\end{eqnarray}}

\renewcommand{\[}{\left[}

\newcommand{\nn}{\nonumber}

\usepackage[capitalise]{cleveref}
\usepackage{nicefrac}
\usepackage{subcaption}
\usepackage{multirow} 		 
\usepackage{rotating}		 

\renewcommand{\O}{\mathcal{O}}

\begin{document}

\vspace{1.5cm}

{\flushright
{\blue{ \hfill}\\
\blue{ \hfill}\\
\blue{DESY-22-082}\\
\blue{IFT-UAM/CSIC-20-144}\\
\blue{FTUAM-20-21}}\\
\hfill 
\blue{IPPP/22/31}\\
\hfill 
}

\begin{center}
{\LARGE\bf\color{blus} 
Discrete Goldstone Bosons\\
}
\vspace{1cm}

{\bf Victor Enguita-Vileta$^{a}$, Belen Gavela$^{a}$, Rachel Houtz$^{b}$, Pablo Quilez$^{c, d}$, }\\[7mm]

{\it $^a$Departamento de Fisica Teorica, Universidad Autonoma de Madrid, \\ 
and IFT-UAM/CSIC, Cantoblanco, 28049, Madrid, Spain}\\[1mm]
{\it $^b$Institute for Particle Physics Phenomenology, Durham University, \\
Durham DH1 3LE, U.K.}\\[1mm]
{{\it $^c$Department of Physics, University of California, San Diego,
9500 Gilman Drive, La Jolla, CA 92093, USA
}}\\[1mm]
{{\it $^d$Deutsches Elektronen-Synchrotron DESY,
Notkestr. 85, 22607 Hamburg, Germany}}

\vspace{0.5cm}

\begin{abstract}

Exact discrete symmetries, if non-linearly realized, can reduce the ultraviolet sensitivity of a given theory.  
The  scalars stemming from spontaneous symmetry breaking are massive without breaking the discrete symmetry, and those masses are protected from divergent quadratic corrections. 
This is in contrast to   non-linearly realized continuous symmetries,  for which  the masses of pseudo-Goldstone bosons require an explicit breaking mechanism. The  symmetry-protected
 masses  and potentials of  those {\it discrete Goldstone bosons} offer promising physics avenues, both theoretically and in view of the blooming experimental search for ALPs. We develop this theoretical setup using invariant theory and focusing on the maximally natural minima of the potential.  For these, we show that typically a subgroup of the ultraviolet discrete symmetry remains explicit  in the spectrum, i.e. realized ``{\it \`a la Wigner}''; this subgroup can be either abelian or non-abelian. This suggests tell-tale experimental signals for those minima: at least two (three) degenerate scalars produced simultaneously if abelian (non-abelian), while the specific ratios of multi-scalar amplitudes provide a hint of the full ultraviolet discrete symmetry.  Examples of exact ultraviolet $A_4$ and $A_5$ symmetries are explored in substantial detail.

\vspace{0.3cm}

\end{abstract}

\vspace{1cm}
{\emph{Email:} \url{victor.enguita@uam.es}, \url{belen.gavela@uam.es}, \linebreak \url{pablo.quilez@desy.de}, \url{rachel.houtz@durham.ac.uk}}

\thispagestyle{empty}
\bigskip

\end{center}

\setcounter{footnote}{0}

\newpage
\tableofcontents

\newpage

\newpage
\section{Introduction } 
\label{sec:intro} 
How is it possible to obtain  scalar particles which are naturally much lighter than the overall scale of the theory? This question is  
at the heart of the electroweak hierarchy problem and of other open fine-tuning issues in particle physics. We address it here using exact hidden (aka spontaneously broken)  non-linearly realized discrete symmetries.  

Outstanding tensions and conundra of the Standard Model of Particle Physics (SM)  have  often  instead been confronted using  exact  hidden continuous symmetries.  
 The delicate issue is then that, for the theory to be realistic, the resulting massless Nambu-Goldstone bosons (GBs) must  somehow acquire  a small mass: 
 they must become pseudo Nambu-Goldstone bosons (pGBs). Examples of the latter 
include the axion that may solve the strong CP problem~\cite{Peccei:1977hh, Weinberg:1977ma, Wilczek:1977pj}, the Majoron if the neutrino masses are dynamical Majorona ones~\cite{Gelmini:1980re}, theories of extra-dimensions in which the Wilson loop around a compact dimension acts like an axion in four-dimensions~\cite{Witten:1984dg,Svrcek:2006yi}, or  the plethora of U(1) pGBs which often appear in string-inspired phenomenological models~\cite{Conlon:2006tq,Arvanitaki:2009fg}. Furthermore, the Higgs itself may be a pGB of some strong dynamics at very high energies~\cite{Kaplan:1983fs, Georgi:1984af,Dugan:1984hq}. 

Small masses of pGBs in continuous --non-linearly realized-- symmetries must arise from an {\it arbitrary}  explicit breaking of the continuous symmetry.
 Those masses are therefore necessarily proportional to the parameters of the explicit breaking mechanism, and in this sense the construction may be technically natural if those parameters are small.  In practice, this often leads to fine-tunings and new hierarchy problems, spoiling the naturalness and beauty of the solutions. Tensions  arise because, in most constructions, data require the pGBs masses to be extremely small compared to the overall scale of the theory. 

Exact but non-linearly realized discrete symmetries have the potential to  substantially ameliorate and restabilize that situation. The point is that discrete symmetries, when non-linearly realized, can render trivial the only possible  invariant quadratic  of the scalar fields  involved in the non-linear constraint.  The low-energy theory  is then protected from being quadratically sensitive to  very high --ultraviolet(UV)-- scales, even  in the presence of marginal couplings. Nevertheless, the discrete symmetry does admit scalar invariants of  higher dimension.  In other words, it permits non-zero masses for the scalars which implement the non-linear scenario, without breaking the discrete symmetry.  We will denote these massive scalar fields ``discrete Nambu-Goldstone bosons'' (dGBs). This name underscores the relation of these scalars with the pGBs of  the continuous group(s) in which  the discrete group can be embedded, to be developed below. 
 The relevant point is that  while the GBs of exact continuous symmetries are exactly massless, the dGBs  get masses even if the discrete symmetry is exact.

Furthermore,   the discrete invariance has the potential to strongly separate the dGB masses from the high-energy scale of the theory.  
 A key ingredient to assess the smallness of dGB masses is the dimensionality of the first scalar invariant in their --discrete-symmetry invariant-- potential.
  We will argue that (in a large class of UV completions of the mechanism) the higher its dimension, the smaller their coefficients are expected to be, and thus the smaller the dGB masses.

  It is illustrative to consider the  discrete symmetry groups that can be embedded in a given continuous group and their fate under explicit breaking. 
 Under explicit breaking  of the continuous group,
 either all its discrete subgroups undergo as well explicit breaking (as in the customary constructions of pGBs) or alternatively some discrete subgroup may remain untouched.  It is in this second option that that  preserved exact   ultraviolet (UV) discrete symmetry  provides the enhanced UV protection. Furthermore, it can also 
  provide the strong suppression of the pGBs masses, that is,  of the dGB masses, as we will develop.  From this point of view, the discrete symmetries under discussion  are not introduced {\it ad hoc} as an extra ingredient, but are part of a continuous symmetry.  Simply, the potential that breaks explicitly the continuous group is required to be invariant under one of its discrete subgroups, instead of being the most general one. The higher the mass dimension of the first term in the potential which breaks the continuous symmetry but is allowed by the discrete invariance,  the lighter the pGB can be.  A global continuous symmetry $G$ may then be viewed as an approximate symmetry induced by an exact discrete invariance $D$ (which may be a subgroup of $G$).\footnote{This view matches well with the perspective that no global symmetry is ultimately fundamental, as they are expected to be broken and destabilized at least by gravity~\cite{Hawking:1975vcx,Banks:1988yz,Coleman:1988cy,Kallosh:1995hi}, while exact discrete ones become more easily harmless, for instance via gauging.}  It is in this context that the concept of dGB's makes sense (i.e. they are a subset of the set of possible pGB's). 
   This may also have a bearing on the analysis of an immediate consequence of spontaneously broken discrete symmetries: the existence of topological defects, in particular domain walls which typically have important cosmological consequences.\footnote{The case of spontaneously broken $Z_2$  symmetry has been extensively studied in the literature~\cite{Kibble:1976sj,Zeldovich:1974uw,Vilenkin:1981zs}, and non abelian domain walls have been analyzed in Refs.~\cite{Kubotani:1991kw,Battye:1999eq,Riva:2010jm,Antusch:2013toa, Gelmini:2020bqg}). }

The fact that spontaneously broken discrete symmetries can improve the UV convergence of theories with pGBs was first argued in Ref.~\cite{Hook:2018jle}.   The case of 
 abelian discrete  $Z_N$ groups (for which $U(1)$ is  the obvious approximate continuous symmetry) has been explored in Refs.~\cite{Hook:2018jle,DiLuzio:2021pxd,DiLuzio:2021gos}, where it was shown that a lighter-than-usual QCD axion is a valid solution to the strong CP problem. Here we develop instead the much richer case of  {\it non-abelian UV discrete symmetries}, for which a first attempt appeared in Ref.~\cite{Das:2020arz}. The latter considered a scalar triplet of $A_4$, showing that no low-energy physical effects can stem from the $A_4$  invariant quadratic in the scalar fields,  and instead  the first relevant invariant is cubic.  Our analysis of this particular scenario is different and provides novel results: we will show that  explicit symmetries may survive in the low-energy spectrum and interactions.  Next, the bulk of the paper explores various new realizations of  $A_4$, $A_5$ and other discrete symmetries. One general question to also be addressed here is how to naturally obtain very high dimensionality for the first discrete-symmetry scalar invariants, that is, how to obtain even further suppressed dGB masses.

   As theoretical tool  we will use Effective Field Theory (EFT) and invariant analysis rather than commit to specific UV models. The basis of the game is to find the minima of the discrete symmetry invariants  that can be built  out of the scalar representation(s) at hand. The most general potential will be an arbitrary function of all such possible invariants. We focus  here on the natural minima of the potential, that is, the minima  whose locations are less or not at all dependent of the values of the parameters of the potential~\cite{michel1971properties, Cabibbo:1970rza}. Nevertheless, for illustrative purposes we will also show in detail the  cancellation of UV divergences among different diagrams, within two specific UV complete models.

 In addition,  the identification of which abelian or  non-abelian discrete symmetries remain exact at low-energies  (i.e. realized {\it \`a la Wigner} in the spectrum) for the maximally natural minima will be shown to have a major bearing  on the possible tell-tale experimental signals of the mechanism. Specifically, the degeneracies of the pGB spectra and the predicted ratios of multi-scalar amplitudes will be analyzed. 
 
The structure of the paper can be easily inferred from the Table of Contents.
  
\newpage
\section{Non-linearly realized discrete symmetries}
We will analyze in this work scalar fields which belong to irreducible real representations of discrete symmetries.   Consider a generic scalar field $\Phi$ in an irreducible $m$-dimensional real representation of the discrete symmetry group $D$,  $\Phi\equiv (\phi_1, \phi_2....... \phi_m)$. A non-linearity constraint  can be expressed as the requirement that, at low energies, the 
fields satisfy the quadratic restriction
\begin{equation} 
\Phi^T \Phi = \phi_1^2 + \phi_2^2 + \ldots + \phi_m^2 = f^2\,,
\label{non-linear}
\end{equation}
where $f$ is a constant with mass dimension one. Eq.~(\ref{non-linear}) reduces by one the number of independent degrees of freedom of the low energy theory.  These are $m-1$ spin zero particles, which would be the massless GBs  of spontaneously broken continuous symmetries, and are the dGBs of discrete symmetries, which will be shown to be massive while the discrete symmetry remains exact.

 Let us consider the model-independent tool of EFT.  
 The dGB scale is defined by $f$ in Eq.~(\ref{non-linear}), and its associated scale  $\Lambda\sim 4\pi f$ will weigh down effective operators of mass dimension larger than four, for instance 
 all those containing  momentum insertions other than kinetic energy terms. 
 Naive dimensional analysis (NDA)~\cite{Manohar:1983md,Jenkins:2013sda,Gavela:2016bzc} will be used to formulate the EFT. To this end, the non-linear constraint above can be expressed in terms of a dimensionless functional $U$, which is a function of the $m-1$ low-energy degrees of freedom,
\begin{equation}
U\equiv \frac{\Phi}{f}\,, \qquad  \text{with} \quad U^T U =1\,,
\label{non-linear-2}
\end{equation}
and one can apply the CCWZ construction~\cite{Coleman:1969structure1,Callan:1969structure2} to parametrize the dGB inside $U$. This function is the real equivalent of the customary pion field parametrization of non-linearly realized continuous symmetries.\footnote{For instance, in spontaneously global $SU(N)$  the function $U= \exp{i\vec{\pi_i}\lambda_i/f}$ where $\pi$ denote the pion fields, $\lambda_i$ the group  generators and $f$ de pGB scale, is customarily used,  i.e. $U^\dagger U=1$.} 
  A possible choice  for   $U$ reads 
 \begin{equation}
 f\,U\equiv 
    \Phi(\pi_1,\dots ,\pi_{m-1}) = 
    \left(
    \begin{array}{c} 
        \phi_1 \\ \phi_2\\ \vdots \\ \phi_m
    \end{array}\right)=\exp\left[\frac{1}{f} \left(
    \begin{array}{ccccc}
        0       & \dots     & 0       & \pi_1 \\
        \vdots       & \vdots     & \vdots       & \vdots \\
        0       & \dots     & 0       & \pi_{m-1} \\ 
        -\pi_1  & \dots &-\pi_{m-1} & 0
    \end{array}\right)
    \right]\left(
    \begin{array}{c} 
        0 \\ 0 \\\vdots \\ f
    \end{array}\right)\,,
    \label{eq:pions  generic definition}
\end{equation}
where $\pi_1\dots \pi_{m-1}$ denote the $m-1$ physical dGB degrees of freedom. Whatever the parametrization, the important point is that we focus on the dynamics of the dGB degrees of freedom including the impact of the possible terms in their symmetry-invariant scalar potential. The dGB Lagrangian can be parametrized as  
\begin{align}
\mathcal{L}_{\text{dGB}} &=  f^2 \Lambda^2\,\tilde{\mathcal{L}} \left(\frac{\partial}{\Lambda}, U, c_\alpha, L_\alpha \right)\,= \,   \frac{\Lambda^4}{(4\pi)^2}\,\tilde{\mathcal{L}} \left(\frac{\partial}{\Lambda}, U, c_\alpha, L_\alpha \right)\,,
\label{LdGB}
\end{align} 
where $\Lambda =4\pi f$ has been assumed,\footnote{Strictly speaking $\Lambda \le4\pi f$~\cite{Manohar:1983md}.} the tilde signals here and in what follows dimensionless functions, and  the compact notation recalls that derivative terms are suppressed by powers of $\Lambda$ in NDA, while $U$ does not scale. The dimensionless $c_\alpha$ and $L_\alpha$ coefficients will weigh down respectively the invariants in the dGB potential and terms involving derivatives,  i.e. 
\begin{equation}
\mathcal{L}_{\text{dGB}} =  \frac{f^2}{4} \,   \left [ \partial^\mu U^T  \partial_\mu U  \right] -   V_{\text{dGB}}  + {\mathcal L}_{\slashed{\partial}}\,,
\label{LdGB-2}
\end{equation}

where  $V_{\text{dGB}}$ denotes the potential
 \begin{equation}
 V_{\text{dGB}}\equiv {f^2} \Lambda^2\, \widetilde{V}(U, c_\alpha)= \frac{\Lambda^4}{(4\pi)^2} \,\widetilde{V}(U, c_\alpha) \,,
  \label{Vreduced-general}
  \end{equation} 
  with $\widetilde{V}(U, c_\alpha)$ 
  a dimensionless function. All terms containing $\partial_\mu U$ components --except the kinetic term--  are encoded in ${\mathcal L}_{\slashed{\partial}}$, 
\begin{equation}
{\mathcal L}_{\slashed{\partial}}=   L_1\,\left[\partial_\mu U^T \partial^\mu U\right]^2 +  L_2\,\left[\partial_\mu U^T \partial_\nu U\right] \,\left[\partial^\mu U^T \partial^\nu U\right]+\cdots 
\label{quartic-derivative}
\end{equation}
where dots stand for both higher order  terms built up purely of $U$ derivatives, and mixed terms made out of both $U$ derivatives and non-derivative insertions of $U$ (e.g. $ \prod_{{\scriptscriptstyle i<j<k}} U_i \partial_\mu U_j\partial^\mu U_k $). For the purely derivative terms, NDA suggests coefficients $\hat{L}_\alpha\sim {\mathcal O}(1)$ or slightly smaller, related to the customary $L_i$ above by powers of $4\pi$, e.g. $\hat{L}_1\equiv (4\pi)^2L_1,\,\hat{L}_2\equiv (4\pi)^2L_2 $.\footnote{
  In NDA the first operator in Eq.~(\ref{quartic-derivative}) would be written as $ \hat{L}_1\,\Lambda^4/{(4\pi)^2}\,\left[(\partial U^T/{\Lambda} )(\partial U/\Lambda)\right]^2= \hat{L}_1/{(4\pi)^2}\,\left[\partial U^T \partial U\right]^2$, from which the relations among the coefficients follow.  The value $L_1\sim 10^{-2}-10^{-3}$ measured from data is thus consistent with $\hat{L}_1\sim 1$, as expected because NDA is  designed precisely so that the the $4\pi$ factors from radiative corrections are consistent with NDA coefficients of order one.}     
 The coefficients of mixed scalar-derivative terms in ${\mathcal L}_{\slashed{\partial}}$, as well as those in the dGB potential $c_\alpha$,  are instead expected to be strongly suppressed because of the constraints stemming from the UV discrete symmetry, though, as argued further below.
 
The  analysis of the terms involving derivatives  will be deferred to subsequent work (except for qualitative comments in the phenomenological analysis further below). We will  focus next instead in the complete  scalar potential invariant under the  UV 
 discrete symmetry, 
 as it sources the dGB masses. Several representative discrete groups and representations will be considered.

   \vspace*{20pt}
   
   \section{Scalars in a triplet of \texorpdfstring{$A_4$}{} }
   \label{triplet-A4}   
   In the example in Ref.~\cite{Das:2020arz}, a triplet  of $A_4$ was studied, $\Phi \equiv(\phi_1, \phi_2, \phi_3)^\intercal$.  This case is analyzed it in detail in this section, where we will also expand on the analysis tools used throughout this paper. Details on the group theory considerations and subsequent analysis are provided in an ancillary Mathematica file. 

\subsection{The invariants}
For a given representation $\Phi$ of the  UV discrete symmetry, the most general potential $V(\Phi)$  can be expressed as a function   of all possible invariants built out of that representation. Invariant theory of finite groups simplifies our task by noting that, for the discrete groups we will study, the invariants form a polynomial ring~\cite{Chevalley:1955invariants,Shephard:1954finite}. The number of independent invariants must match the number of independent degrees of freedom of the representation.  For an $A_4$ triplet of real scalars, the potential  can be indeed constructed as a function of the following three invariants:
\begin{align}
\mathcal{I}_2 =&  \, \sum \phi_i^2  =   \phi_1^2 + \phi_2^2 + \phi_3^2\,,\hfill \label{eq:invariant_2}\\
\mathcal{I}_3 =&  \, \prod_{{\scriptscriptstyle i<j<k}}^{} \phi_i\phi_j\phi_k
 =\phi_1\phi_2\phi_3\,,\hfill \label{eq:invariant_3}\\
\mathcal{I}_4 =&  \, \sum \phi_i^4  = \phi_1^4 + \phi_2^4 + \phi_3^4\,\,,\hfill \label{eq:invariant_4}
\end{align}
which are often called \emph{primary invariants}, plus  a fourth --{\it secondary}-- invariant of mass-dimension six which is a non-polynomial function of these three. The equation which relates  secondary and primary invariants is often referred to in the literature as a \emph{syzygy}~\cite{hilbert1890ueber,cummins1988polynomial}, which in this case reads 
\begin{equation}
	-4\,\I_6^2=-2\,\I_4^3 + 5\,\I_4^2\,\I_2^2 - 4\,\I_4\,\I_2^4 
	+ 36\,\I_4\,\I_3^2\,\I_2  + \I_2^6 - 20\,\I_3^2\,\I_2^3 + 108\,\I_3^4\,.
	\label{eq:A4:invs:syzigy}
\end{equation}
This sixth-order invariant coincides in fact  with the  determinant of the Jacobian for  the three primary invariants above (up to a trivial normalization),  
\begin{equation}
	\operatorname{det} \J=8\,\I_6(\phi) =  (\phi_1^2 -\phi_2^2)(\phi_1^2 -\phi_3^2)(\phi_2^2 -\phi_3^2)\,,
\end{equation}
where
\begin{equation}
 \J\equiv \frac{\partial (\I_2,\I_3,\I_4)}{\partial (\phi_1,\phi_2,\phi_3)}\,.
 \label{JacobianA4}
\end{equation}
It follows from the above that the understanding of the behavior of $\mathcal{I}_2$, $\mathcal{I}_3$ and $\mathcal{I}_4$ will provide essential information for the characterization of 
the general potential $V(\phi_1, \phi_2, \phi_3)$.

At energies $E< \Lambda$, the non-linear constraint in Eq.~(\ref{non-linear})  implies 
\begin{equation}
\I_2 |_{(|\Phi|^2 = f^2)}= f^2 \, U^T U= \phi_1^2 + \phi_2^2 + \phi_3^2 =f^2\,,
\label{equationf-A4-3}
\end{equation} 
 a constraint which  is crucial  for the UV (in)sensitivity of the low-energy effective theory to quadratic divergences:  the only possible quadratic scalar invariant becomes trivial at low energies. In other words, no quadratic correction to the dGB masses can arise  at any order from physics dynamics at scales  higher than $ \Lambda$.\footnote{Other non-linearity constraints that also preserve the $A_4$ symmetry are possible~\cite{Feruglio:2017spp}. Alternative choices  include  $\phi_1^2 +\exp{(i \omega})\phi_2^2 + \exp{(i 2\omega)} \phi_3^2=0$, where $\omega$ is a phase, or $\phi_1^4 +\phi_2^4 +  \phi_3^4= \text{cte}$. Those choices would not provide protection from quantum quadratic divergences, though. We thank F. Feruglio for this comment.}

  \subsection{The potential} 
The
most general analytic 
 potential   can be expressed as a polynomial of the primary and secondary invariants. Taking into account that $\I_2$ does not contribute to the potential for the dGB, and that even powers of the secondary invariant $\I_6$ can be expressed in terms of the others via the syzigy in \cref{eq:A4:invs:syzigy}, the most general analytic potential for the dGB reads
\begin{align}
V_{\rm dGB}= {f^2} \Lambda^2\,\sum_{a,b,c}^\infty \hat{c}_{abc}\, \left[\frac{\I_3}{f^3}\right]^a\, \left[\frac{\I_4}{f^4}\right]^b\, \left[\frac{\I_6}{f^6}\right]^c \quad \text{ with } a,b\in \mathbb{N}  \text{ and } c= 0,1 \,,
\label{Eq:MostGeneraldGBPotential1}
\end{align}
where the $\hat{c}_{abc}$ coefficients are defined in NDA,\footnote{Recall that in NDA, the weight of fields and couplings is: an overall factor of the Lagrangian $\Lambda^2 f^2= \Lambda^4/(4\pi)^2$, and scalings $\partial/\Lambda$, $4\pi \phi/\Lambda$ for a generic scalar field,  $4\pi A/\Lambda$ for a gauge field $A$,  $4\pi \Psi/(\Lambda)^{3/2}$ for a fermion field.}
 and where the $\I_n/f^n$ dependence stems from the $U$-dependence in Eq.~(\ref{Vreduced-general}). Note that in this case the complete $V_{\text{dGB}}$ potential -i.e. {\it including all possible invariants of any order}--   is a function\footnote{This function may be  non-polynomial  if any $\hat c_{a,b,1}\neq0$, i.e. if \cref{Eq:MostGeneraldGBPotential1} includes the secondary invariant.} of only two non-trivial invariants ($\I_3$ and $\I_4$) after the non-linearity constraint  and the syzygy are taken into account, which is consistent with the fact that a real triplet scalar sources two dGB degrees of freedom.

  The main results in this paper apply to that complete potential, that is, they do not rely on any expansion or truncation of the potential and we will focus on its most natural minima. From this perspective,  the reader can skip the illustrative discussion that comes next and go directly to Sec.~\ref{natural-extrema} if wished. 
   
    It can be argued, though,  that the lower dimension terms are expected to dominate the dGB effective potential of spontaneously broken discrete symmetries~\cite{Das:2020arz}, as we proceed to discuss next.  For illustrative purposes, let us expand Eq.~(\ref{Eq:MostGeneraldGBPotential1}) as follows 
    \begin{equation}
    V_{\text{dGB}} = {f^2} \Lambda^2\, \big[\hat{c}_3 \,\frac{\I_3}{f^3} + \hat{c}_4 \,\frac{\I_4}{f^4} + \hat{c}_6 \frac{\I_6}{f^6}  + \hat{c}_7 \frac{\I_7}{f^7} \cdots\big]\,,
  \label{Vreduced}
  \end{equation}
   which can be rewritten in a more customary notation as 
     \begin{equation}
     V_{\text{dGB}}
     =  c_3  \Lambda\,{\I_3} + c_4 \,\I_4 + c_6 \,\frac{\I_6}{\Lambda^2} + c_7 \,\frac{\I_7}{\Lambda^4} + \cdots\,,
   \label{Vreduced-nohats}
  \end{equation}
 where 
 $c_n\equiv (4\pi)^{n-2}\hat{c}_n$.  In these equations any $\I_{n>6}$  can be expressed in terms of lower dimensional ones, e.g. $\I_7= \I_3\,\I_4$, and dots indicate terms with higher dimension invariants.  The relation with Eq.~(\ref{Eq:MostGeneraldGBPotential1}) is given by $\hat{c}_3= \hat{c}_{100}$,  $\hat{c}_4= \hat{c}_{010}$, $\hat{c}_6= \hat{c}_{011}$,  $\hat{c}_7= \hat{c}_{110}$, etc. 
 
 Were all $\hat{c}_i$ coefficients of the same order, all terms in the series would contribute on equal grounds to the dGB potential.  In particular, dGB masses will be shown to stem from {\it all} terms in the potential which do not contain $\I_6$.  For generic $\hat{c}_i\sim\mathcal{O}(1)$ or slightly smaller, the resulting pion masses would be  $\sim \Lambda$ (instead of much smaller as expected for pGBs).   This is as expected for generic fields in a theory with a given overall scale, unless a symmetry protects the size of the mass terms, as we will analyze next for the case under study. 
    
  Indeed, while NDA is designed so that the coefficients on the EFT are expected to be ${\cal{O}}(1)$ or slightly smaller, this is not yet the case for the $\hat{c}_\alpha$  defined above, though, because a stronger source of suppression can be expected in the presence  of an exact discrete invariance.  Its enforcement in the couplings of $\Phi$ to the BSM fields (which are dynamic at energy scales above $\Lambda$) typically leads to a particular structure of the EFT couplings at  lower energies, for instance a power-like dependence on those BSM couplings. The situation is somewhat alike to that for little Higgs models~\cite{Georgi:1974yw,Georgi:1975tz,Arkani-Hamed:2001nha,Schmaltz:2005ky} or clockwork constructions~\cite{Kim:2004rp,Choi:2014rja,Choi:2015fiu,Kaplan:2015fuy,Giudice:2016yja,Ahmed:2016viu}: a chain of interactions is required to enforce invariant operators so that, while the individual couplings of the BSM theory may be slightly smaller than one, the effective operator coefficient is a power of them, ensuring strong exponential suppressions. A similar situation can be expected in the presence of discrete symmetries. Indeed, the enforcement  of the UV exact discrete symmetry in the couplings of $\Phi$ to the BSM fields  results in a typical dependence of the form 
 \begin{equation}
\hat{c}_n \,\sim \,\epsilon^n\,,
\label{cn}
\end{equation} 
where $\epsilon$ is a small quantity $\epsilon <1$.  For instance, in the scenario with an UV abelian discrete symmetry  explored in Refs.~~\cite{Hook:2018jle,DiLuzio:2021pxd} it was identified $\epsilon\sim (m_u/m_d)$, where $m_u$ and $m_d$ denote respectively the up and down quark masses, and  the dGB mass exhibited a $\sim (m_u/m_d)^n$   suppression.

Such  $n$-dependent exponential suppression via a small parameter has also been shown to appear in specific UV completions of  non-abelian discrete invariance, see Ref.~\cite{Das:2020arz}. In the latter, a real scalar triplet of $A_4$ is considered, $\Phi$, with  Yukawa couplings $y$  to exotic heavy fermions with mass $M\ge \Lambda$: $n$ is then the number of exotic fermion exchanges among $\Phi$ fields needed to obtain an effective scalar operator invariant under the discrete symmetry, and in consequence 
the effective operator coefficients obey  $\hat{c}_n \sim y^n$. More in detail, the NDA analysis of that case indicates a potential of the form
  \begin{equation}
  V= \frac{M^4 }{(4\pi)^2} \sum_n  \left(  \hat{y} \frac{4\pi\Phi}{M}   \right)^n =  \frac{M^4 }{(4\pi)^2} \sum_n  \left(  y \frac{\Phi}{M}   \right)^n = \Lambda^2f^2  \sum_n y^n \left(  \frac{\Lambda}{M}\right)^{n-4} \left( \frac{\Phi}{\Lambda}   \right)^n\,,
  \label{eq:M-pcounting}
  \end{equation}
  with $y\equiv  4\pi\hat{y}$. It follows that in this example $\hat{c}_n$ in Eq.~(\ref{Vreduced}) is given by
  \begin{equation}
\hat{c}_n \sim y^n\,\left(  \frac{\Lambda}{M}\right)^{n-4}\,,
\label{c_n}
\end{equation}
which shows that  all operator coefficients   are suppressed by powers of the Yukawa coupling as $y^n$ as far as $y<1$, and the operators with $n\ge 4$ are additionally suppressed by the ratio of scales whenever $\Lambda<M$, the larger $n$ the stronger the suppression. For the particular case of scenarios which admit cubic scalar invariants --as in the case studied in this section-- it also follows that $y<  4\pi\left(  {\Lambda}/{M}\right)^{1/3}$ is necessary and sufficient  to obtain dGB masses  $m_{\text{dGB}}$ smaller than the EFT scale,  i.e. $m_{\text{dGB}}^2 < \Lambda^2$, as desired.  This requirement does not apply  in other scenarios to be discussed below, where the UV discrete symmetry will forbid cubic scalar invariants. The above is only one specific example of how the (hierarchical) suppression of the coefficients of the EFT can arise when an exact discrete symmetry is present. The suppression by the scale that generates the invariant operators, $M$, can also be of interest in cases explored in later sections, in which the leading invariant is a non-renormalizable operator. If $M$ is even mildly larger than $\Lambda$, then $\hat c_n$ naturally inherits a large suppression without relying on $y<1$. Whether the scale $M$ is separated from or coincides with $\Lambda$ is a detail of the UV model building, however, and from here on we will keep our discussion more general.

An important question is that of the quantum stability of small $\hat{c}_n$ coefficients. They are indeed necessarily stable if assumed small, because in the limit in which they are taken to zero the Lagrangian exhibits a larger --continuous-- symmetry, see Sec.~\ref{continuous}. 

In summary, a hierarchical ordering with $n$ of the coefficients of the invariant scalar operators may be naturally at play in the presence of  a UV discrete symmetry which remains exact at all scales --albeit spontaneously broken. The consequence is that the lowest order scalar invariants should provide a good approximation to the whole EFT potential. This property also underlies the interest of scenarios in which the lowest-order non-trivial scalar invariant corresponds to very large $n$, for which an even stronger suppression of the dGB masses can be expected. A simple scenario of this type will be presented in the next section.

\subsubsection{Embedding in a continuous group: \texorpdfstring{$SO(3)$}{}}
 \label{continuous}  
  It is illuminating to consider the embedding of the discrete groups under consideration in  continuous ones. In a nutshell, the results in this paper could be presented for pedagogical reasons as a study of spontaneously broken continuous symmetries supplemented with an explicit breaking potential  which is invariant under some of the discrete subgroups of the continuous symmetry, instead of being  the customary most general one.

From this point of view, all terms in the dGB potential  preserve the discrete symmetry but break explicitly the continuous symmetry of the embedding group. 
 That is, all coefficients $\hat{c}_\alpha$ in     $V_{\text{dGB}}$ (Eq.~(\ref{Vreduced}))  are parameters that break that continuous symmetry.\footnote{Analogous considerations apply to the coefficients of the mixed derivative-non-derivative operators.}  From this perspective, it is natural that they are smaller than one, because they herald the explicit breaking of a continuous global symmetry, alike to the usual breaking parameter of chiral Lagrangians which is proportional to the quark masses: $m_q/\Lambda <1$.  In consequence,  the putative GBs acquire small masses.  The crucial advantage of these dGBs with respect to generic pGBs is that, even if  there are marginal operators breaking the continuous symmetry in the UV, their contribution to the low-energy dGB potential will amount to contributions only to $c_{n>2}$ coefficients: no quadratic divergences can arise through radiative contributions. This will hold as long as the first invariant that breaks the continuous symmetry while respecting the discrete one has dimension larger than two, which is precisely what happens in the UV discrete symmetry scenarios studied here.  In this context, whenever the dGB potential is solely generated via these suppressed radiative corrections, one can argue that the exact discrete symmetry is inducing the existence of an approximate continuous one.  In other words, the restrictions
  that discrete symmetries impose on the allowed operators of the Lagrangian  induce the existence of approximate continuous symmetries.

  The smallest continous group that can contain $A_4$ as subgroup with a scalar triplet is $SO(3)$, see  Table~\ref{table:finite_groups_SO3}, with    
   $\Phi$ being thus  a triplet of both groups.  The only possible pattern  of spontaneous symmetry breaking (SSB) of $SO(3)$ with scalar fields in the $\boldsymbol{3}$ representation is  $SO(3)\to SO(2)$. 
\begin{table}[h!]\centering
\begin{align*}
\begin{array}{|c|c|c|c|c|c|c|}
\hline
\multicolumn{7}{|c|}{\text{Finite subgroups of $SO(3)$}}\\
\hline\hline
 & Z_2 & Z_{N>2} & D_{2n} & A_4 & S_4 & A_5\\
\hline
\multirow{6}{*}{\begin{sideways}\text{Irreducible reps.}\end{sideways}} &1    & 1 & 1  & 1       & 1       & 1     \\
  &\ 1' & 2 & 2  & \ 1'    & \ 1'    & 3     \\
  &     &   &    & \ \ 1'' & \ \ 1'' & \ 3'  \\
  &     &   &    & 3       &     2   & 4     \\
  &     &   &    &         & 3       & 5     \\
  &     &   &    &         & \ 3'    &       \\
\hline
\end{array}
\end{align*}
\caption{Finite subgroups of $SO(3)$ and their irreducible representations~\cite{Etesi:1996urw}.}
\label{table:finite_groups_SO3}
\end{table}
This would source two GB degrees of freedom below the SSB scale, two ``pions''.  The only possible quadratic term invariant under the $A_4$ discrete symmetry, $\I_2$, is {\it also}  $SO(3)$ invariant and it will yield no contribution to the pion masses. In contrast,  $\I_3, \I_4, \I_6...$ are invariants of $A_4$ only: they break explicitly the continuous $SO(3)$ symmetry. That is, by requiring that the potential that  breaks explicitly $SO(3)$ is invariant under one of its discrete subgroups, the quadratic sensitivity to high scales is absent, as advertized, while the  pions of the continuous theory acquire a mass from operators with dimension three~\cite{Das:2020arz} (with coupling $c_3$) and larger.

A pertinent question is whether all finite groups allow an embedding in a continuous group.  Indeed, any finite group $D$ is isomorphic to a subgroup of  the symmetric group of $m$ elements, $S_m$, where $m$ is precisely the order of $D$\,\cite{cayley1854vii}. Therefore, $D$ is also a subgroup of the continuous Lie groups GL$_m(\mathds{R})$ or GL$_m(\mathds{C})$ and can in fact be realized as a subgroup of $O(m)$, and thus also as a subgroup of $SO(2m)$. There are, on the other hand, cases in which the smallest embedding continuous group turns out to be smaller. A familiar example is that of the alternating group of $4$ elements, $A_4$, which is obviously a subgroup of the permutation group $S_4$: the smallest continuous group which contains $A_4$ is then $SO(3)$.  In the presence of spontaneous symmetry breaking, both the discrete symmetry and its continuous  embedding group are spontaneously broken.
 Our analysis, however, does not rely on a specific choice of continuous group embedding.

\subsubsection{Understanding the nature of quantum quadratic protection}
\label{UVprotection}

In order to better grasp the origin of the absence of quadratic sensitivity of dGB, it is interesting to consider specific examples of UV complete models  where the cancellation of the quadratic divergences becomes explicit.

\subsubsection*{A UV model with exotic fermions}
Let us consider a model with a fundamental scalar triplet of $A_4$, $\Phi$, and a triplet of fermions, $\mathbf{\Psi} \equiv(\Psi_1, \Psi_2, \Psi_3)^\intercal$ (which was first studied in Ref.~\cite{Das:2020arz}). There are two possible $A_4$-symmetric Yukawa interactions among the scalars and the fermions\footnote{See Ref.~\cite{ishimori2010non} for details on how to construct $A_4$ singlets} which can be written as: 
\begin{equation}
\begin{split}
\mathcal{L}_{\text {int }}
&=\left[y_{G}\left(\hspace{-5pt}\begin{array}{c}
{[\bar{\Psi}_{2}, \Psi_{3}]} \\
{[\bar{\Psi}_{3}, \Psi_{1}]} \\
{[\bar{\Psi_{1}}, \Psi_{2}]}
\end{array}\hspace{-5pt}\right)
+y_{\slashed{G}}\left(\hspace{-5pt}\begin{array}{l}
\{\bar{\Psi}_{2} ,\Psi_{3}\}\\
\{\bar{\Psi}_{3} ,\Psi_{1}\}\\
\{\bar{\Psi}_{1} ,\Psi_{2}\}
\end{array}\hspace{-5pt}\right) \right]^\top \cdot\, \Phi = \, \\
&= y_{G}\left(\left[\bar{\Psi}_2, \Psi_3\right] \phi_1+\left[\bar{\Psi}_1, \Psi_3\right] \phi_2+\left[\bar{\Psi}_1, \Psi_2\right] \phi_3\right)\\
&\hspace{0.3\textwidth}+\ybreak\left(\{\bar{\Psi}_2, \Psi_3\} \phi_1+\{\bar{\Psi}_1, \Psi_3\} \phi_2+\{\bar{\Psi}_1, \Psi_2\} \phi_3\right)\,,
\label{Eq: Yukawas hook model}
\end{split}
\end{equation}
which for small Yukawa couplings presents an approximate continuous $G\equiv SO(3)$ symmetry.
All interactions  in this Lagrangian have mass-dimension four with a dimensionless coefficient, i.e. they are marginal couplings, and one could naively assume that these two Yukawa couplings  would generate UV sensitive pion masses. Nevertheless: 
\begin{itemize}
\item The term proportional to $\yG$ preserves $SO(3)$, and thus cannot source a potential for the GBs of the theory due to the Goldstone theorem. In other words, any scalar potential would need to be a function of the $SO(3)$ primary invariant $\I_2=\phi^T\phi$ which does not depend on the pions and this leads to a vanishing pion potential.
\item  The term proportional to $\ybreak$ explicitly breaks $SO(3)$ but is invariant under $A_4$. This term does source a potential and masses for the pions of the theory, but it does not lead to quadratic sensitivity to scales heavier than that of the dGB Lagrangian.
\end{itemize}
\begin{figure}[h]
  \centering\begin{minipage}{.44\textwidth}
        \centering \includegraphics[width=\linewidth]{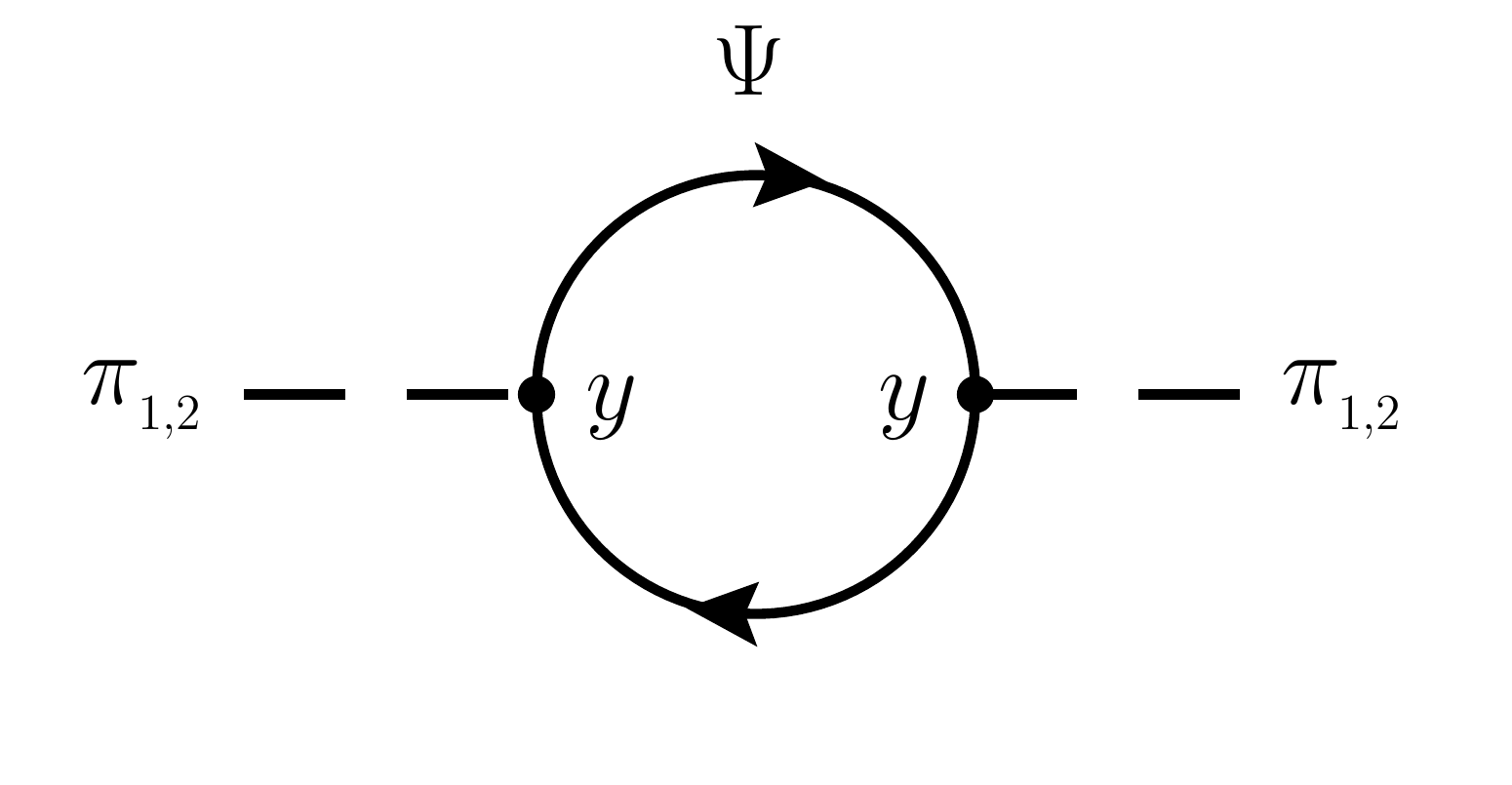}
     \end{minipage}\begin{minipage}{.01\textwidth} {\large $+$}\end{minipage} 
    \begin{minipage}{.44\textwidth}
        \centering \includegraphics[width=\textwidth]{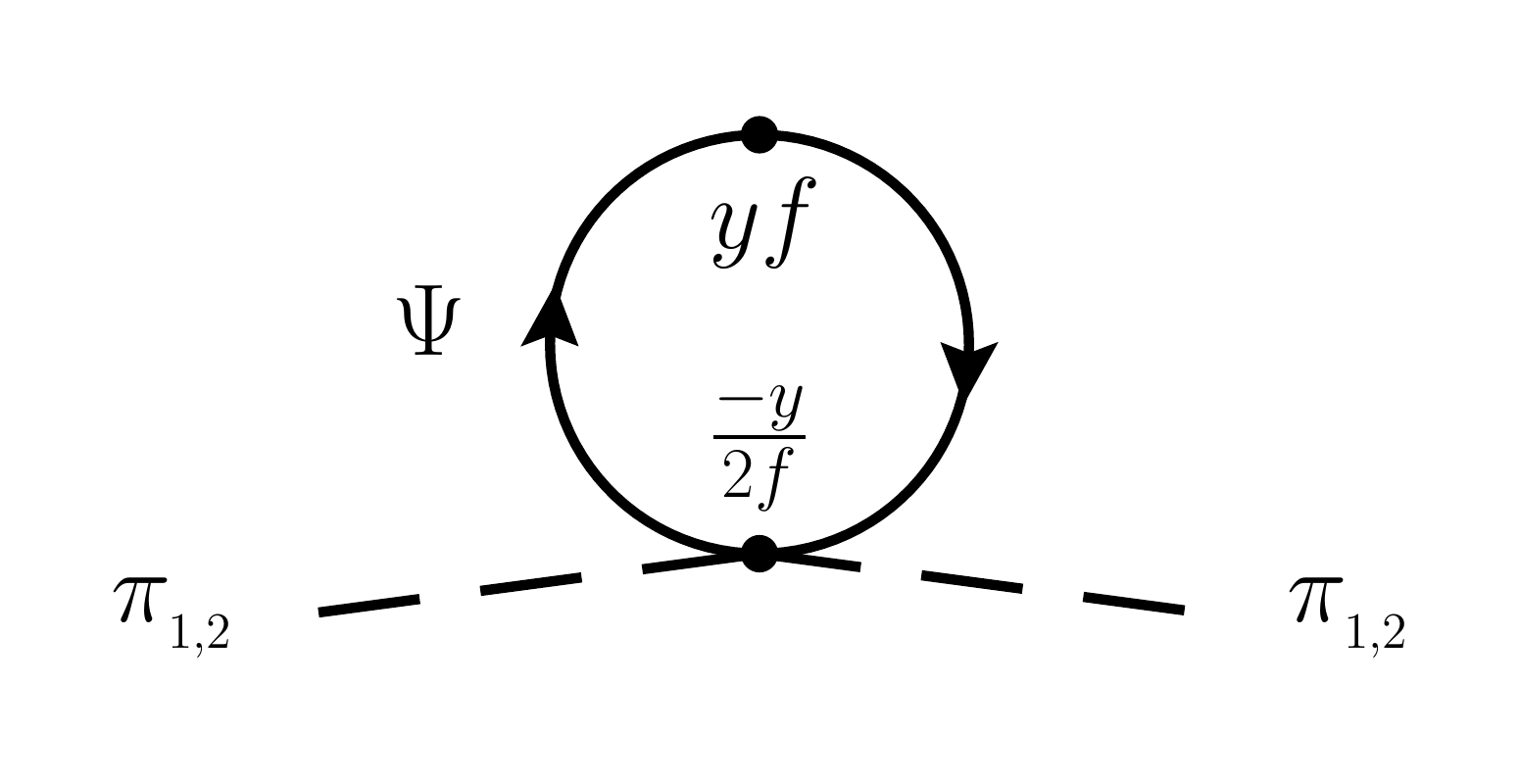}
     \end{minipage} 
  \caption{\small Diagrams {\it a priori } sourcing quadratic divergences, with $y=\yG$ in the $SO(3)$ symmetric case in Eq.~(\ref{SO3sym}), and $y=\ybreak$ in the SO(3)-breaking but $A_4$-preserving case in Eq.~(\ref{A4sym}).   Only the diagram on the left for $\pi_1$ is present instead for the explicit breaking in Eq.~(\ref{nosym}) with $y=y_{\slashed{A}_4}$. }
 \label{Fig:CancelationHook}
\end{figure}

 In order to illustrate in detail the last statement, let us compute explicitly the possible loops induced by $\yG$ and $\ybreak$ and obtain the cancellation of the divergences in both cases.  Expanding to second order in the pion fields,~\footnote{$ \phi_{1}\simeq \pi_{1}, \phi_{2}\simeq \pi_{2}, \phi_{3}\simeq f\big[1-\frac{1}{2} {(\pi_{1}^{2}+\pi_{2}^{2})}/{f^{2}}\big]$.} the $SO(3)$-symmetric Yukawa interactions  read 
\begin{multline}
 \mathcal{L}_{\text {int, G }}  =\yG\, \pi_{1}\left(\bar{\Psi}_{2} \Psi_{3}-\bar{\Psi}_{3} \Psi_{2}\right)+\yG\, \pi_{2}\left(\bar{\Psi}_{3} \Psi_{1}-\bar{\Psi}_{1} \Psi_{3}\right) \\
 +\yG\, f\left(1-\frac{1}{2} \frac{\pi_{1}^{2}+\pi_{2}^{2}}{f^{2}}\right)\left(\bar{\Psi}_{1} \Psi_{2}-\bar{\Psi}_{2} \Psi_{1}\right) + \O(\pi^3)\,, \label{SO3sym}
 \end{multline}
 and a very similar expression holds for the $SO(3)$-breaking  but $A_4$-preserving Yukawa couplings:
\begin{multline}
\mathcal{L}_{\text{int. }{\slashed{G}}} 
= \quad   y_{\slashed{G}}\Big[ \pi_{1}\left(\bar{\Psi}_{2} \Psi_{3}+\bar{\Psi}_{3} \Psi_{2}\right)+ \pi_{2}\left(\bar{\Psi}_{3} \Psi_{1}+\bar{\Psi}_{1} \Psi_{3}\right) \nn\\
 + f\left(1-\frac{1}{2} \frac{\pi_{1}^{2}+\pi_{2}^{2}}{f^{2}}\right)\left(\bar{\Psi}_{1} \Psi_{2}+\bar{\Psi}_{2} \Psi_{1}\right)\Big] + \O(\pi^3)\,.\label{A4sym}
\end{multline}
 The relevant Feynman diagrams for both Yukawa Lagrangians are depicted in Fig.~\ref{Fig:CancelationHook}.  We can now compute the \emph{a priori} divergent loop contributions to the mass of the pions: these are guaranteed by symmetry to vanish for the $\yG$ contributions, but they turn out  to vanish  for the $y_{\slashed{G}}$ contributions as well. Indeed, in both cases,  the correlation between the  $\O\left(\pi_i \Psi\Psi\right)$ interaction and the $\O\left(\pi_i^2 \Psi\Psi\right)$ interaction  implies that the quadratic divergences stemming from those two Feynman diagrams combine as 
 \begin{align}
\delta m_{\pi_{1,2}}^2\propto \frac{1}{2}y\Lambda^2 - \frac{y}{2f}\, y f\Lambda^2 =0\,,
\end{align}
where $y$ is either $\yG$ or  $y_{\slashed{G}}$.   In the case of the ($SO(3)$-breaking ) $y_{\slashed{G}}$ terms, the cancellation of quadratic divergences is a consequence of the unbroken discrete $A_4$ symmetry. This property can be traced back to the fact that the first $A_4$-invariant that breaks the continuous symmetry is of dimension three and thus requires the insertion of at least three Yukawa couplings~\footnote{If the fermions have a bare mass term $m_\psi$, the mass of the dGBs is $m_{\pi_{1,2}}^2\propto \ybreak^3 \,m_\psi f$. Otherwise an enhanced $A_4\times Z_2$ symmetry arises for which the first $SO(3)$-breaking invariant has dimension four and thus $m_{\pi_{1,2}}\propto \ybreak^4 f^2$ \cite{Das:2020arz}. },  i.e. $m_{\pi_{1,2}}^2\propto \ybreak^3$  while all contributions $\propto \ybreak^2$ must cancel out.

The conclusion is that, in order for the quadratic divergences to cancel, one does not need to impose the full $SO(3)$ symmetry, but instead it is enough to  impose ultraviolet $A_4$ invariance. In the latter case, other mass terms survive for the dGBs, though.    
This is in contrast with generic pGB with arbitrary $SO(3)$-breaking terms. For instance, an explicit $SO(3)$ breaking which also breaks the $A_4$ subgroup  is given by
\begin{equation}
\mathcal{L}_{\slashed{A}_4}= y_{\slashed{A}_4}\phi_1\,\bar{\Psi}_{1} \Psi_{1} =
y_{\slashed{A}_4}\, \pi_{1}\bar{\Psi}_{1} \Psi_{1} + \O(\pi^3)\,, \label{nosym}
\end{equation}
in which case only the UV divergent diagram on the left of Fig.~\ref{Fig:CancelationHook} is present and only for $\pi_1$, and thus a divergent contribution to the pion mass does remain,
 \begin{align}
     \delta m_{\pi_{1}}^2\propto \frac{1}{2}y_{\slashed{A}_4}^2\Lambda^2 \neq 0\,.
  \end{align}


\subsubsection*{A UV model with heavy scalars} 
\begin{figure}[h]
  \begin{minipage}{.24\textwidth}
       \centering \includegraphics[width=\linewidth]{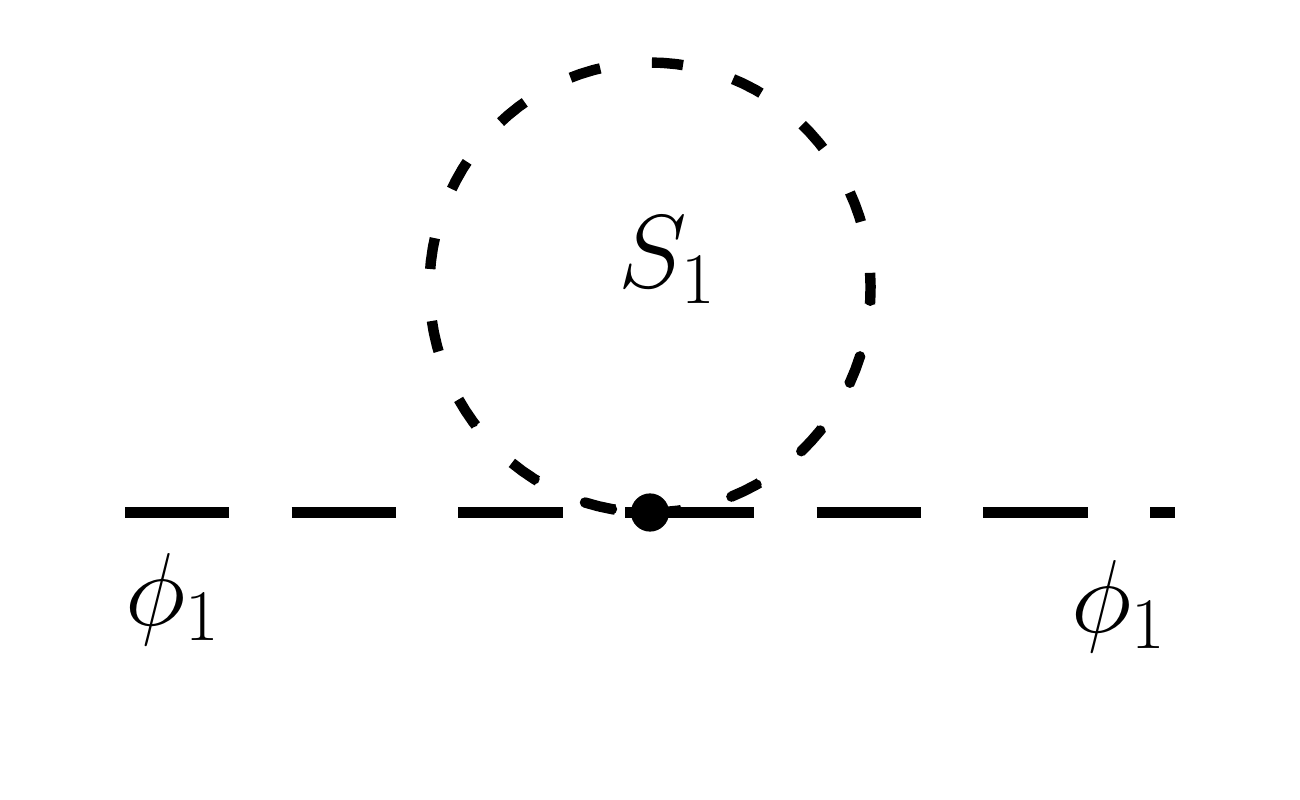}
    \end{minipage}\begin{minipage}{.01\textwidth} $+$\end{minipage} 
   \begin{minipage}{.24\textwidth}
       \centering \includegraphics[width=\textwidth]{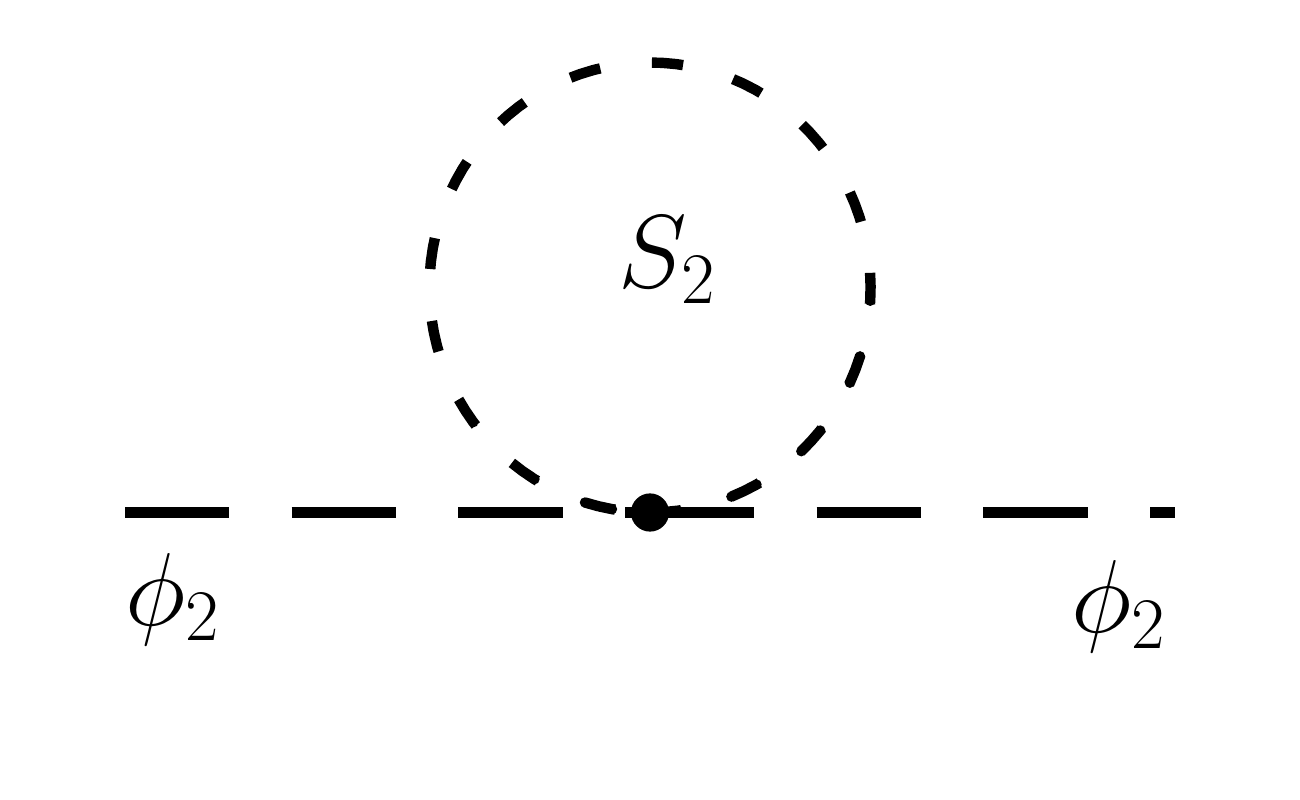}
    \end{minipage} 
      \begin{minipage}{.01\textwidth} $+$\end{minipage} 
     \begin{minipage}{.24\textwidth}
       \centering \includegraphics[width=\textwidth]{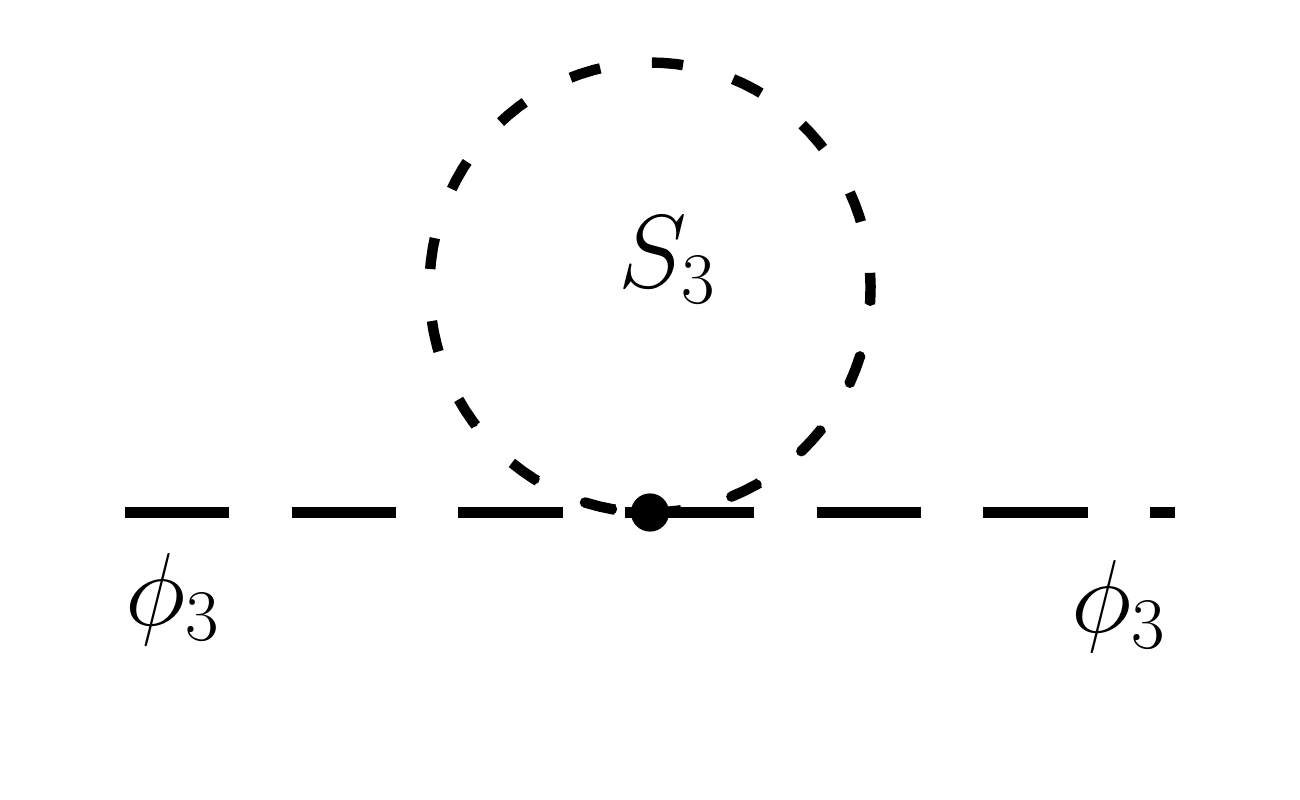}
    \end{minipage}\begin{minipage}{.22\textwidth} $\qquad \propto \quad\lbreak \,\I_2 \Lambda_{UV}^2$\end{minipage}
  \caption{Insensitivity of the dGB masses to the cutoff in the model with heavy scalars.}
  \label{Fig:CancellationHeavyScalar}
\end{figure}

The example above with exotic fermions is not unique. We consider next a BSM scalar sector  invariant under an exact ultraviolet $A_4$ symmetry and inducing  effective $\phi^n$ couplings. An example of this kind 
where one can see at play the mechanism of cancellation of quadratic sensitivity to high scales  involves the interaction of the $A_4$ triplet $\phi$ with another scalar triplet $S$ which does not take a vacuum expectation value (vev) and has a large mass $m_S\gg f$. Let's assume that these scalars interact through the following $A_4$-symmetric and $SO(3)$-violating quartic couplings,\footnote{We have repeated the exercise for all possible $A_4$ invariant couplings of the scalars $S$ and confirmed that the same pattern of cancellations holds.}
\begin{align}
\mathcal{L}_{\text{int } \slashed{G}} =\frac{\lbreak}{4}\left(\phi_1^2 S_1^2 + \phi_2^2 S_2^2 + \phi_3^2 S_3^2\right)\,. \label{heavyscalars}
\end{align}
Assuming that the theory has a high energy cut-off $\Lambda_{UV}$, one can easily check that the quadratically divergent diagrams in \cref{Fig:CancellationHeavyScalar} arrange themselves in a way that they only contribute to the $\I_2$ invariant, leaving the dGB masses insensitive to the UV scale,  as illustrated in Fig.~\ref{Fig:CancellationHeavyScalar}.

\begin{figure}[h]
  \begin{minipage}{.25\textwidth}
        \centering \includegraphics[width=\linewidth]{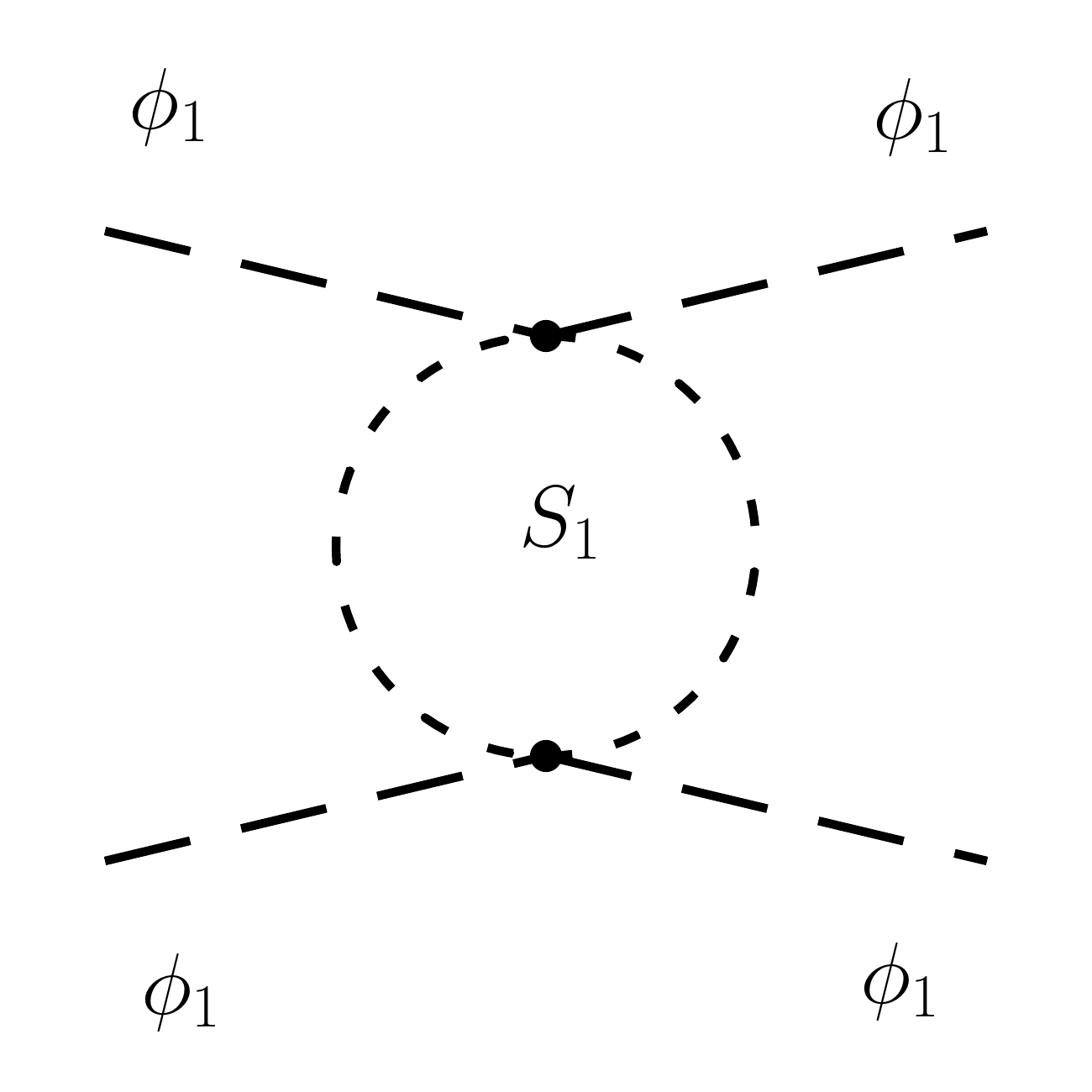}
     \end{minipage}\begin{minipage}{.01\textwidth} $+$\end{minipage} 
    \begin{minipage}{.25\textwidth}
        \centering \includegraphics[width=\textwidth]{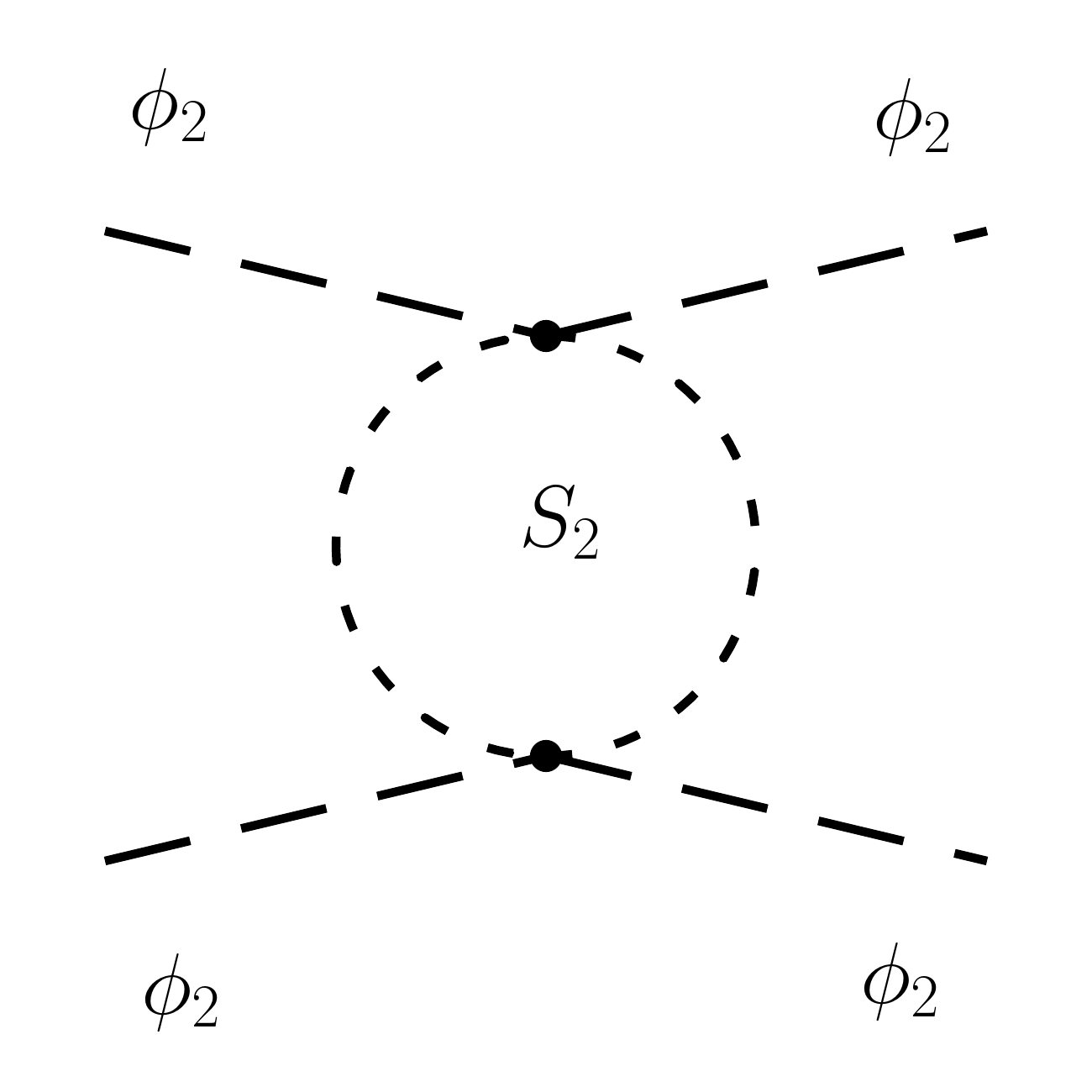}
     \end{minipage} 
       \begin{minipage}{.01\textwidth} $+$\end{minipage} 
      \begin{minipage}{.25\textwidth}
        \centering \includegraphics[width=\textwidth]{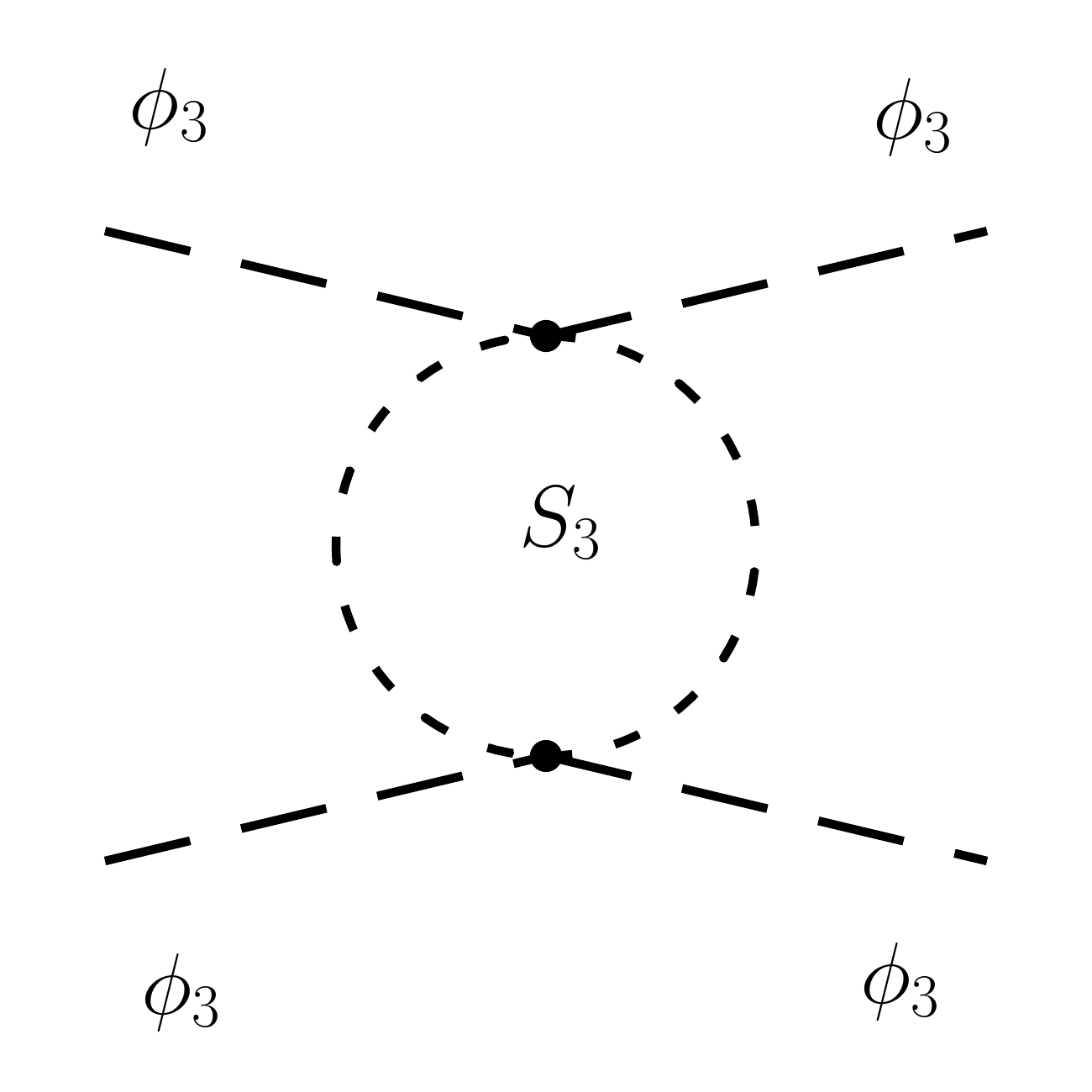}
     \end{minipage}    \begin{minipage}{.2\textwidth} $\qquad \propto \quad \lbreak^2\,\I_4
     $\end{minipage}
  \caption{Finite contribution to the dGB mass through the quartic invariant.}
  \label{Fig:FirstContributionHeavyScalar}
\end{figure}

The dGBs do not remain massless though, and a mass term $\propto \lbreak^2$ is generated via loops through the invariant $\I_4$, see \cref{Fig:FirstContributionHeavyScalar}.

\begin{figure}[h]
  \begin{minipage}{.3\textwidth}
        \centering \includegraphics[width=\linewidth]{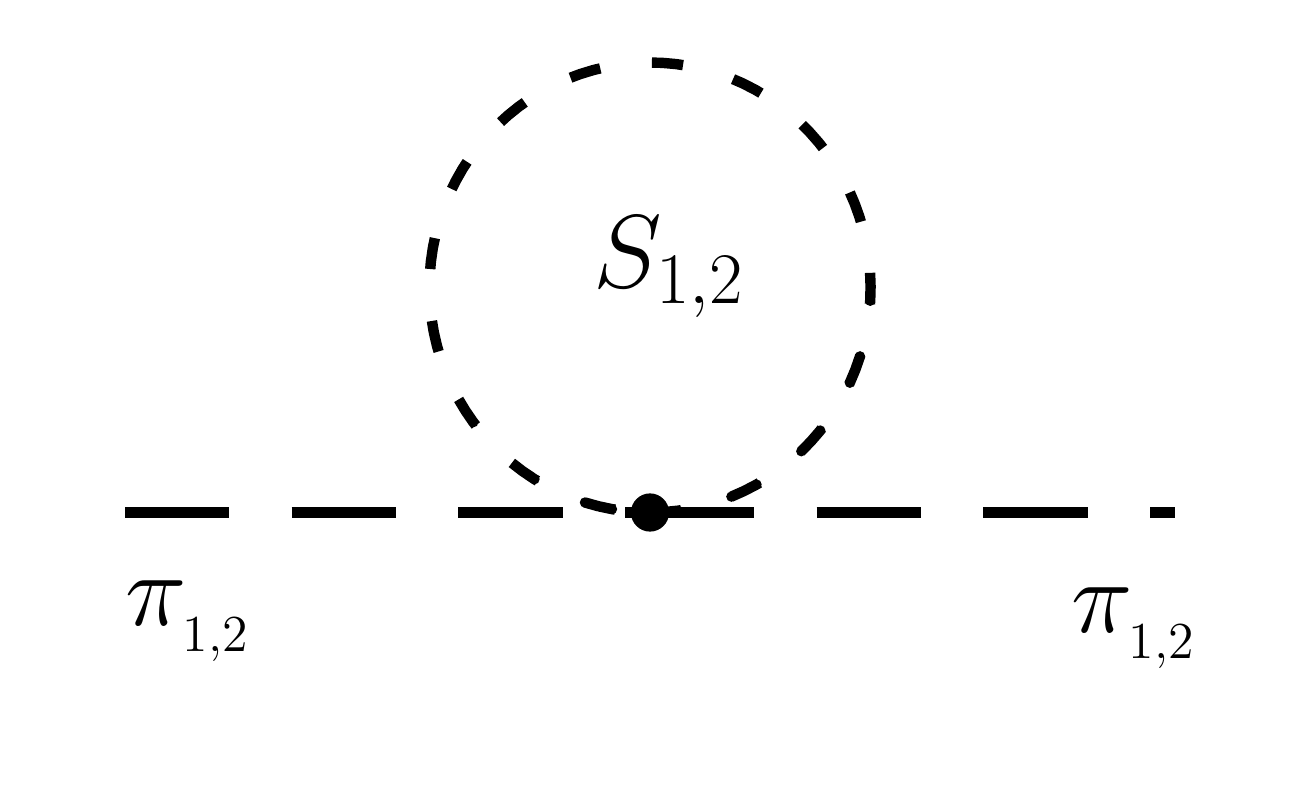}
     \end{minipage}\begin{minipage}{.01\textwidth} $+$\end{minipage} 
    \begin{minipage}{.3\textwidth}
        \centering \includegraphics[width=\textwidth]{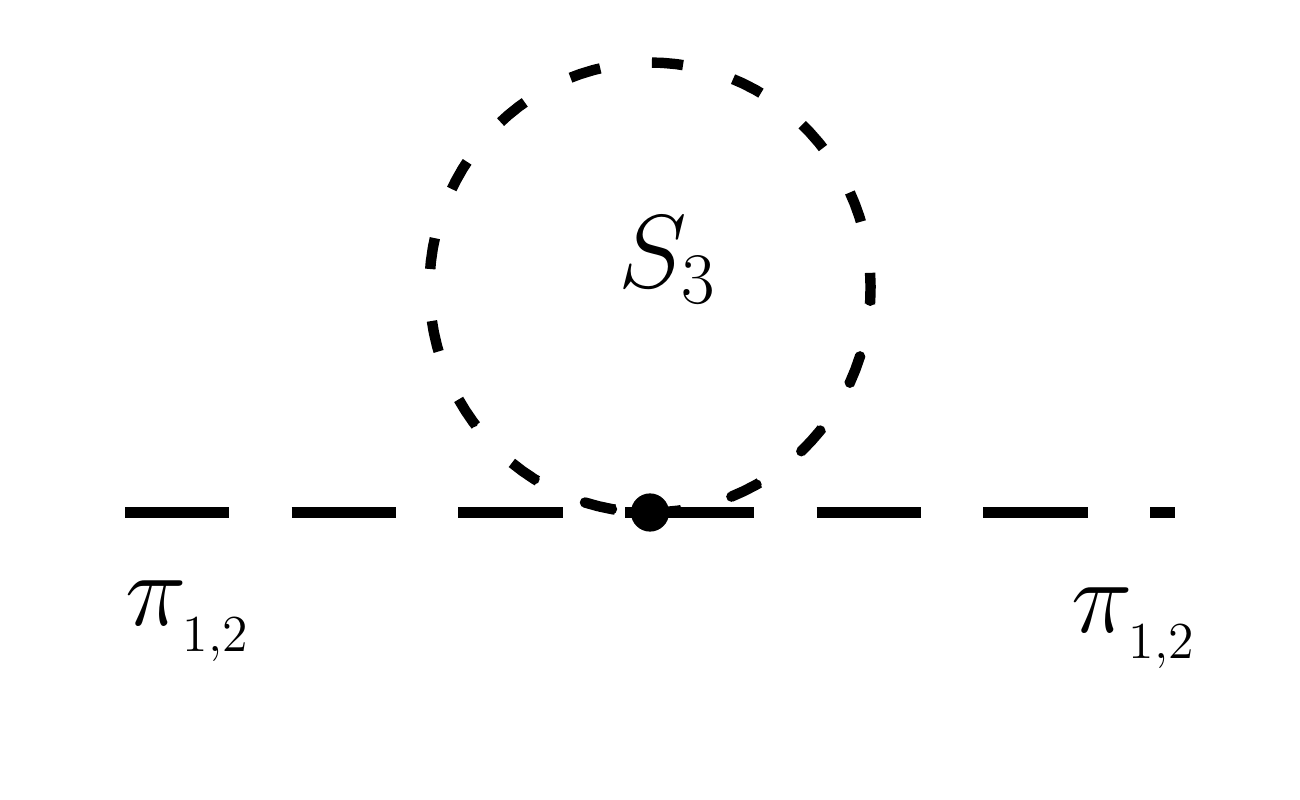}
     \end{minipage} 
        \begin{minipage}{.4\textwidth} $\qquad \propto \quad\lbreak \Lambda_{UV}^2 - \lbreak \Lambda_{UV}^2=0
     $\end{minipage}
  \caption{Cancellation of the UV divergences resulting from diagrams that could have contributed to the dGB masses in the model with heavy scalars, Eq.~(\ref{scalar-model-I}).}
  \label{Fig:CanbcelationPionsHeavyScalar}
\end{figure}

An even more explicit manifestation of the cancellation arises when one computes the interaction of the scalars $S_i$ with the dGB themselves. Expanding again at second order in the dGBs, Eq.~(\ref{heavyscalars}) results in 
\begin{align}
  \mathcal{L}_{\text{int } \slashed{G}}\simeq\lbreak\left[\pi_1^2 S_1^2 + \pi_2^2 S_2^2 + f^2S_3^2-\left(\pi_1^2 + \pi_2^2\right)S_3^2\right]+\O(\pi_i^3)\,.
  \label{scalar-model-I}
\end{align}
The computation of the corresponding loops, depicted in \cref{Fig:CanbcelationPionsHeavyScalar}, shows that in this case the cancellation arises due to a correlation among the quartic interactions with $S_{1,2}$ ensured by the $A_4$ symmetry.

  \subsubsection{Maximally natural extrema of an invariant potential}
  \label{natural-extrema}
  The description and discussion in this subsection will apply as well to the other discrete representations and groups to be explored later on.

 In order to study the properties of the dGB and identify the physical states, it is necessary to study the minima of the potential. Looking at the most general potential in Eqs.~(\ref{Eq:MostGeneraldGBPotential1}) or (\ref{Vreduced}), this may seem a colossal task given that the coefficients are arbitrary. Nevertheless, one can study and classify the {\it natural extrema}, those that are less or not at all dependent on a specific combination of coefficients in the potential, but rather arise as a consequence of the symmetry. A subset of the natural extrema we explore below will not depend at all on the specific form or tuning of coefficients in the potential, and these we will call maximally natural ({\it MaNa}) extrema. The MaNa extrema are of particular interest for the analysis of the minima of a physics potential. These points are guaranteed to be (absolute or local) extrema of the complete potential since they are extrema of {\it all} the primary and secondary invariants.


Consider a generic potential $V(x)$ for a set of $n$ independent scalar fields $x_i$. If the potential preserves a symmetry, it can  be expressed  as a function of just $n$ independent functions $I_i(x)$ invariant under the symmetry:  the {\it primary invariants} $I_i(x)$.\footnote{The inversion of the relation of the invariants in terms of the variables  is unique. One may then identify the inverse relation $x_i=x_i(I_j)$ and express any new invariant $I'$ in terms of the independent set $\{I_j\}$; $I'=I'(x_i)=I'(x_i(I_j))$.} The most general potential  will be a function of all such possible invariants, $V(x)=V[I_i(x)]$.  It follows that all the extrema of the potential (and among them the true vacuum) can be obtained imposing
\begin{equation}
\frac{\partial V}{\partial x_j}=\sum_i \frac{\partial V}{\partial I_i}\frac{\partial I_i}{\partial x_j} =\sum_j \frac{\partial V}{\partial I_i}\,J_{ij}=0\,.
\label{EqsMin}
\end{equation}
\begin{figure}[h!]
\centering
\includegraphics[width=0.7\textwidth]{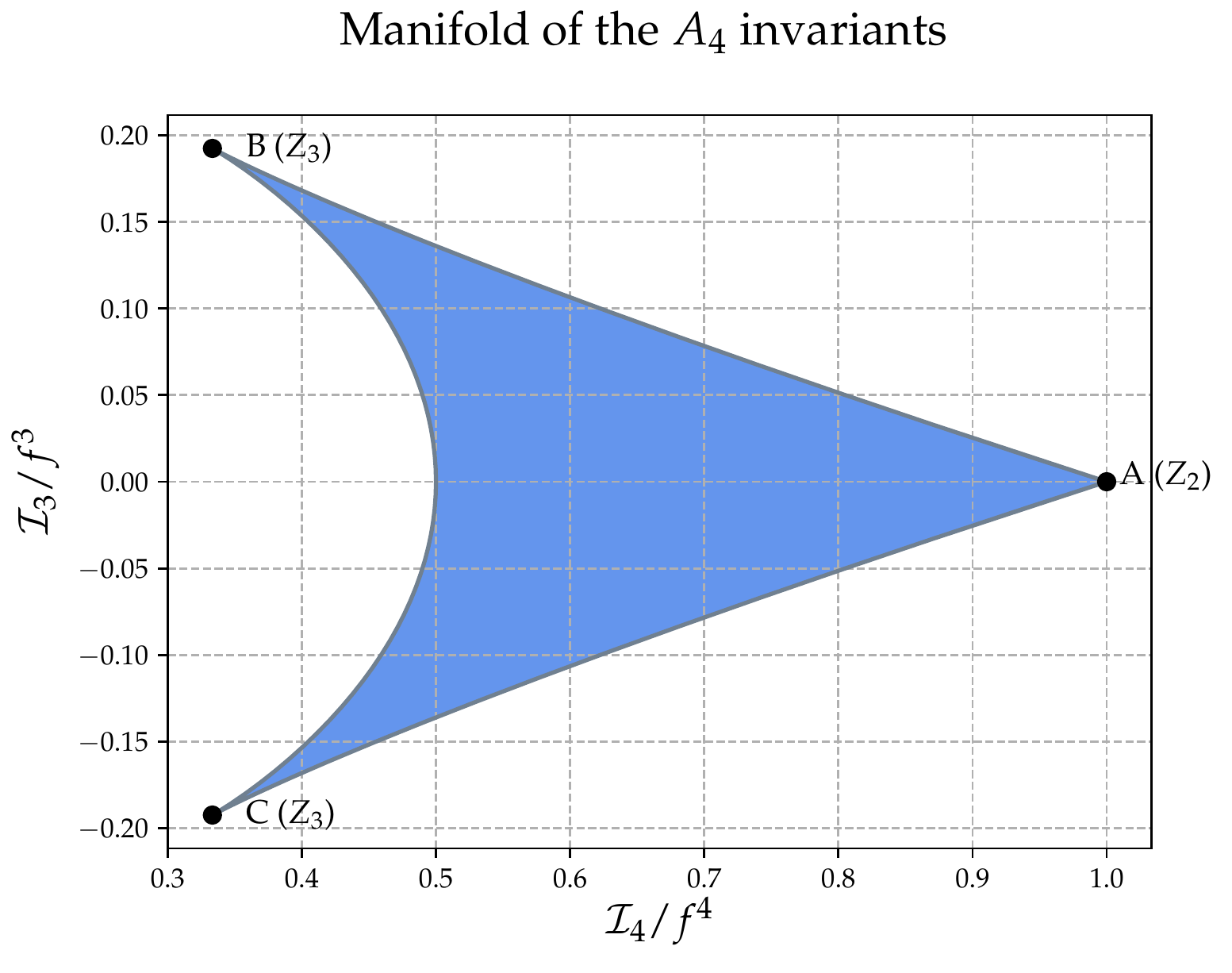}
\caption{Manifold defined by the $\mathcal{I}_3(\Phi)$ and $\mathcal{I}_4(\Phi)$ invariants of $A_4$, with $\Phi$  a triplet fulfilling $\Phi^T\Phi=\phi_1^2 + \phi_2^2 + \phi_3^2 = f^2$. The little group which remains invariant at each natural extrema is indicated.}
\label{fig:A4:invariant_manifold}
\end{figure}
This equation can be seen as the Jacobian matrix of the change of variables from the fields to the invariants, $J_{ji}\equiv{\partial I_{i}}/{\partial x_{j}}$ times the vector ${\partial V}/{\partial I_{i}}$, and allows one to identify two kinds of extrema of a potential: 
\begin{enumerate}
\item[1)] Model-dependent extrema: If the rank of the Jacobian is maximal, then the extremal points necessarily correspond to a vanishing vector ${\partial V}/{\partial I_{i}}=0$. In consequence, the extrema will depend on the specific parameters of the combination of invariants that build the potential. This is the case, for example, of the Higgs potential of the SM.
\item[2)]  Natural extrema: Those points that are extrema of the invariants $\I_i$ themselves and therefore correspond to 
\begin{equation}
\det[J_{ji}]=0\,,
\label{natural_condition}
\end{equation}
 that is, the rank of the Jacobian is less than maximal.  They are called natural extrema as they  are set by the symmetry (that determines the invariants) and therefore depend less on the specific coefficients of the potential. In general terms,
the reduction of the rank implies the appearance of symmetries left explicit, i.e. unbroken in the spectrum.
 \end{enumerate}

The essential point of the analysis of the natural extrema is that the space of the physical variables $x$ has no boundary, while the manifold spanned by $ I_i(x)$ does have boundaries.  Fig.~\ref{fig:A4:invariant_manifold} illustrates the manifold and its boundaries for the particular case of a non-linearly realized $A_4$  symmetry with a scalar triplet. Extrema of $V$  in the bulk of the manifold correspond to case 1) above and are always model-dependent extrema, while extrema  on the boundaries  are natural extrema~\cite{Cabibbo:1970rza}:  they fall in category 2) above. The boundaries are characterized by the rank of the Jacobian matrix being less than maximum and are thus described by $n-1$ dimensional manifolds (e.g. surfaces, for $n=3$), each characterized by a different little group (i.e. the set of transformations that leave the corresponding extremum invariant). Such manifolds  meet along  $n-2$ dimensional manifolds (e.g. lines) which in turn meet along even lower dimensionality manifolds (e.g. singular points), etc. Each of these boundaries corresponds to a particular little group, with again the smaller the rank of the Jacobian, the larger the little group.  

Indeed, those points whose little group is a maximal subgroup\footnote{ Given a group, a subgroup is maximal if the only subgroup that contains it is the complete group itself, i.e. a subgroup that can be included  only in the full group.} of $D$ are expected to be the most natural minima~\cite{michel1971properties}. 
Extrema of $V$ on a given boundary~\cite{Cabibbo:1970rza} are dubbed natural precisely because they are more natural than the generic extrema in the interior of the manifold. The former  require the vanishing of only Rank$(J)=n-k$ derivatives of $V$ with respect to the invariants given that, on the boundary, the Jacobian $J$ has $k$ vanishing eigenvectors (orthogonal to the boundary).   Note as well that the derivatives of the invariants must vanish  on the boundaries in the directions orthogonal to them, which forces the existence of a singular point wherever  $n$ boundaries intersect. Therefore one should always find an extremum at such intersections, which is where all derivatives vanish. 
 The natural minima at the boundary intersections, that is, the zero-dimensional vertices of the boundary, are what we have called the MaNa minima. The MaNa minima are maximally natural in that their position (the vevs) do not rely on the values of the parameters of the potential. 

Each point $P$ in the  manifold of the invariants corresponds in general to several possible vacuum expectation values of the vectors $\Phi$, which we will denote $\Phi_P$: they constitute an orbit. Each element of the  exact UV discrete invariance group 
 $D$ 
either leaves the vector invariant (and therefore belongs to its little group) or transforms the vector into another element of the orbit. In consequence,  the following counting rule applies:
\begin{align}
	\text{order} (L)\times \# \Phi_P=\text{order} (D)\,,
	\label{counting-rule}
\end{align}
where $\# \Phi_P$ is the number of  MaNa extrema for the representation studied, and $L$ denotes the little group at $\Phi_P$.

In this work, we focus on the exploration of the  MaNa minima for each of the --spontaneoulsy broken-- discrete groups to be considered.  

\subsubsection{MaNa extrema for the triplet of \texorpdfstring{$A_4$}{}\, }
In this scenario, considering the three degrees of freedom at high energies,  $\Phi \equiv(\phi_1, \phi_2, \phi_3)^\intercal$, the invariant manifold would be three-dimensional. This is reduced to a two-dimensional manifold when the non-linear constraint Eq. (\ref{equationf-A4-3}) is taken into account, which is the setup of interest to explore the dGB potential. This invariant manifold is  that spanned by the set of two invariants $\{ \I_3,\I_4\}$ which source all possible terms in the dGB potential. This is depicted in Fig.~\ref{fig:A4:invariant_manifold}: the points inside the bulk of the manifold correspond to the rank  two Jacobian,  the boundaries are lines, and the invariant little groups at the MaNa extrema points --the vertices $A$, $B$ and $C$-- are indicated. 

The coordinates of the MaNa points in field space are shown in Table~\ref{tab:A4:critical_points}, together with the  value of the invariants and their exact --explicit-- invariances. Note that the coordinate values enforce the non-linear constraint in Eq.~(\ref{equationf-A4-3}), as required.  It is easy to check that Eq.~(\ref{counting-rule}) is satisfied: $2 \times 6 \,\text{(for A)}= 3 \times 4 \,\text{(for B)} = 3 \times 4 \,\text{(for C)}= 12$, and the number of extrema is $6+4+4=14$. The three  MaNa  points $A$, $B$ and $C$ are related by $A_4$ transformations.  The discrete symmetries that may remain explicit are typically a maximal subgroup of the UV discrete group $A_4$: this is the case for $Z_3$ identified for the  extrema $B$ and $C$  in Table~\ref{tab:A4:critical_points}, which are both MaNa minima\footnote{Depending on the sign of the coefficient $\hat c_i$, these extrema are either maxima or minima.}, while $A$ is a saddle point and its  little group is $Z_2$ (which is not a maximal subgroup of $A_4$).

\begin{figure}[h!]
    \centering
    \includegraphics[width = 0.6\textwidth]{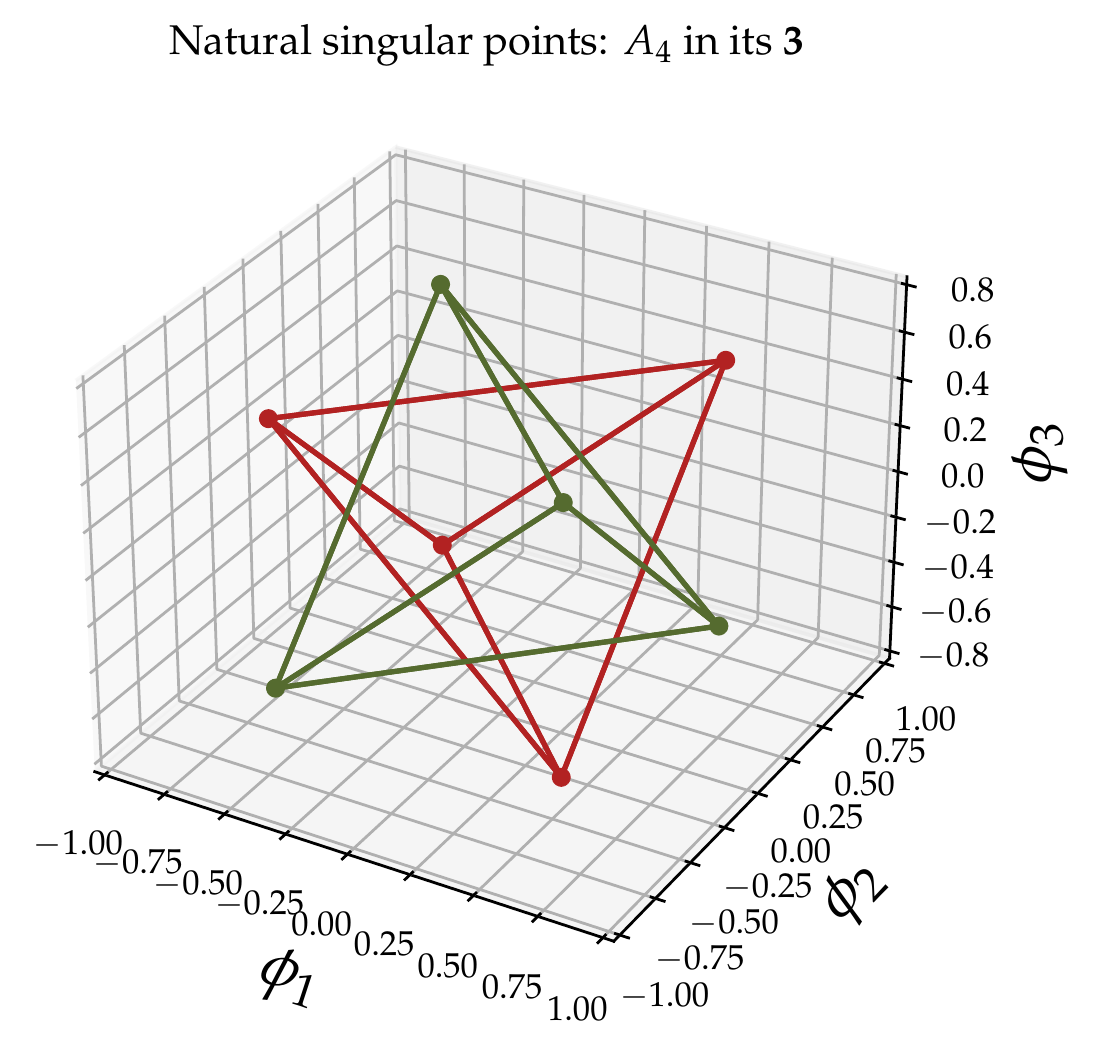}
    \caption{Geometrical distribution in field space of the two $Z_3$-symmetric MaNa extrema of the dGB potential for a scalar triplet of $A_4$. Points B of Table~\ref{tab:A4:critical_points} are depicted in green and points C are in red.}
    \label{fig:A4:singular_points}
 \end{figure}

 The discrete symmetry is reflected in the geometrical locations of the  MaNa extrema in field space. For instance, the $Z_3$-symmetric minima arrange themselves in the shape of two tetrahedrons, as depicted in Fig.~\ref{fig:A4:singular_points}.

\begin{table}[h!]
\centering
\begin{align}
	\begin{array}{|c||cc|ccc|c|c|}
		\multicolumn{8}{c}{\text{\textbf{ MaNa extrema for a triplet of} $\boldsymbol{A_4}$}}\\
		\hline
		\text{Point} &\I_3 & \,\I_4 & \phi_{1}\, &\phi_{2}\, &\phi_{3} & \text{Little group}& \text{Nature}\\[1pt]
		\hline\hline
		\multirow{3}{*}{A} & \multirow{3}{*}{0} & \multirow{3}{*}{1} & 0 & 0 & \pm1 & \multirow{3}{*}{\text{$Z_2$}}& \multirow{3}{*}{\text{Saddles}}\\
							  & 					  &						  & 0 & \pm1 & 0 & 						  & \\
							  & 					  &						  & \pm1 & 0 & 0 &						  & \\
		\hline	\rule{0pt}{2.3ex}
		\multirow{3}{*}{B} & \multirow{3}{*}{$\frac{1}{3\sqrt{3}}$} & \multirow{3}{*}{$\frac{1}{3}$} & \frac{1}{\sqrt{3}} & \frac{1}{\sqrt{3}} & \frac{1}{\sqrt{3}} & \multirow{3}{*}{\text{$Z_3$}} & \multirow{3}{*}{\text{Minima}}\\
							  & 										&								  & -\frac{1}{\sqrt{3}} & \frac{1}{\sqrt{3}} & -\frac{1}{\sqrt{3}} &  & \\
							  &   										& 								  & \pm\frac{1}{\sqrt{3}} & -\frac{1}{\sqrt{3}} & \mp\frac{1}{\sqrt{3}} & & \\[0.8ex]
		\hline  \rule{0pt}{2.3ex}
		\multirow{3}{*}{C} & \multirow{3}{*}{$-\frac{1}{3\sqrt{3}}$} & \multirow{3}{*}{$\frac{1}{3}$}  & -\frac{1}{\sqrt{3}} & -\frac{1}{\sqrt{3}} & -\frac{1}{\sqrt{3}} & \multirow{3}{*}{$Z_3$} &\multirow{3}{*}{\text{Minima}}\\
							  & 										   &	  								 & \frac{1}{\sqrt{3}} & -\frac{1}{\sqrt{3}} & \frac{1}{\sqrt{3}} & 	 & \\
							  &											   &	  								 & \pm\frac{1}{\sqrt{3}} & \frac{1}{\sqrt{3}} & \mp\frac{1}{\sqrt{3}} &  & \\[0.8ex]
	\hline
	\end{array}
\end{align}
\caption{Location and symmetries of the  MaNa extrema of the manifold spanned by the invariants that build up the dGB potential, for  the case of a triplet scalar of $A_4$. The values of the invariants and the locations of the extrema are given in units of the dGB scale $f$. The location in field space of the  MaNa minima is depicted in Fig.~\ref{fig:A4:singular_points}. The manifold is  illustrated in Fig.~\ref{fig:A4:invariant_manifold}.}
\label{tab:A4:critical_points}
\end{table}

\paragraph{Expansion in terms of low-energy degrees of freedom.}
The parametrization of the real scalar field $\Phi$ in terms of physical dGBs in Eq.~(\ref{eq:pions  generic definition})  spans in this case two low-energy fields $\pi_{i=1,2}$,
\begin{equation}
	\Phi(\pi_1,\pi_{2}) =  \left(
	\begin{array}{c} 
		\phi_1 \\ \phi_2 \\ \phi_3
	\end{array}\right)=\exp\left[\frac{1}{f} \left(
	\begin{array}{cccc}
		0 & 0 & \pi_1 \\
		0 & 0 & \pi_{2} \\ 
		-\pi_1  & -\pi_{2} & 0
	\end{array}\right)
	\right]\left(
	\begin{array}{c} 
		0 \\ 0 \\ f
	\end{array}\right)\,.
	\label{eq:pions definition}
\end{equation} 
The expansion of the low-energy dGB Lagrangian in Eq.~(\ref{LdGB-2}) around the  MaNa extrema in Table~\ref{tab:A4:critical_points}, in terms of these physical pions fields (dGBs), must be done carefully. The point is that the minima in that table are rotated with respect to the direction around which the pions are defined in  Eq.~(\ref{eq:pions definition}). In order to maintain the latter parametrization consistently,  it is necessary to apply a three-dimensional real rotation $R\in SO(3)$ that aligns each  MaNa  point with the $z$-direction in field-component space, which corresponds to the unbroken generator.  The invariants can then be simply redefined as 
\begin{equation}
\I'_{i}(\Phi) = \I_{i}(R^{-1}\Phi)\,,
\label{tilted-minima}
\end{equation}
where $\Phi$ is as defined by Eq.~(\ref{eq:pions definition}); any MaNa extremum $\Phi_s$ is therefore transformed into  $\Phi'_s = R\,\Phi_s = (0, 0, f)^\top$. The absence of this step has prevented Ref.~\cite{Das:2020arz} to identify the surviving explicit symmetry of the spectrum, that we unveil here to be $Z_3$.\footnote{If one keeps the parametrization in Eq.~(\ref{eq:pions definition}) and expands around the vaccuum expectation values (vevs) of the pion fields   in   \cref{tab:A4:critical_points},  $\langle\pi_1\rangle=\langle \pi_2\rangle=( f/\sqrt{2})\cos ^{-1}\left(1/\sqrt{3}\right)$, as done in Ref.~\cite{Das:2020arz}, the resulting kinetic terms of the pions would not be canonically normalized. Upon redefinition  to canonically normalize the fields, our results are recovered.} Therefore, $Z_3$ is the symmetry that should appear realized {\it  \`a  la Wigner} in the low-energy spectrum for the MaNa minima of the potential, as discussed next in further detail.

The expansion around any of the $Z_3$ symmetric vevs  (for instance $\Phi_C = -\frac{1}{\sqrt{3}}(1,1,1)$) of the $A_4$ invariants for the scalar triplet in terms of  physical pion fields, parametrized as in Eq.~(\ref{eq:pions definition}), yields 
\begin{align}
    \I_{3} &= \frac{f}{\sqrt{3}}\left[-\frac{f^2}{3}+\left(\pi _1^2+\pi _2^2\right) 
    - \frac{1}{3\sqrt{2}f}\left(\pi _1^3 -3\pi_1\pi_2^2\right) 
    - \frac{17}{24f^2}\left(\pi _1^2+\pi _2^2\right)^2\,\right] + \cdots \label{eq:A4:invariants_pions_3}\\
    \I_4 &=   \frac{4f^2}{3} \left[
    \frac{f^2}{4}+
    \left(\pi _1^2+\pi _2^2\right)
    +\frac{1}{\sqrt{2} f} \left(\pi _1^3 -3\pi_1\pi_2^2\right)
    -\frac{29}{24}\left(\pi _1^2+\pi _2^2\right)^2\, \right] \, + \cdots \label{eq:A4:invariants_pions_4}\\
    \I_6 &=  -\frac{4}{3}\sqrt{\frac{2}{3}} f^3 \left(\pi_2^3 - 3\pi_1^2\pi_2\right)\, + \cdots
    \label{eq:A4:invariants_pions_6}
\end{align}
 where dots indicate  terms  of higher order in the pion fields, and the non-linear constraint $(|\Phi|^2 = f^2$) in Eq.~(\ref{equationf-A4-3}) is automatically implemented  (here and all through the rest of the paper, the non-linear constraint will be implicit whenever the dGB dynamics is analyzed). These equations show that the dGB fields appear arranged in an irreducible representation of the little group of the  MaNa minima: a $\boldsymbol{2}$  of $Z_3$. Indeed, two primary invariants exist for the latter which, in the pion parametrization chosen read 
\begin{align}
\I_{2}^{(\boldsymbol{2}, Z_3)} &= \pi_1^2 + \pi_2^2\,,
\label{eq:A4:invariants_Z3_2}\\
\I_{3}^{(\boldsymbol{2}, Z_3)}&= \pi_1^3 - 3\pi_1\pi_2^2\,, 
\label{eq:A4:invariants_Z3}
\end{align} 
and those are  precisely the field combinations exhibited by Eqs.~(\ref{eq:A4:invariants_pions_3})-(\ref{eq:A4:invariants_pions_4}). There is also a secondary $Z_3$ invariant (obtained via a  syzygy)
\begin{equation} 
\I_{3'}^{(\boldsymbol{2}, Z_3)} = \pi_2^3 - 3\pi_1^2\pi_2\,,
\label{eq:A4:sizygy_Z3}
\end{equation}
whose pion dependence equals that for the secondary $A_4$ invariant $\I_6$ in Eq.~(\ref{eq:A4:invariants_pions_6}). In fact, the role of the two cubic invariants $\I_{3}^{(\boldsymbol{2}, Z_3)}$ and $\I_{3'}^{(\boldsymbol{2}, Z_3)}$ can be exchanged, or substituted by combinations of them, by changing the pion parametrization in Eq.~(\ref{eq:pions definition}).  Note that we have denoted here the $Z_3$ invariants as 
\begin{equation}
\I_{n}^{(\boldsymbol{m}, G)}
\label{notation-invariants}
\end{equation}
where $n$ indicates the mass-dimension of the invariant, $G$ the invariance group and $\boldsymbol{m}$ the field representation. This notation will be implemented all through the rest of the paper, except for the triplet of $A_4$, for which we will maintain the simplified notation above, i.e. $\I_2\equiv \I_{2}^{(\boldsymbol{3}, A_4)}$, $\I_3\equiv \I_{3}^{(\boldsymbol{3}, A_4)}$, $\I_4\equiv \I_{4}^{(\boldsymbol{3}, A_4)}$ and $\I_6\equiv \I_{6}^{(\boldsymbol{3}, A_4)}$.

\subsection{Phenomenology of \texorpdfstring{$A_4$}{} \ dGBs}
\label{pheno-A4}
We have shown that an unbroken $Z_3$ symmetry will be realized {\it \`a la Wigner} in the dGB spectrum as long as the vevs of the scalar correspond to the MaNa extrema. Some experimental consequences follow. We focus below in those signals associated with the terms in the potential discussed above. The detailed phenomenological analysis for the interaction terms in the Lagrangian Eq.~(\ref{LdGB-2})  which involve derivatives will be left for later work, although the pattern of signals will have strong  similarities to those from the potential terms, as argued further below.\footnote{ To keep our discussion general enough, note that the phenomenological analysis presented here assumes that the global minimum of the potential is one of the MaNa extrema identified above. It is possible that a deeper global minimum arises in Nature which depends instead on a precise combination of couplings, though, and the analysis would depend in that case on the value of the potential parameters.}

 \subsubsection{Degenerate spectrum} 
 The first physical prediction stems from the pions being in an irreducible doublet representation of $Z_3$: 
 {\it two degenerate dGBs} are expected, 
\begin{equation}
	\begin{split}
	m_{\pi_1}^2=m_{\pi_2}^2.
	\label{eq:A4:ratios:masses}
	\end{split}
\end{equation}
Degeneracy introduces freedom into the definition of the dGBs, though. They can be redefined by an arbitrary rotation of angle $\beta$, 
\begin{equation}
\begin{split}
\left\{\begin{array}{c}
\pi_1\\ \pi_2 
\end{array}\right.  \longrightarrow
\left\{\begin{array}{l}
 \hat\pi_1 = \pi_1\cos\beta  + \pi_2\sin\beta\\
 \hat\pi_1 = \pi_1\sin\beta  - \pi_2\cos\beta  \,,
\end{array}\right.
\end{split}
\label{beta-redef}
\end{equation}
which in turn redefines the $Z_3$ invariants. For instance, for the cubic ones, 
\begin{align}
&\!
\I_{3}^{(\boldsymbol{2}, Z_3)} 
	&& \!\!\!\!\!
	\rightarrow c_{3\beta} \,\I_{3}^{(\boldsymbol{2}, Z_3)}  +s_{3\beta}\, \I_{3'}^{(\boldsymbol{2}, Z_3)}
	&& \!\!\!\!\!
	= (\pi_1^3 - 3\pi_1\pi_2^2)\cos3\beta +(\pi_2^3 - 3\pi_1^2\pi_2)\sin3\beta\,,\\
 &\!
 \I_{3'}^{(\boldsymbol{2}, Z_3)}
	&& \!\!\!\!\!
	\rightarrow-s_{3\beta}\,\I_{3}^{(\boldsymbol{2}, Z_3)} + c_{3\beta}\,\I_{3'}^{(\boldsymbol{2}, Z_3)} 
	&& \!\!\!\!\!
	= -(\pi_1^3 - 3\pi_1\pi_2^2)\sin3\beta +(\pi_2^3 - 3\pi_1^2\pi_2)\cos3\beta\,,
\label{beta-cubic-invariant}
\end{align}
where the short-hand notation $c_{3\beta}\equiv \cos (3\beta)$ and $s_{3\beta}\equiv \sin (3\beta)$ has been used. This means that the cubic interactions correspond to an arbitrary linear combination of the little group invariants $\I_{3}^{(\boldsymbol{2}, Z_3)}$ and $\I_{3'}^{(\boldsymbol{2}, Z_3)}.$ The predictions for the physical observables are reparametrization invariant, though, as expected and illustrated further below.

\subsubsection{Simultaneous production} 
The second prediction is that, unlike for usual ALPs, 
dGBs are not expected to be produced alone. This is an unavoidable consequence of the fact that the pion fields belong to  non-trivial irreducible multiplets, as long as  
the SM fields do not carry charges under the discrete symmetry. In any collision or decay sourcing dGBs, {\it at least two dGBs are then expected to be simultaneously produced}.

\begin{figure}[h!]
\centering
	\begin{subfigure}[b]{0.49\textwidth}
         \centering
         \includegraphics[width=\textwidth]{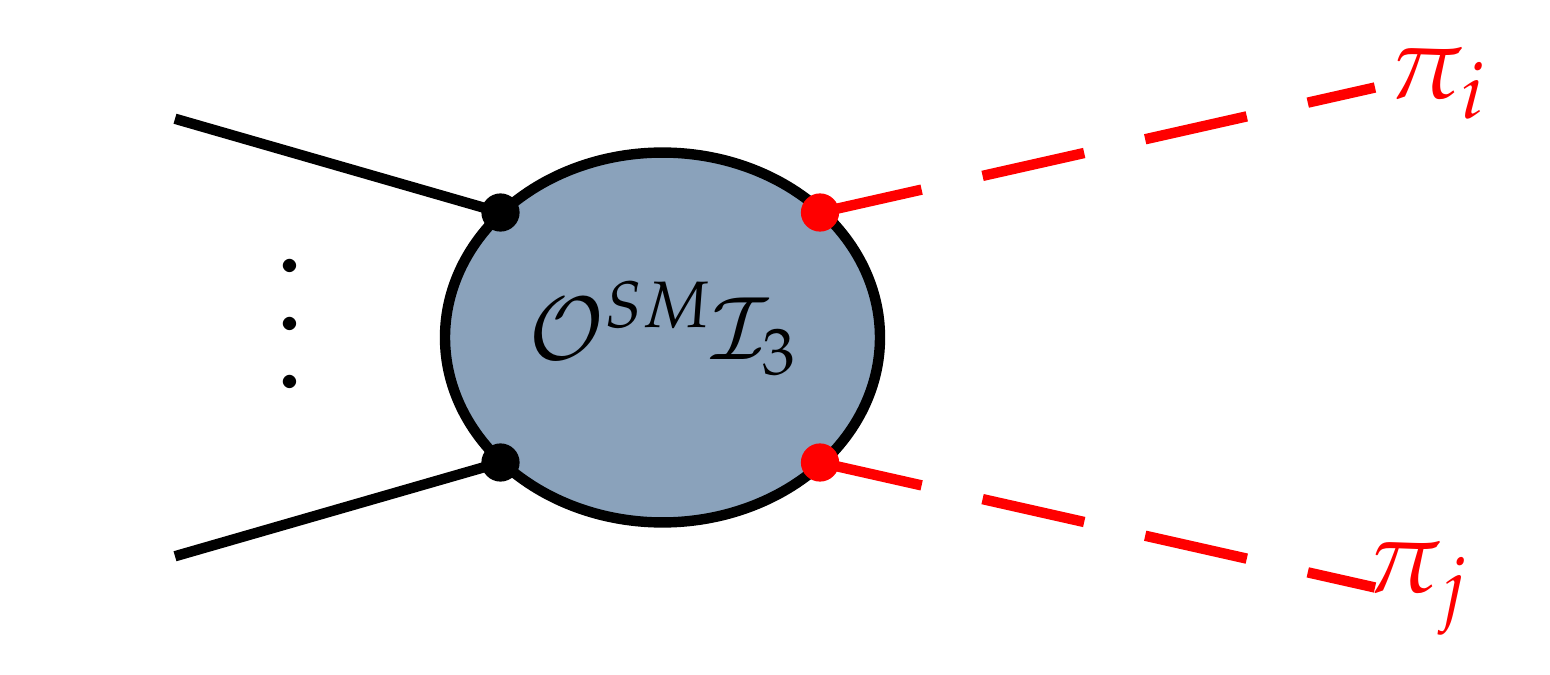}
         \caption{}
         \label{fig:feynman:pion_production-2pi}
    \end{subfigure}
	\begin{subfigure}[b]{0.49\textwidth}
         \centering
         \includegraphics[width=\textwidth]{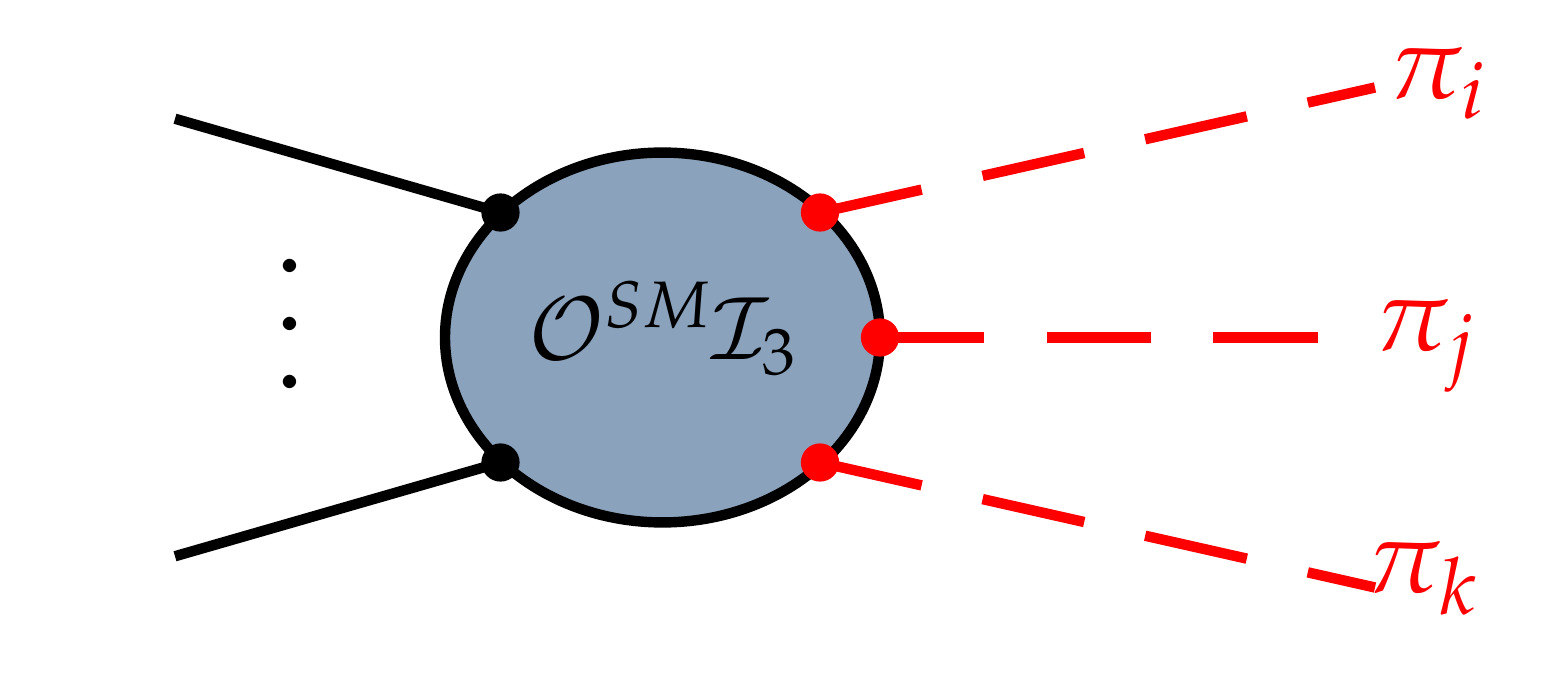}
         \caption{}
         \label{fig:feynman:pion_production-3pi}
 	\end{subfigure}
	\caption{Production of dGBs from SM collisions for a triplet of non-linearly realized $A_4$. The SM fields and interactions appear in black, the dGBs are shown in red.}
	\label{fig:feynman:pion_production}
\end{figure}

 Consider for illustration a generic interaction Lagrangian between SM fields which are singlets of the discrete symmetry --encoded in an operator $\mathcal{O}^{SM}$-- and the dGBs stemming from the $A_4$ invariant in the scalar potential with lowest dimension, $\I_3$,
\begin{equation}
\mathcal{L}^{\text{int}}\propto \frac{1}{M^m}\,\mathcal{O}^{SM} \I_3\,,
\label{effective-SM-dGB}
\end{equation}
where $m$ is some integer power which depends on the dimensionality of $\mathcal{O}^{SM}$, and $M$ denotes here the high-energy scale of the UV complete theory where both sectors would be connected.  Some expected topologies for dGB production from SM processes which follow from  Eq.~(\ref{eq:A4:invariants_pions_3}) are illustrated in Fig.~\ref{fig:feynman:pion_production}. The simultaneous emission of two or three degenerate dGB are thus the simplest processes. 

While the two dGBs are physically distinct fields, they are degenerate. The latter means that they cannot be individually distinguished in experiments.  The dGB observables will be given by the sum of  the probabilities for each possible individual final dGB state. This is alike to QCD in which the three colors of a quark cannot be experimentally separated even in the perturbative regime, and thus the perturbative QCD cross sections typically require one to compute the {\it probabilities} for each possible channel with given initial and final colors, to average  over color in the initial state, and to sum over all possible final color configurations.
 Therefore, for the example in Eq.~(\ref{effective-SM-dGB}) the probability will be proportional to  
 \begin{equation}
 \text{Prob}(\text{SM} \rightarrow n\, \pi) \propto\, \sum_{i,j....l=1,2} |\mathcal{A}(\text{SM}\rightarrow   \underbrace{\pi_i \pi_j....\pi_l}_n )|^2\,,
 \label{probability}
 \end{equation}
 where $n$ denotes the number of dGBs produced (i.e. $n=2, 3$ or $4$ for the example  in Eq.~(\ref{effective-SM-dGB})). Upon this proviso, it is easy to check that the physical observables are invariant under field reparametrizations, as they should,  i.e. be independent of the arbitrary parameter $\beta$ in Eqs.~(\ref{beta-redef}) and (\ref{beta-cubic-invariant}). While this is trivial for the two-dGB processes in Fig.~\ref{fig:feynman:pion_production-2pi},  the 3-dGB processes in Fig.~\ref{fig:feynman:pion_production-3pi} are shown to be $\beta$-invariant only upon implementation of the sum in Eq.~(\ref{probability}).

The third prediction depends on a particular structure of the relative probabilities for multi-dGB channels, which may point to the full UV discrete symmetry. 
For the real scalar triplet of $A_4$ and assuming that, to a good approximation, Nature's dGB interactions can be described  by the lowest dimension non-trivial operator,  i.e. as in Eq.~(\ref{effective-SM-dGB}), it follows that the cross sections $\sigma(\text{SM}\rightarrow n \pi)$ for 2-, 3-, and 4-dGB production processes can be calculated using the interactions of Eq.~(\ref{eq:A4:invariants_pions_3}) and
	\begin{align}
	\sigma( A_\text{SM} B_\text{SM} \to n\pi )
		&= \displaystyle\int \frac1{ 2 E_A 2 E_B | v_A - v_B | } \left| \mathcal{A} ( A_\text{SM} B_\text{SM} \to n \pi ) \right|^2 d\Pi_n \,,
	\end{align} 
where $E_{A, B}$ and $v_{A, B}$ are the energies and velocities respectively of the initial SM particles, and $d\Pi_n$ is the differential $n$-body phase space. Interestingly, the expansion of the $\I_3$ invariant in terms of the pions in \cref{eq:A4:invariants_pions_3} predicts certain specific relations among the n-pion interactions. Neglecting momentum-dependent $\mathcal{O}^{SM}$ in Eq.~(\ref{effective-SM-dGB}), the integral reduces to calculating the $n$-body final state phase space, and we can write:
	\begin{align}
	\qquad 
	\frac{\sigma(\text{SM} \rightarrow 2 \pi)}{\sigma(\text{SM} \rightarrow 3 \pi)} 
		= 2 f^2 \frac{ \Pi_2 }{ \Pi_3 }  
		\,,\qquad 
	\frac{\sigma(\text{SM} \rightarrow 3 \pi)}{\sigma(\text{SM} \rightarrow 4 \pi)} 
		= \frac{36 f^2 }{19(17)^2}\frac{ \Pi_3 }{\Pi_4}\,,\qquad 
	\label{ratios:Amp4-3}	
	\end{align}
In the $m_\pi\to0$ limit, we find:
	\begin{align}
	\qquad 
	\frac{\sigma(\text{SM} \rightarrow 2 \pi)}{\sigma(\text{SM} \rightarrow 3 \pi)} 
		= 64 \pi^2  \frac{f^2}{E_\text{CM}^2}\,,\qquad 
	\frac{\sigma(\text{SM} \rightarrow 3 \pi)}{\sigma(\text{SM} \rightarrow 4 \pi)} 
		= \frac{6(24\pi)^2}{19(17)^2}\frac{f^2}{E_\text{CM}^2}\,.\qquad
	\label{ratios:A4-3}	
	\end{align}
Information about the initial state cancels out when taking the above ratios, and so Eq~(\ref{ratios:A4-3}) holds for any number of particles in the initial state, and also gives the ratios of decay rates $\Gamma(\text{SM} \to n \pi )$. These ratios are specific to the $\I_3$ invariant under consideration, see \cref{eq:A4:invariants_pions_3}, and thus may constitute an interesting tool to disentangle the full UV discrete symmetry.  Remarkably, the invariants of the potential provide a handle to infer (at least partially) the UV discrete symmetry from the low energy observables. This is an improvement over the case of nonlinearly realized continuous symmetries, in which the full UV symmetry cannot be extracted from the dynamics of pGBs below the SSB scale~\cite{Low:2014nga}.

In this illustration, the SM sector is assumed to be a singlet of the unbroken UV discrete symmetry $D$. Were some SM fields to carry charges under that symmetry,  single dGB production and other channels would open. The theoretical construction would require one to embed SM fields in representations of the discrete group, which is a major task left for future work.

\subsubsection{Experimental signals from interactions with derivative couplings}
\label{signals-derivative-couplings} 
Although we have focused above on the signals stemming from the terms in the (discrete-symmetry invariant) potential, further signals are expected from the  terms in the dGB Lagrangian Eq.~(\ref{LdGB-2}) containing $\Phi$ derivatives, such as  for instance the interaction term
\begin{equation}
\mathcal{L}^{\text{int}}\propto  \frac{1}{M^m}\,\mathcal{O}^{SM} \partial_\mu \Phi^T \partial^\mu\Phi\,.
\label{effective-SM-dGB2}
\end{equation}
 It should be not expected {\it a priori } for these terms to  have coefficients suppressed due to the UV discrete symmetry, as they are also allowed by the embedding continuous symmetry, see Eq.~(\ref{quartic-derivative}). Nevertheless, they will exhibit the differentiating characteristics of its momentum dependence, which is a tool to disentangle them from those stemming from scalar invariants discussed in the previous subsection. They will share, though, the main characteristic, i.e. to lead to the simultaneous production of at least two dGBs, with no single dGB event, as far as the SM fields are uncharged under the discrete symmetry. Analogous considerations hold for the interactions involving mixed derivative/non-derivative dGB terms.  We leave to future work a detailed phenomenological analysis of the interaction terms involving dGB derivatives. These comments apply as well to the rest of the UV discrete symmetry scenarios to be discussed  below.

\subsubsection{Counting the number of dGBs produced in experiment } 
\label{counting-dGBs}
A typical signal of an ALP (or generic dark matter particle) is the observation of events with missing energy in excess of the SM background,  assuming that neutral particle is stable or can escape the detector before decaying.  The question of how to count how many degenerate invisible particles are simultaneously produced in a collision has been already addressed~\cite{Giudice:2011ib,Herrero-Garcia:2017vrl} in two cases: degenerate particles with  mass negligible compared to the collision energy, and degenerate but massive particles. 

It was shown that the distribution of visible (SM) particles suffices to infer the number of invisible particles ejected, without the need to reconstruct specific characteristics of the invisible sector. The key is the end-point behavior of certain observables, which has a strong power-law dependence on the number of invisible particles in the event.
 This method has the strength and clarity of a purely kinematic analysis, as it relies on the property that, in the end point of observables for the visible particles, the invisible ones must be produced either parallel to each other (if massless) or at rest. Possible  discriminators include the end point of the invariant mass and/or the invariant transverse mass of the visible particles, among others. In particular, the distributions for one versus more than one emitted invisible particles differ widely. 
 
 This approach, designed to hunt for multicomponent dark matter, can be translated to the setup of this work.  In summary, the absence of events with just one invisible (e.g. missing energy) track and the appearance instead of multi-invisible tracks is the first experimental tell-tale signal consistent with a BSM theory  UV protected by the mechanism explored in this paper.\footnote{The case in which some SM particles would carry discrete symmetry charges may lead to a different pattern. } This is the case whenever the non-trivial irreducible representations of the theory are lighter than the possible singlet ones, a pattern that holds through all the examples studied in this work.

\subsubsection{Inferrence of the underlying discrete symmetry}
\label{subsec:Inferrence}
It may not be possible  to infer the complete UV discrete symmetry  from only the low-energy part of the spectrum and interactions. The question is whether one can identify the explicit {\it \`a la Wigner} symmetry remaining, which by consistency would allow one to at least delineate the set of possible UV invariances. More importantly, they can be the lighthouse signal of an UV stable BSM theory with scalars.

Once the number of degenerate dGBs is experimentally established, it may be possible to identify the final discrete symmetry of the spectrum because the precise relative weights of the  terms cubic in dGB fields, and of each set of terms of higher order, is a trademark of the final explicit symmetry and representation. In the case discussed of a doublet of $Z_3$, these are the combinations in parenthesis in Eqs.~(\ref{eq:A4:invariants_pions_3})-(\ref{eq:A4:invariants_pions_6}). 

Under the further assumption that the lowest-dimension invariant dominates the potential, the relative weights of the two/three/four/... dGB amplitudes can provide hints of the complete invariant at play, as illustrated above. The point is that those ratios are a feature of the complete UV discrete symmetry D, not of the explicitly realized one. That is, although there is not a one-to-one correspondence between the amplitudes that follow from the first terms of a leading invariant and a particular UV discrete symmetry, the possibilities typically narrow down to a very reduced set of UV symmetries. This type of analysis has been illustrated for the real triplet of $A_4$ in Eqs.~(\ref{ratios:A4-3})  above and will be further illuminated below, upon the comparison  with the signals expected from alternative UV discrete scenarios.

\subsection{A related symmetry: \texorpdfstring{$S_4$}{} }

The case of the non-abelian UV discrete symmetry $S_4$ is close to that of $A_4$, as $S_4 = A_4\times Z_2$. For a  triplet  of real scalars, there will again be three primary invariants (corresponding to two for the dGB potential, taking into account the non-linearity constraint). For $S_4$ there are  two possible triplet representations, $\boldsymbol{3}$ and $\boldsymbol{3'}$. The interest of a triplet of $S_4$ as compared with a triplet of $A_4$ is two-fold:
\begin{itemize}
\item Further suppressed dGB masses for the $\boldsymbol{3'}$ representation. In this case, the three primary invariants are  of dimension two, four and six,  $\{\I_2, \I_4, \I_6\}$.
For the dGB potential, as $\I_2=f^2$ upon imposing the non-linearity condition, the first operator contributing to the dGB masses is $\I_4$, and furthermore the most general renormalizable potential reduces to this invariant. 
\item The predicted  dGB spectrum can have non-abelian properties.  Indeed,  the discrete symmetries exhibited by the  MaNa minima (for both the $\boldsymbol{3}$ and $\boldsymbol{3'}$ cases), include (very simple) non-abelian  ones: e.g. $S_3 = Z_3\times Z_2$. 
\end{itemize}
The $S_4$ theory also allows us to showcase how different exact UV discrete symmetries may be disentangled at low energies. While both  $S_4$ and $A_4$ predict the same number of degenerate dGBs (two), the assumption that the leading term of the dGB potential corresponds to the lowest dimensional invariant leads to different ratios among the multi-dGB amplitudes stemming from pure scalar invariants: for $S_4$ the dGB potential is expected to be dominated by $\I_4$ and therefore the ratios depend on the coefficients in \cref{eq:A4:invariants_pions_4} (the cubic invariant is absent) while for $A_4$  those ratios are fixed by $\I_3$, as studied in \cref{ratios:A4-3}.   For instance, the following cross-section ratios hold for the triplet of $S_4$ developed around its  $Z_3$-symmetric vacuum:
\begin{align}
\frac{\sigma\left(\text{SM}\rightarrow 2\pi\right)}{\sigma\left(\text{SM}\rightarrow 3\pi\right)} = 2(8\pi)^2\frac{f^2}{E_\text{CM}^2}\,,\qquad
\frac{\sigma\left(\text{SM}\rightarrow 3\pi\right)}{\sigma\left(\text{SM}\rightarrow 4\pi\right)} = \frac{27(8\pi)^2}{53371}\frac{f^2}{E_\text{CM}^2}\,,\label{eq:S4:ratios34}
\end{align}
to be compared with the analogous ones in Eq.~(\ref{ratios:A4-3}) for the triplet of $A_4$.

We refer to App.~\ref{App:S4} for the technical analysis and the detailed classification of the $S_4$  MaNa extrema. The next two chapters will instead focus on examples that demonstrate automatically very suppressed dGB masses or non-abelian symmetric spectra.

\section{Scalars in a triplet of \texorpdfstring{$A_5$}{} }
   This section analyzes the case of an exact (albeit non-linearly realized) $A_5$ invariance with scalars in a  $\boldsymbol{3}$ or a $\boldsymbol{3}'$ representation. It will be shown how this  discrete symmetry may further protect the dGB masses,  as the first possible discrete invariant contributing to the generic potential scalar is  of higher dimension than that for the $A_4$ (and $S_4$) triplets discussed in the previous section.

\subsection{Invariants and potential}
As the high-energy (the low-energy) theory has three (two) dynamical scalar degrees of freedom, we expect again three (two) primary invariants to describe the scalar physics at those energies. 

For a real scalar triplet, the most general scalar potential invariant under the UV $A_5$ symmetry will be expressed as a function of all possible $A_5$ invariants built out of that representation.  The dimension and quantity of the primary and secondary invariants follows from the Molien generating function~\cite{Molien:1897invarianten,Burnside:1911theory}.\footnote{ For comprehensive reviews of this formalism see Refs.~\cite{Patera:1978qx, Merle:2011vy}.} All other analytical invariants will be polynomial combinations of the primary and secondary invariants. The Molien functions for the $\boldsymbol{3}$ and $\boldsymbol{3}'$ of $A_5$ are identical, and read
 \begin{equation}
	\F_{A_5}(\boldsymbol{1}, \boldsymbol{3}, \lambda) = \frac{1+ \lambda^{15}}{(1-\lambda^2)(1-\lambda^6)(1-\lambda^{10})}\,,
	\label{eq:A5:Molien}
\end{equation}
which implies three primary invariants of order 2, 6, and 10, together with a syzygy that allows one to construct a 15-th order secondary invariant in terms of  the primary invariants. 
 For their explicit construction, one can profit from the fact that $A_4$ is a subgroup of $A_5$ \cite{ishimori2010non}. In consequence, the three primary   invariants for 
 the  $\boldsymbol{3}$ of $A_5$  can be written in a compact way as combinations of the three primary $A_4$ invariants ($\I_2$, $\I_3$, $\I_4$) and the secondary $A_4$ invariant ($\I_6$)  
 given in Eqs.~(\ref{eq:invariant_2})-(\ref{eq:A4:invs:syzigy}),  
\begin{align}
	\I_2^{(\boldsymbol{3}, A_5)}&= \I_2\,,\\
	\I_6^{(\boldsymbol{3}, A_5)}
	&= 22\,\I_3^2 + \I_2\,\I_4 -2\sqrt{5}\,\I_6\,,\\
	\I_{10}^{(\boldsymbol{3}, A_5)} &= \I_2^2\,\I_4 + 38\,\I_3^2\,\I_4 - \frac{7}{11}\,\I_2^3\,\I_4 - \frac{128}{11\sqrt{5}}\,\I_2^2\,\I_6 + \frac{6}{\sqrt{5}} \,\I_4\,\I_6\,.
\label{eq:A5:invs:integrity_basis}
\end{align}
The invariants for the $\boldsymbol{3'}$ representation of $A_5$ are exactly the same as for the $\boldsymbol{3}$. 

Once again, the quadratic invariant  $\I_2^{(\boldsymbol{3}, A_5)}$ will not contribute to the dGB potential, given that the non-linearity constraint in Eq.~(\ref{equationf-A4-3}) also holds here. This theory is therefore protected from quadratic instabilities via quantum corrections induced by physical scales in the theory heavier than $\Lambda$.  Furthermore, the first invariant which can induce a mass for  the dGBs now has dimension six, 
\begin{equation}
    V_{\text{dGB}}  = {f^2} \Lambda^2\, \big[ \hat{c}_6 \frac{\I_6}{f^6}  + \hat{c}_{10} \frac{\I_{10}}{f^{10}}
    		+ \hat{c}_{12} \frac{\I_{6}^2}{f^{12}} 
		 + \hat{c}_{15} \frac{\I_{15}}{f^{15}}\cdots\big]\,,
  \label{Vreduced-A5-3}
  \end{equation}
and thus the dGB masses are expected to be weighted down by a strong $\epsilon^6$ suppression factor due to the UV discrete symmetry, see Eq.~(\ref{cn}) and the subsequent discussion. Therefore, the dGB masses are much more suppressed than in the case of $A_4$ with a scalar triplet (which exhibited an $\epsilon^3$ suppression). 

Furthermore, note that in this case {\it the physical impact of the  complete potential should be well approximated by just one invariant, $\I_6$,} given the strong hierarchy of the potential coefficients: the impact of the other primary invariant, $\I_{10}^{(\boldsymbol{3}, A_5)}$,  the secondary invariant $\I_{15}^{(\boldsymbol{3}, A_5)}$,  and the rest of the terms, can be disregarded here altogether given their very high suppression.

 \subsubsection{Embedding in a continuous group: \texorpdfstring{$SO(3)$}{}.}
 Alike to $A_4$,  $A_5$ is a subgroup of $SO(3)$,  for which two irreducible triplet representations  are possible, see  Table~\ref{table:finite_groups_SO3}. Again, the only possible pattern of spontaneous symmetry breaking (SSB) of the continuous group with those representations  is  $SO(3)\to SO(2)$, and consequently two physical pion fields, $\pi_1$ and $\pi_2$, would survive at low energies, see Eqs.~(\ref{equationf-A4-3}) and (\ref{eq:pions definition}).  
   
   The highly suppressed potential coefficients and their strong hierarchy imply that an exact (and spontaneously broken) UV $A_5$ symmetry induces the existence of an approximate continuous (spontaneously broken) $SO(3)$ invariance, which is explicitly broken to a very mild extent. This  leads to highly suppressed pGBs masses as compared with the overall effective scale $\Lambda$.

As stated before, in what follows we do not rely on any particular choice of continuous group among those in which the discrete group can be  embedded.

\subsubsection{ MaNa extrema}
\label{extrema_triplet_A5}
\begin{figure}[h!]
    \centering
     \begin{subfigure}{0.49\textwidth}
    \includegraphics[width = \textwidth]{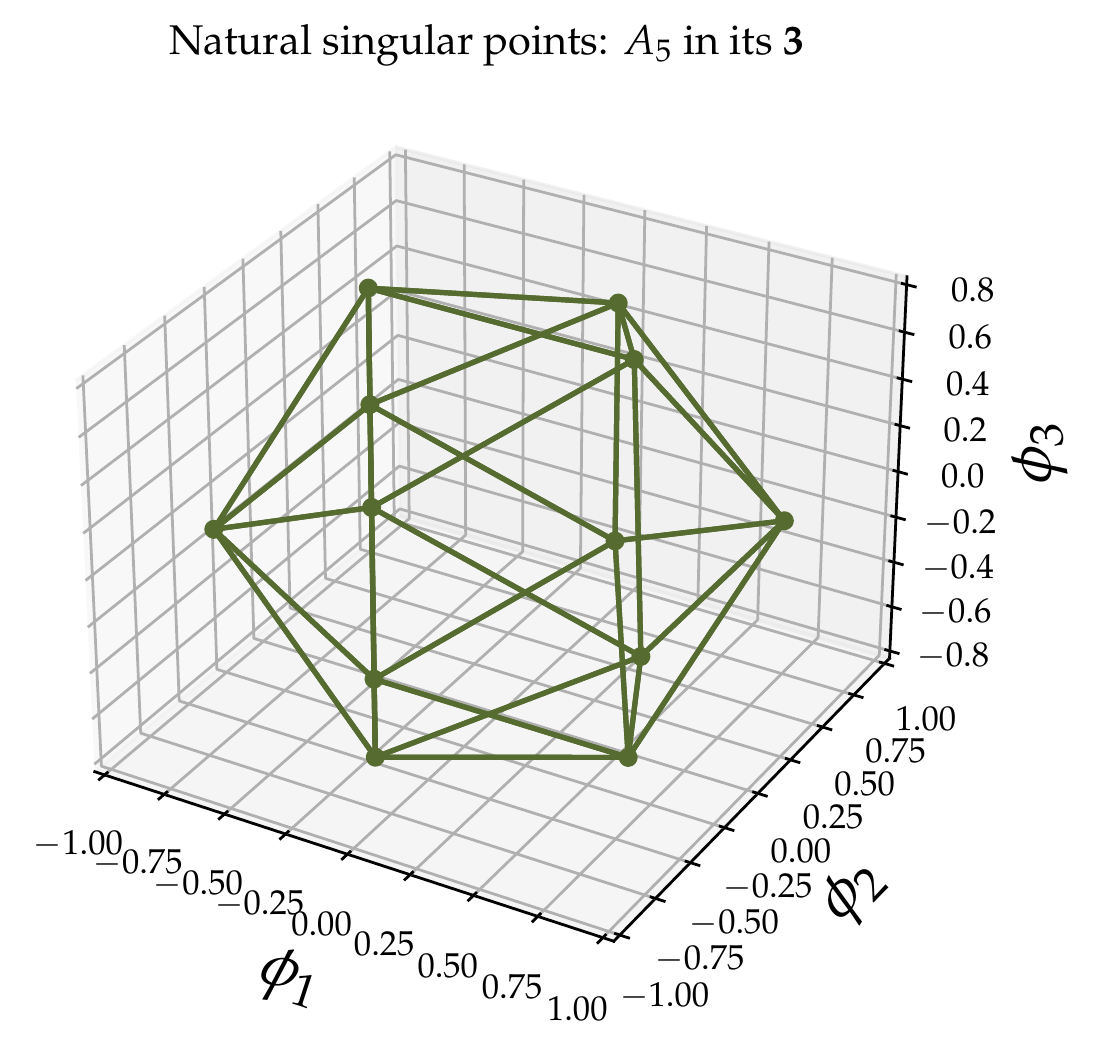}
    \caption{}
    \label{fig:A5:singular_points_z5}
    \end{subfigure}
    \begin{subfigure}{0.49\textwidth}
    \includegraphics[width = \textwidth]{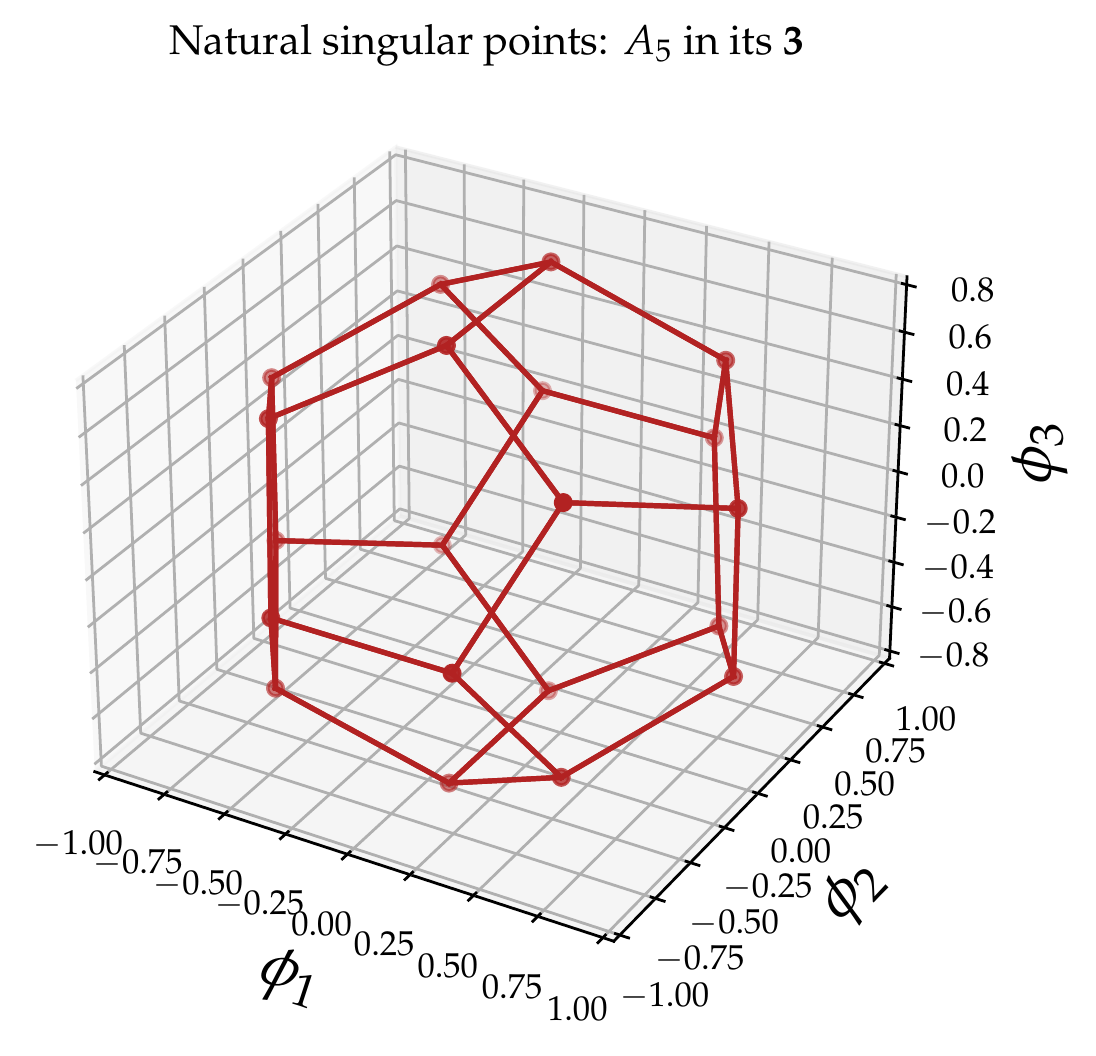}
    \caption{}
    \label{fig:A5:singular_points_z3}
    \end{subfigure}
    \caption{Geometrical distribution in field space of the MaNa extrema which are invariant under $Z_5$ (a) and $Z_3$ (b). The field values are given in units of $f$. }
    \label{fig:A5:singular_points}
 \end{figure}

There are a total of 62 MaNa extrema of the  $A_5$ invariants described above. They can be classified as: 
\begin{itemize}
	\item 12 stable extrema with $Z_5$ as their little group. They are the vertices of a dodecahedron in \cref{fig:A5:singular_points_z5}.
	\item 20 stable extrema with $Z_3$ as little group. Geometrically, they are the vertices of an icosahedron in \cref{fig:A5:singular_points_z3}.
	\item 30 saddle points with $Z_2$ as their little group. A total of 20  lie at the  midpoints of the edges of the icosahedron. 
\end{itemize}
The counting constraint in Eq.~(\ref{counting-rule}) works out nicely, $60 = 12\times5= 20\times3 = 30\times2$, and also the number of extrema is consistent, $62=12+20+30$.
Table~\ref{tab:A5:critical_points} shows the location in field space of some representatives of the MaNa extrema, together with the  value of the invariants at those points and their exact (explicit) invariances. 

\begin{table}[h!]
\centering
\begin{align*}
    \begin{array}{|c||cc|ccc|c|c|}
        \multicolumn{8}{c}{\text{\textbf{MaNa extrema for a triplet of} $\boldsymbol{A_5}$}}\\
        \hline
        \multirow{3}{*}[0.23cm]{\text{Point}} &\multirow{3}{*}[0.23cm]{$\I_6$} & \,\multirow{3}{*}[0.23cm]{$\I_{10}$} & \phi_{1}\, &\phi_{2}\, &\phi_{3} & \multirow{3}{*}[0.23cm]{\text{Little group}}& \multirow{3}{*}[0.23cm]{\text{Nature}}\\
         & & & \multicolumn{3}{c|}{\textit{(Representatives)}} & & \\[1pt]
        \hline\hline
        \multirow{1}{*}{A} & \multirow{1}{*}{$\frac{1}{5}$} & \multirow{1}{*}{$-\frac{472}{1375}$}  & \pm\sqrt{\frac{5-\sqrt{5}}{10}} & 0 & \mp \sqrt{\frac{5+\sqrt{5}}{10}} & \multirow{1}{*}{$Z_5$} &\multirow{1}{*}{\text{Minima}}\\[0.8ex]
        \hline  \rule{0pt}{2.3ex}       
        \multirow{2}{*}[-0.15cm]{B} & \multirow{2}{*}[-0.15cm]{$\frac{31}{27}$} & \multirow{2}{*}[-0.15cm]{$\frac{328}{891}$} & \frac{1}{\sqrt{3}} & \frac{1}{\sqrt{3}} & \frac{1}{\sqrt{3}} & \multirow{2}{*}[-0.15cm]{\text{$Z_3$}} & \multirow{2}{*}[-0.15cm]{\text{Minima}}\\[0.8ex]
                              &                                         & & \sqrt{\frac{3+\sqrt{5}}{6}} & 0 &\sqrt{\frac{3-\sqrt{5}}{6}} &  & \\[0.8ex]
        \hline  \rule{0pt}{3.5ex}
        \multirow{2}{*}[-0.15cm]{C} & \multirow{2}{*}[-0.15cm]{$1$} & \multirow{2}{*}[-0.15cm]{$\frac{4}{11}$} & 0 & 0 & \pm1 & \multirow{2}{*}[-0.15cm]{$Z_2$}& \multirow{2}{*}[-0.15cm]{\text{Saddles}}\\
                              &                         &               & \frac{\left(-1+\sqrt{5}\right)}{4} & -\frac{1}{2} & \frac{\left(1+\sqrt{5}\right)}{4} &    & \\[0.8ex]
    \hline
    \end{array}
\end{align*}
\caption{Location and symmetries of a few representative  MaNa extrema for  the case of a triplet scalar of $A_5$. The values of the invariants and the locations are normalized to $f=1$. The location in field space of the ensemble of MaNa minima is depicted in Fig.~\ref{fig:A5:singular_points}. The manifold is  illustrated in Fig.~\ref{fig:A5:invariant_manifold}.}
\label{tab:A5:critical_points}
\end{table}

\begin{figure}[h!]
\centering
\includegraphics[width=0.7\textwidth]{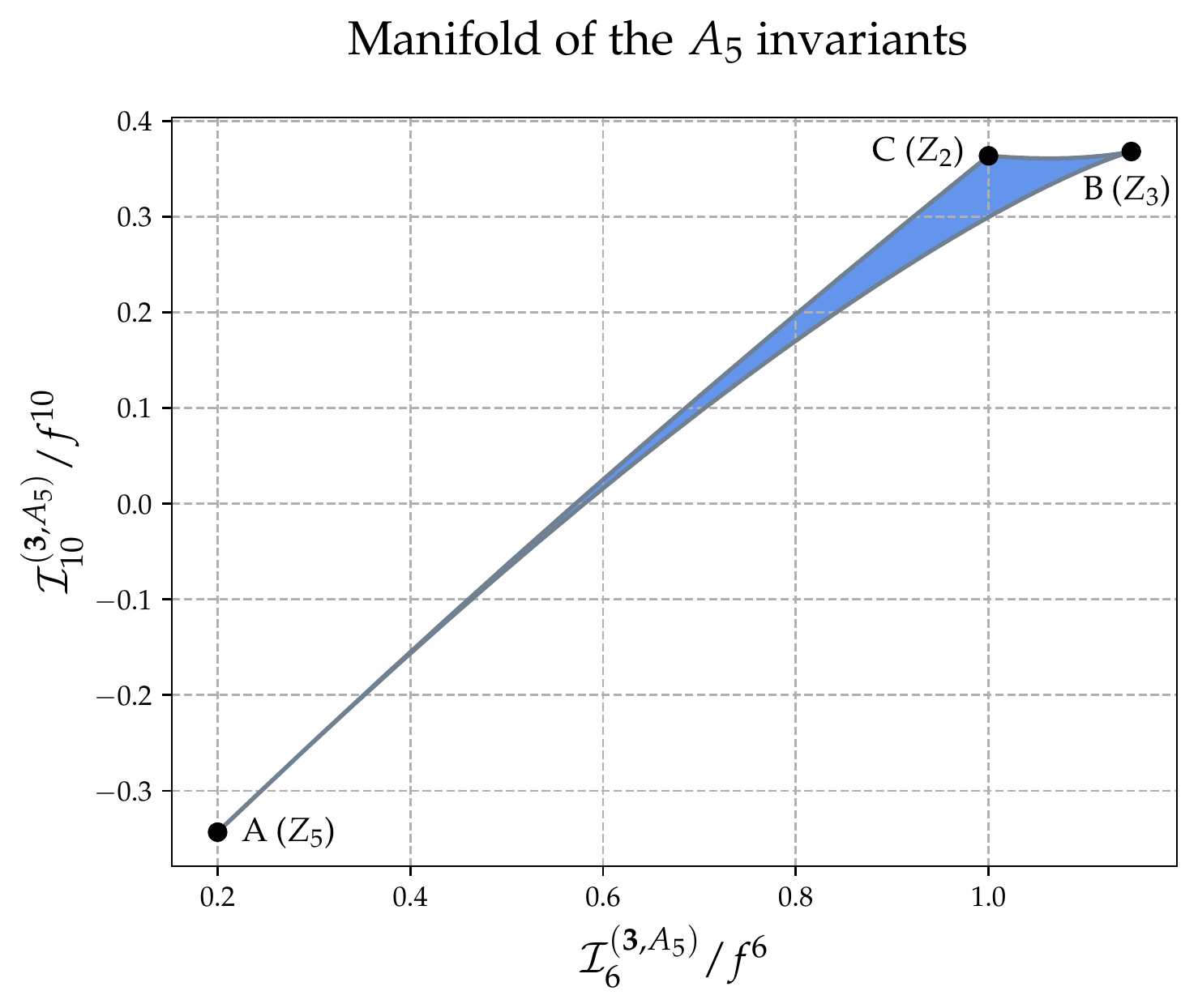}
\caption{The manifold defined by the $\mathcal{I}^{(\boldsymbol{3}, A_5)}_6(\Phi)$ and $\mathcal{I}^{(\boldsymbol{3}, A_5)}_{10}(\Phi)$ invariants of the non-linearly realized $A_5$ 
with a scalar triplet. The MaNa extrema correspond to the singular points of this manifold: $A,B \text{ and }C$.}
\label{fig:A5:invariant_manifold}
\end{figure}

The locations in field space of the ensemble of the  MaNa extrema is depicted in Fig.~\ref{fig:A5:singular_points}. The manifold defined by the two $A_5$ primary invariants relevant for the potential is depicted in Fig.~\ref{fig:A5:invariant_manifold}.

The set of $Z_3$-symmetric points and the set of $Z_5$-symmetric points cannot be simultaneously minima or maxima.
Depending on the sign of the $\hat c_i$ coefficients, either one of the other are minima, while the $Z_2$-symmetric extrema are always saddle points. It is interesting to note that the larger the symmetry at a MaNa minima, the deeper  the potential well at that point.

The expansion of the invariant $\I_6^{(\boldsymbol{3}, A_5)}$ in terms of dGBs around the $Z_3$-symmetric MaNa minimum   $\Phi_B\,=\,\frac{1}{\sqrt{3}}\left(1,1,1\right)^\top$ yields the following contribution to the potential:
\begin{multline}
	\I_6^{(\boldsymbol{3}, A_5)}= \frac{32}{9}\,f^4\,\left[
	\frac{31}{96}f^2
	- \left(\pi_1^2+\pi_2^2\right) 
	+\frac{10\sqrt{2}}{24f} \left(\pi_1^3-3\pi_1\pi_2^2\right) \right.\\
	\ \ \ \ \ \left.+ \frac{\sqrt{30}}{4 f}\left(\pi_2^3-3\pi_1^2\pi_2\right) 
	+\frac{31}{12 f^2}\left(\pi_1^2+\pi_2^2\right)^2 \right]\,,
	\label{Z3minimum}
\end{multline}
which is a combination of the quadratic and the two cubic $Z_3$  invariants for a doublet representation, see Eqs.~(\ref{eq:A4:invariants_Z3_2})-(\ref{eq:A4:sizygy_Z3}). If the potential can be approximated by this lowest invariant  (with the appropriate sign $V_{\rm dGB}\sim -\I_6^{(\boldsymbol{3}, A_5)}$), this $Z_3$-symmetric point will correspond to a minimum.

In contrast, the expansion of $\I_6^{(\boldsymbol{3}, A_5)}$ around the $Z_5$-symmetric MaNa minimum $\Phi_A$ yields 
\begin{equation}
	\begin{split}
	\!
	\I_6^{(\boldsymbol{3}, A_5)}&= \frac{32}{5} f^4
	\left[ \frac{f^2}{32}
	+ \left(\pi_1^2+\pi_2^2\right) 
	\!-\!\frac{31}{12 f^2}\left(\pi_1^2+\pi_2^2\right)^2
	\!-\!\frac{1}{4 f^3}\left(\pi_1^5 - 10\pi_1^3\pi_2^2 + 5\pi_1\pi_2^4\right)\right],
	\end{split}
	\label{Z5minimum}
\end{equation}
showing that the nature of this family of  MaNa extrema is revealed by terms which are of fifth order in the pion dependence, reflecting the 5-point symmetry of the theory.  Indeed, the Molien function for the irreducible 
doublet representation of $Z_5$ reads
 \begin{equation}
	\F_{Z_5}(\boldsymbol{1}, \boldsymbol{2}, \lambda) = \frac{1+ \lambda^{5}}{(1-\lambda^2)(1-\lambda^5)}\,,
	\label{eq:A5:Molien}
\end{equation}
which implies  two primary invariants of mass dimension two and five,
\begin{align}
\I_2^{(\boldsymbol{2}, Z_5)} &= \pi_1^2 + \pi_2^2\\
\I_5^{(\boldsymbol{2}, Z_5)}&= \pi_1^5 - 10\pi_1^3\pi_2^2 + 5\pi_1\pi_2^4\,,
\end{align}
plus a secondary invariant also of dimension five, 
\begin{align}
\I_5^{(\boldsymbol{2}, Z_5)} &= \pi_2^5 - 10\pi_1^2\pi_2^3 + 5\pi_1^4\pi_2\,.
\end{align}
The expansions in terms of dGBs  for $\I_6^{(\boldsymbol{3'}, A_5)}$ are identical to those for $\I_6^{(\boldsymbol{3}, A_5)}$. Again, the physical results can be easily proven to be independent of changes of parametrization of the dGB fields, see Eq.~(\ref{beta-redef}),  as they should be. 

\subsection{Phenomenological signals with an \texorpdfstring{$A_5$}{} \ triplet}
\begin{figure}[h!]
\centering
	\begin{subfigure}[b]{0.325\textwidth}
         \centering
         \includegraphics[width=\textwidth]{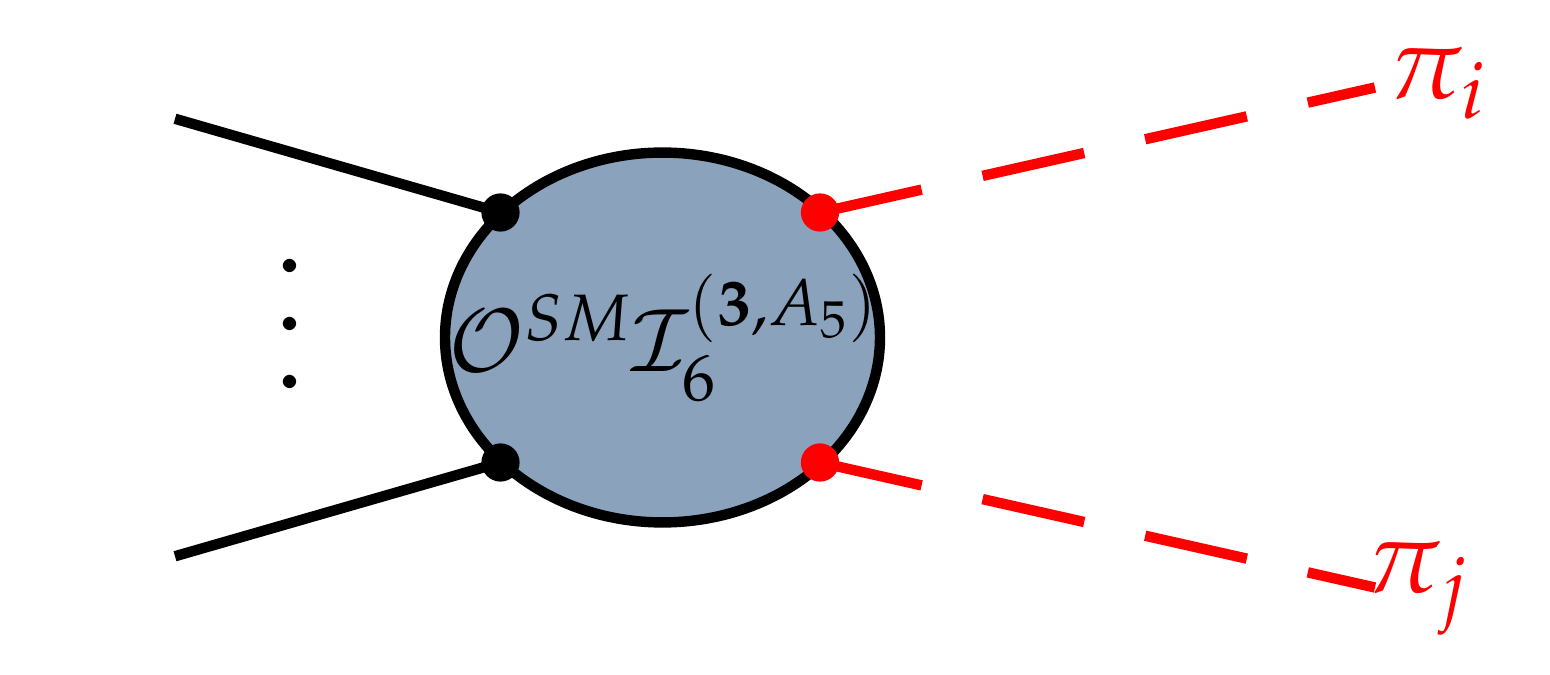}
         \caption{}
    \end{subfigure}
    \begin{subfigure}[b]{0.325\textwidth}
         \centering
         \includegraphics[width=\textwidth]{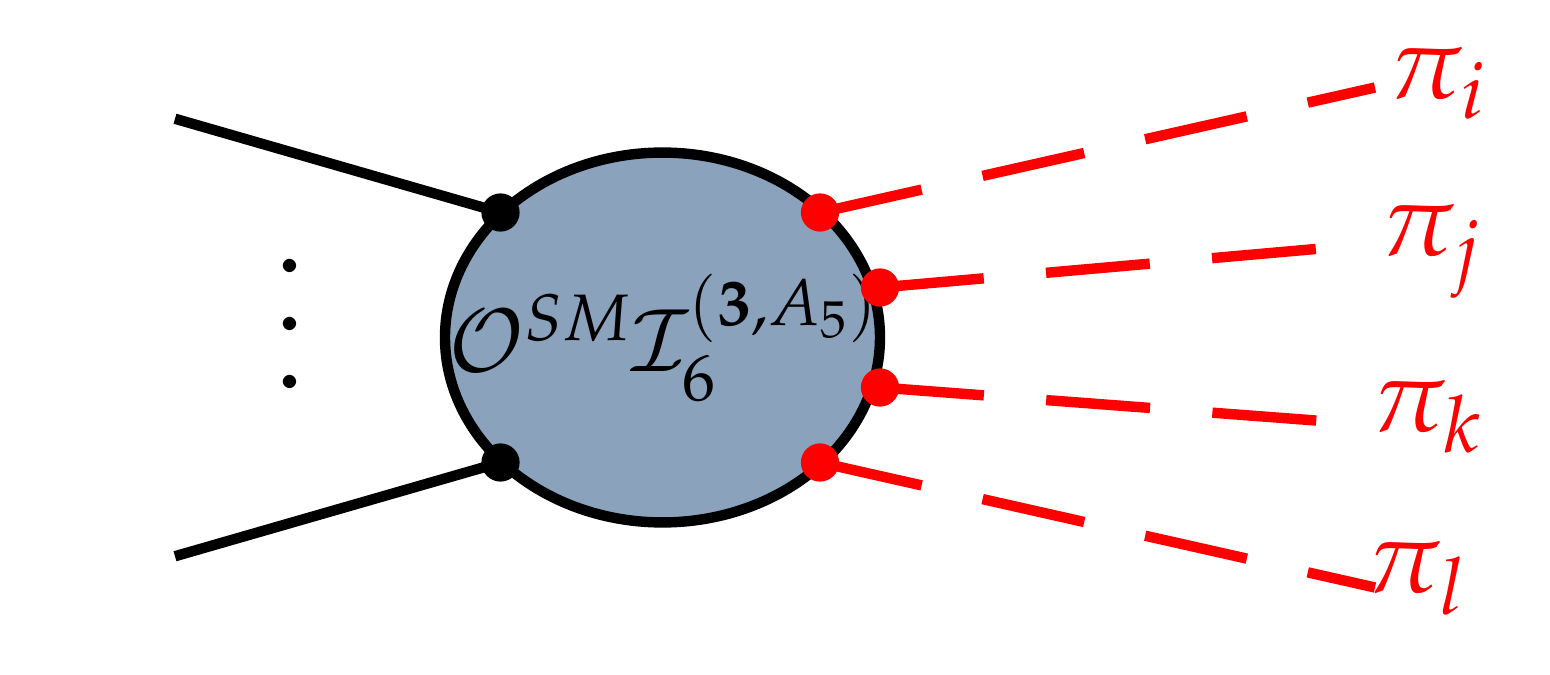}
         \caption{}
 	\end{subfigure}
	\begin{subfigure}[b]{0.325\textwidth}
         \centering
         \includegraphics[width=\textwidth]{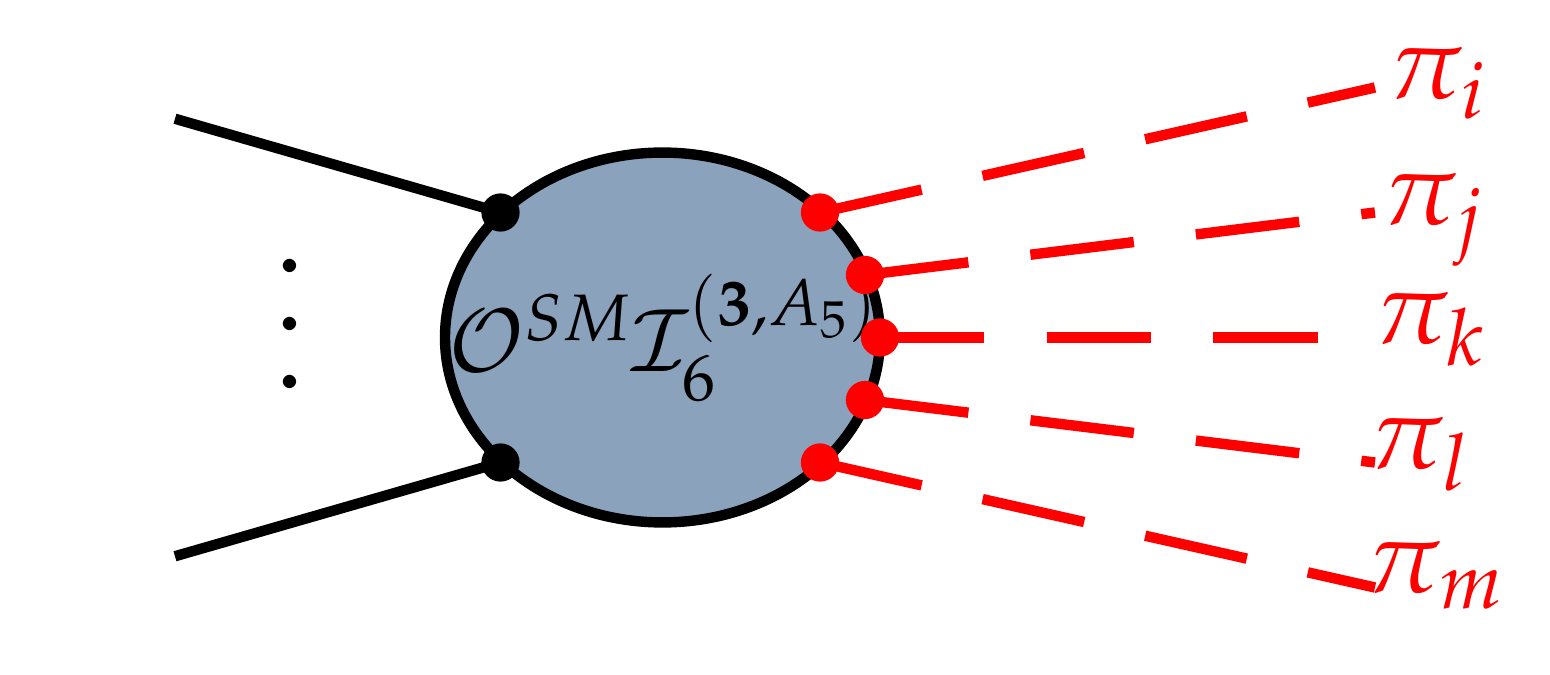}
         \caption{}
 	\end{subfigure}
	\caption{Production of dGBs from SM collisions, for a triplet of non-linearly realized $A_5$. While events with three dGBs are absent, those with five dGBs are characteristic of this kind of dGBs whenever the $Z_5$-symmetric MaNa minimum are the absolute minimum. The color code is that of Fig.~\ref{fig:feynman:pion_production}.}
	\label{fig:feynman:pion_production_Z5}
\end{figure}

Given the $Z_5$ symmetry at the MaNa minimum with the largest little group, the signals to be expected mirror those obtained earlier for the $A_4$ case,  see Sec.~\ref{pheno-A4}, that is:
\begin{itemize}
\item Two degenerate dGBs, as befits the only irreducible representation of $Z_N$ with real fields and dimension larger than one (the $\boldsymbol{2}$).  Their masses are expected to be naturally more suppressed than those for the $A_4$ case, though, as compared with the overall UV scale,  as explained above. 
\item  The simultaneous emission of  two, four or five degenerate dGBs in SM initiated processes is expected (as long as the SM fields are singlets of the UV discrete symmetry). No events with only one dGB emitted are then expected, nor with three (in contrast to the $A_4$ scenario).
\end{itemize}
The latter property can be illustrated with the generic interaction Lagrangian,
\begin{equation}
\mathcal{L}^{\text{int}}\propto
\frac{1}{M^m}\,\mathcal{O}^{SM} \I_6^{(\boldsymbol{3}, A_5)}  + \cdots \,,
\label{effective-SM-dGB_A5}
\end{equation}
 expanded around the dGB fields at the MaNa minimum with the largest little group, which we showed above to exhibit a {\it \`a la Wigner} $Z_5$ symmetry. The expansion  around this minimum --see Eq.~(\ref{Z5minimum})-- shows  that no tree-level production topology is expected with three dGBs emitted (unlike for the UV $A_4$ case), and {\it the first vertex with an odd number of dGBs involves five dGBs}. This is illustrated in Fig.~\ref{fig:feynman:pion_production_Z5}.   The considerations about reparametrization invariance and the production probability in Eq.~(\ref{probability}) apply as well in this case. 

In consequence, this scenario yields a $Z_5$-symmetric spectrum resulting in tell-tale relative probabilities for multi-dGB channels from 
the interaction in Eq.~(\ref{effective-SM-dGB_A5}), which are null for three dGB production. Using Eq.~(\ref{probability}), we find:
\begin{equation}
\frac{\sigma(\text{SM} \rightarrow 2 \pi)}{\sigma(\text{SM} \rightarrow 4 \pi)} 
	= \frac{ 18  f^4 }{ 19(31)^2 } \frac{ \Pi_2 }{ \Pi_4 } \,,\qquad 
\frac{\sigma(\text{SM} \rightarrow 4 \pi)}{\sigma(\text{SM} \rightarrow 5 \pi)}
	= \frac{19(31)^2 f^2 }{3 (45)^2 } \frac{ \Pi_4 }{ \Pi_5 }\,.
\label{ratios:A5-3}
\end{equation}
Evaluating the final state phase space gives explicitly:
\begin{equation}
\frac{\sigma(\text{SM} \rightarrow 2 \pi)}{\sigma(\text{SM} \rightarrow 4 \pi)} = \frac{ 216 (4\pi)^4}{ 19(31)^2 } \frac{ f^4 }{ E_\text{CM}^4 } \,,\qquad \frac{\sigma(\text{SM} \rightarrow 4 \pi)}{\sigma(\text{SM} \rightarrow 5 \pi)}= \frac{19(31)^2 (8\pi)^2}{(45)^2 } \frac{ f^2 }{ E_\text{CM}^2 }\,.
\label{ratios:A5-3}
\end{equation}
These are to be compared with Eq.~(\ref{ratios:A4-3})    for the  $Z_5$-symmetric spectrum (stemming from a scalar triplet of a UV $A_5$ symmetry) developed in the previous section.
 
Analogous remarks to those in Sect.~\ref{signals-derivative-couplings} for the signals expected from interaction terms containing $\partial_\mu U$ components apply in this case.
 
 \subsubsection{Generalization to low-energy \texorpdfstring{$Z_N$}{} symmetric spectra}
 
 The examples above have been constrained to triplet (real) scalar representations of simple UV discrete symmetries, non-linearly realized,\footnote{ Which happen to be contained in $SO(3)$.} which we showed lead to MaNa minima with explicit $Z_3$ or $Z_5$ symmetry (or involving them).  Larger irreducible representations of  various UV discrete symmetries   
 could be considered along the same lines, which may leave a low-energy spectra invariant under some $Z_N$ symmetry. 
 
 The important point is that, whatever the value of $N$, the only irreducible representations of $Z_N$ --aside from singlets-- are doublets,\footnote{  Rotations of angle $2\pi/N$ in the plane $\mathds{R}^2$ are always a representation of $Z_N$  with real fields.  Such a rotation has no invariant subspace and thus it is irreducible and of dimension two. This is not the case rotations in $\mathds{R}^{n> 2}$. In consequence, the only irreducible representations  for any $Z_N$ are doublets and singlets.} whose lowest dimensional invariant has dimension two.  This implies that whenever the final preserved discrete symmetry of the spectrum is abelian, i.e.  $Z_N$,  two degenerate dGBs (or several sets of them) are to be expected, and no events with just one invisible (e.g. missing energy) track should be found among the leading signals, as far as the SM is a singlet of the discrete symmetry. Furthermore, the characteristic tell-tale production topology for the absolute minimum is that of $N$-point dGB interactions. 

The experimental methods discussed in Sec.~\ref{counting-dGBs} to disentangle how many invisible particles are ejected in a collision or decay  are valid for any $N$.  The relative probabilities of events with different numbers of dGBs contain information which {\it a priori} can allow one to  identify the surviving discrete symmetry of the low-energy spectrum and interactions, as exemplified above. This in turn may delimitate by consistency the set of possible UV discrete invariances which are ultimately responsible for the increased UV convergence of the BSM theory.

A qualification is pertinent here for the case of large UV discrete groups. The UV protection under study relies on a non-linear realization of the discrete symmetry $D$ for an irreducible $m$-dimensional real representation of the group, and this leads to $m-1$ dGBs at low energies.  Large discrete groups may not have irreducible representations of dimension smaller than four, though (e.g. $m\ge5$ for $A_6$). If the surviving explicit symmetry of the spectrum is abelian, it follows then that the $m-1$ low-energy dGB degrees of freedom will belong to a reducible representation. The latter is expected to decompose as a combination of dGB doublets (for $m$ odd),  or a combination of doublets plus a singlet (for $m$ even).  The examples that we have analyzed suggest that the singlet may be typically heavier than the doublets. In summary, the leading experimental signals (with SM fields uncharged under $D$) are still characterized then by the absence of single ALP emission, and the simultaneous emission of  (sets of) two degenerate ALPs, as a tell-tale of an abelian-symmetric spectrum.

A pertinent question is whether the prediction of simultaneous dGB production  could be broken in favor of emission of a single dGB instead, for instance  if the UV symmetry  were to be classically exact but anomalous (assuming the SM sector is uncharged under the discrete group).\footnote{We thank Anson Hook for a question prompting this consideration.}  This cannot happen, though, as  the non-abelian UV symmetries discussed are subgroups of continuous ones, which cannot be anomalous because the non-abelian generators are traceless. This remark only depends on the non-abelian character of the UV discrete symmetry and in consequence applies as well to the next section.
\newpage
\section{Scalars in a quadruplet of \texorpdfstring{$A_5$}{} : non-abelian symmetry of the spectrum }
\label{sec-A5-4}

The analysis in the two previous sections led to predict low-energy spectra  explicitly symmetric under abelian groups, e.g. $Z_3$ and $Z_5$ (except for the case of  $S_3$  from $S_4$  which led to  low-energy consequences similar to those for abelian little groups), as long as the scalar vevs correspond to the MaNa minima. The physical dGBs to be detected are then degenerate ones belonging to the largest irreducible representation of any $Z_N$ group: the $\boldsymbol{2}$, while the interactions could allow one to disentangle the precise abelian group, i.e. to identify $N$.

We address here whether the  explicit (i.e.\,{\it\`a la Wigner})  discrete  symmetry of the spectrum can be non-abelian instead, and what would then be the differentiating signals.
 A non-trivial case shown below to realize this scenario is an EFT built out of a real scalar field in a $\boldsymbol{4}$ of $A_5$.

\subsection{Invariants and Potential}
For four independent degrees of freedom at high-energies, four primary $A_5$ invariants are expected, with all other possible invariants being a function of those.
 Indeed, the Molien function for a quadruplet of $A_5$ reads 
\begin{equation}
\mathcal{F}_{A_5}(\boldsymbol{1}, \boldsymbol{4}, \lambda) = \frac{1+\lambda^{10}}{(1-\lambda^2)(1-\lambda^3)(1-\lambda^4)(1-\lambda^5)}\,,
\label{eq:A5:Molien_function}
\end{equation} 
which indicates four primary invariants with mass dimension two, three, four and five. 
 We choose the particular realization of the  $\boldsymbol{4}$ representation of $A_5$, $\{\phi_1, \phi_2, \phi_3, \phi_4\}$,  to be that for which one of its $A_4$ subgroups  lives
 in the three first components of the quadruplet, i.e. $\{\phi_1, \phi_2, \phi_3\}$ forms a triplet of the $A_4$ subgroup. The $A_5$ invariants can then be written in terms of  polynomials constructed with the invariants $\I_2$, $\I_3$ and $\I_4$ for the $\boldsymbol{3}$ of $A_4$, see Eqs.~(\ref{eq:invariant_2})-(\ref{eq:invariant_4}), plus $\phi_4$, 
\begin{align}
\I_2^{(\boldsymbol{4}, A_5)} & = \I_2 + \phi_4^2\,,\\
\I_3^{(\boldsymbol{4}, A_5)} & = \I_3
- \frac{\phi_4}{2\sqrt{5}}\I_2 + \frac{\phi_4^3}{2\sqrt{5}}\,,
\\
\I_4^{(\boldsymbol{4}, A_5)} & = \I_4 
+ \frac{12}{\sqrt{5}}\I_3\phi_4 + \frac{12}{5}\I_2\phi_4^2 + \frac{\phi_4^4}{5}\,,
\\
\I_5^{(\boldsymbol{4}, A_5)} & = 
\I_4\phi_4 
-\frac{1}{2}\I_2\phi_4 
- \frac{4}{\sqrt{5}}\I_3\phi_4^2 
- \frac{\phi_4^3}{5}\I_2+ \frac{\phi_4^5}{50}\,.
\label{eq:A5_4:invariants_bonitos}
\end{align}
The non-linear constraint reads in this case,
\begin{equation}
 \I_2^{(\boldsymbol{4}, A_5)} = \Phi^T \Phi= \phi_1^2+\phi_2^2+\phi_3^2  + \phi_4^2= f^2\,,
 \label{non-linear-A5-4}
\end{equation} 
and thus the first explicit breaking of the continuous symmetry will appear through a cubic $A_5$  invariant. In other words, as in all previous scenarios,  the theory is protected from quadratic corrections at the quantum level.

In the spirit of EFT,  the generic potential can thus be written as  
  \begin{equation}
    V_{\text{dGB}} = {f^2} \Lambda^2\, \left[\hat{c}_3 \,\frac{\I_3}{f^3} 
    		+ \hat{c}_4 \,\frac{\I_4}{f^4} 
		+ \hat{c}_5 \frac{\I_5}{f^5}  
		+\hat{c}_{6} \frac{\I_6}{f^6}  
		\cdots\right]\,,
  \label{Vreduced-A5-4}
  \end{equation} 
where $\I_{6}$ is formed out of primary invariants $\I_6 = \I_3^2$. 
 The manifold of invariants spanned by  the set  of primary invariants 
 $\{\I_3^{(\boldsymbol{4}, A_5)}, \I_4^{(\boldsymbol{4}, A_5)}, \I_5^{(\boldsymbol{4}, A_5)}   \}$   is analyzed next, subject to the non-linearity constraint Eq.~(\ref{non-linear-A5-4}). 
\begin{table}[h!]
\centering
\begin{align*}
	\begin{array}{|c||ccc|cccc|c|c|}
		\multicolumn{10}{c}{\text{\textbf{ MaNa extrema for a quadruplet of} $\boldsymbol{A_5}$}}\\
		\hline
		\multirow{2}{*}{\text{Point}} &\multirow{2}{*}{$\I_3$} &\multirow{2}{*}{$\I_4$} & \,\multirow{2}{*}{$\I_{5}$} & \phi_{1}\, &\phi_{2}\, &\phi_{3} &\phi_{4} & \multirow{2}{*}{\text{Little group}}& \multirow{2}{*}{\text{Nature}}\\
		&	&	&	& \multicolumn{4}{c|}{\textit{(Representatives)}} & & \\[1pt]
		\hline\hline
		\multirow{2}{*}[-0.15cm]{A} & \multirow{2}{*}[-0.15cm]{$\frac{1}{2\sqrt{5}}$} & \multirow{2}{*}[-0.15cm]{$\frac{1}{5}$} &\multirow{2}{*}[-0.15cm]{$\frac{1}{50}$} & 0 & 0 & 0 & 1 & \multirow{4}{*}[-0.15cm]{$A_4$}& \multirow{4}{*}[-0.15cm]{\text{Minima}}\\
		&	&	&	&	-\frac{\sqrt{5}}{4}   &	-\frac{\sqrt{5}}{4}	& -\frac{\sqrt{5}}{4}   & \frac{1}{4} & 	&	\\[0.8ex]
		\cline{1-8}
		\multirow{2}{*}[-0.15cm]{B} & \multirow{2}{*}[-0.15cm]{$-\frac{1}{2\sqrt{5}}$} & \multirow{2}{*}[-0.15cm]{$\frac{1}{5}$} &\multirow{2}{*}[-0.15cm]{$-\frac{1}{50}$} & 0 & 0 & 0 & -1 & & \\
		&	&	&	&	\frac{\sqrt{5}}{4}   &	\frac{\sqrt{5}}{4}	& \frac{\sqrt{5}}{4}   & -\frac{1}{4} & 	&	\\[0.8ex]
		\hline	\rule{0pt}{3ex}	
		\multirow{2}{*}[-0.15cm]{C} & \multirow{2}{*}[-0.15cm]{$\frac{1}{3\sqrt{30}}$} & \multirow{2}{*}[-0.15cm]{$\frac{31}{30}$} &\multirow{2}{*}[-0.15cm]{$-\frac{4}{25}\sqrt{\frac{2}{3}}$} & \sqrt{\frac{5}{24}} & \sqrt{\frac{5}{24}} & \sqrt{\frac{5}{24}} &  \sqrt{\frac{3}{8}}	& \multirow{4}{*}[-0.4cm]{$S_3$}& \multirow{4}{*}[-0.4cm]{\text{Saddles}}\\[0.8ex]
		&	&	&	& - \sqrt{\frac{5}{6}}   &			0				&	   		0			 &  -\frac{1}{\sqrt{6}} & 	&	\\[0.8ex]
		\cline{1-8}	\rule{0pt}{3.3ex}	
		\multirow{2}{*}[-0.15cm]{D} & \multirow{2}{*}[-0.15cm]{$-\frac{1}{3\sqrt{30}}$} & \multirow{2}{*}[-0.15cm]{$\frac{31}{30}$} &\multirow{2}{*}[-0.15cm]{$\frac{4}{25}\sqrt{\frac{2}{3}}$} & -\sqrt{\frac{5}{24}} & -\sqrt{\frac{5}{24}} & -\sqrt{\frac{5}{24}} & -\sqrt{\frac{3}{8}}	& & \\
		&	&	&	& \sqrt{\frac{5}{6}}   &			0				&	   		0			 &  \frac{1}{\sqrt{6}} & 	&	\\[0.8ex]
		\hline
	\end{array}
\end{align*}
\caption{Location and symmetries of some representative MaNa extrema for  the case of a real scalar fields in a quadruplet of $A_5$. All dimensional  entries are normalized to $f=1$.  The complete manifold is  illustrated in Fig.~\ref{fig:A5_4:invariant_manifold}.}
\label{tab:A5_4:critical_points}
\end{table}

\subsubsection{Embedding in a continuous group: \texorpdfstring{$SO(4)$}{} }
While the scenarios explored in previous sections with scalar triplets could all be embedded in $SO(3)$, for a quadruplet of $A_5$ the smallest continuous embedding corresponds to $SO(4)$.  From the point of view of the continuous symmetry, this theory is again insensitive to quadratic divergences, while the first contributions to the pion masses stem from the dimension three $A_5$-symmetric invariant.

\subsubsection{MaNa extrema}
\label{extrema_quadruplet_A5}

There are a total of 30 MaNa extrema for the invariant potential built out of the $\boldsymbol{4}$ representation of $A_5$:
\begin{itemize}
	\item 10 $A_4$-symmetric MaNa extrema\,, and 
	\item 20  MaNa extrema with $S_3$ as their little group,
\end{itemize}
i.e. all MaNa extrema are  non-abelian as desired. 
The counting rule in Eq.~(\ref{counting-rule}) works out nicely after one realizes that all the $A_4$- symmetric points do not belong to the same orbit, since transformations under $A_5$ are only able to relate among themselves half of them. The same applies to the set of $S_3$-symmetric points.  In consequence,   there are four different orbits  in total which all comply neatly with the counting criterium.

\begin{figure}[h!]
\centering
\includegraphics[width=1\textwidth]{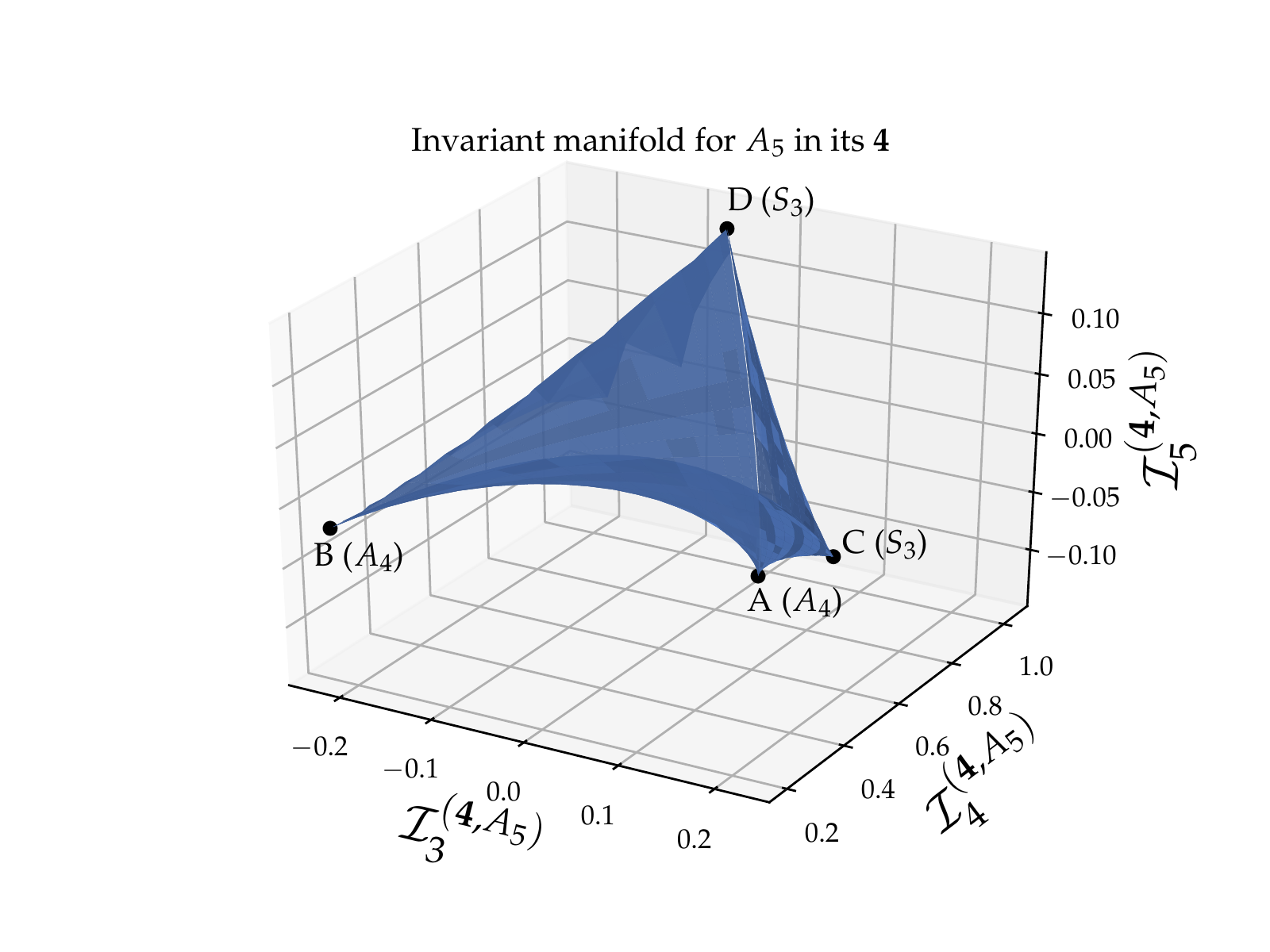}
\caption{Manifold defined by $\mathcal{I}^{(\boldsymbol{3}, A_5)}_3(\Phi)$, $\mathcal{I}^{(\boldsymbol{3}, A_5)}_4(\Phi)$ and  $\mathcal{I}^{(\boldsymbol{3}, A_5)}_{5}(\Phi)$ invariants of $A_5$ when $\Phi$ is a $SO(4)$ triplet fulfiling $\phi_1^2 + \phi_2^2 + \phi_3^2  + \phi_4^2 = f^2$.}
\label{fig:A5_4:invariant_manifold}
\end{figure}

 Representatives of each  MaNa extremum are shown in Table~\ref{tab:A5_4:critical_points}.  The manifold spanned by the three primary invariants of this scenario  is depicted in Fig.~\ref{fig:A5_4:invariant_manifold}.

\paragraph{Expansion in terms of low-energy degrees of freedom.} 
We will use again the standard pion parametrization in Eq.~(\ref{eq:pions  generic definition})  to project the results in Table~\ref{tab:A5_4:critical_points} in terms of the three low energy dGBs, 
\begin{equation}
	\Phi(\pi_{1},\pi_{2}, \pi_{3}) = \exp\left[\frac{1}{f} \left(
	\begin{array}{cccc}
		0 		 & 	0 		& 0 		& \pi_{1} \\
		0 		 &	0		& 0 		& \pi_{2} \\ 
		0 		 &	0		& 0 		& \pi_{3} \\ 
		-\pi_{1} & -\pi_{2} & -\pi_{3}  &	0
	\end{array}\right)
	\right]\left(
	\begin{array}{c} 
		0\\ 0  \\ 0 \\ f
	\end{array}\right)\,.
	\label{eq:pions_definition}
\end{equation}
 In order to obtain physically meaningful results, one must rotate those points among the  MaNa extrema  in Table~\ref{tab:A5_4:critical_points} which do not point in the $(0,0,0,1)$ direction chosen for the pion parametrization,  as described in  Eq.~(\ref{tilted-minima}) albeit with a four-dimensional rotation in parameter space. Two types of extrema appear, both exhibiting non-abelian invariances, described next. 
\paragraph{ MaNa extrema with $A_4$ as their little group.}The  minima in this class are stable. For them, the  low-energy spectrum exhibits and exact ({\it  \`a  la Wigner}) {\it non-abelian} $A_4$ symmetry.  Consider for illustration the representative  MaNa extremum $\Phi_B = (0, 0, 0, -1)$. In terms of the dGB degrees of freedom, the three primary invariants relevant for the potential  take the form 
\begin{align}
\I_3^{(\boldsymbol{4}, A_5)} 
	&= \frac{\sqrt{5 }f}{4} \left[
		-\frac{2f^2}{5} +  \left(\pi_1^2 + \pi_2^2 + \pi_3^2\right) - \frac{4}{\sqrt{5} f}\,\pi_1\pi_2\pi_3 - \frac{41}{60f^2}\left(\pi_1^2 + \pi_2^2 + \pi_3^2\right)^2\right]\,,
\label{I_4_A5_3}
\end{align}
\vspace*{-20pt}
\begin{multline}
\I_4^{(\boldsymbol{4}, A_5)}
    = 2f^2\left[\frac{f^2}{10} + \left(\pi_1^2 + \pi_2^2 + \pi_3^2\right) + \frac{6}{\sqrt{5}f}\pi_1\pi_2\pi_3 + \frac{1}{2f^2}\left(\pi_1^4 + \pi_2^4 + \pi_3^4\right) 
    \right.\\
    \left.
         - \frac{43}{30f^2} \left(\pi_1^2 + \pi_2^2 + \pi_3^2\right)^2\right]\,,
\label{I_4_A5_4}
\end{multline}
\vspace*{-20pt}
\begin{multline}
\I_5^{(\boldsymbol{4}, A_5)} 
    = \frac{f^3}{4}\left[\frac{2f^2}{25} +\left(\pi_1^2 + \pi_2^2 + \pi_3^2\right) +\frac{16}{\sqrt{5} f}\pi_1\pi_2\pi_3 
        -\frac{4}{f^2}\left(\pi_1^4 + \pi_2^4 + \pi_3^4\right) 
    \right.\label{I_4_A5_5}\\
    \left.
        + \frac{19}{60f^2}\left(\pi_1^2 + \pi_2^2 + \pi_3^2\right)^2\right]\,.
\end{multline}
If the dGB potential can be approximated by the lowest relevant invariant  (with the appropriate sign $V_{\rm dGB}\sim -\I_3^{(\boldsymbol{4}, A_5)}$), the $A_4$-symmetric point will correspond to a minimum. The expressions above show that the low energy theory exhibits an exact  $A_4$ symmetry for a $\boldsymbol{3}$ representation, i.e. 
\begin{itemize}
\item The low-energy spectrum consists of {\it a triplet of degenerate scalars} $\{\pi_1, \pi_2, \pi_3\}$. This is in contrast with the spectrum of two degenerate dGBs found whenever the exact symmetry of the low-energy theory was a discrete abelian group of any order.
\item The  invariants in terms of  $\{\pi_1, \pi_2, \pi_3\}$ in Eqs.~(\ref{I_4_A5_3})-(\ref{I_4_A5_5}) are combinations of those for a $\boldsymbol{3}$ representation of $A_4$, see for instance  Eqs.~(\ref{eq:invariant_2})-(\ref{eq:invariant_4}). 
\end{itemize}

\paragraph{MaNa extrema with $S_3$ as their little group.} These are saddle points. It is worth it to remark, though, that the saddle character stems exclusively from the cubic 
invariant, $\I_3^{(\boldsymbol{4}, A_5)}$.  Specifically, denoting by $\Phi_{A,B}$ and $\Phi_{C,D}$ the extrema in Table~\ref{tab:A5_4:critical_points} with little groups respectively invariant under $A_4$ and $S_3$ ,  the correspondence is 
\begin{align*}
\left.\I_3^{(\boldsymbol{4}, A_5)}\right|_{\Phi_{A,B}} \longrightarrow \ \text{Minima}\\
\left.\I_3^{(\boldsymbol{4}, A_5)}\right|_{\Phi_{C,D}} \longrightarrow \ \text{Saddles}
\end{align*}
Therefore, if some constraint were to forbid $\I_3$, a stable dGB theory constructed around the $S_3$-invariant extrema would be possible. In that case, the low-energy spectrum 
 would arrange itself in a $\boldsymbol{1}+\boldsymbol{2}$ of $S_3$: the triplet degeneracy would be lifted; a doublet and a singlet representation could be then expected. This would open new avenues for model-building and lead to a  very different phenomenology. We do not pursue this path here, as extra symmetries would have to be advocated to sustain such a construction. In their absence, the non-abelian spectra suggest at least three degenerate dGBs to be detected, in contrast with two degenerate dGBs for abelian symmetric spectra.

\subsection{Phenomenological signals from an \texorpdfstring{$A_5$}{} quadruplet}
 Once again, one finds the absence of quadratic sensitivity of the dGB potential to ultraviolet scales, and dGB masses whose size can be further separated from the scale of the effective theory $\Lambda$, as compared with the expectations for generic pGBs from non-linearly realized and explicitly broken symmetries. The essential low-energy property of  this ultraviolet $A_5$ scenario is the explicit non-abelian $A_4$ symmetry of the low-energy theory, realized by a triplet of dGBs, as long as the scalar vev corresponds to the MaNa minimum with the largest little group:

\begin{itemize}
\item Three degenerate dGBs are expected. This is in stark contrast with the abelian- symmetric ($Z_N$) spectra found in previous sections, which predicted in all cases a doublet of degenerate dGBs. 
\item  No events with a single ALP-like particle are expected (as long as the SM fields are singlets of the UV discrete symmetry). The experimental tell-tale signals would be then the simultaneous emission of  two, three or four  degenerate dGBs in SM initiated processes. In number of dGBs emitted, the event topologies are therefore alike to those for the $Z_3$ symmetric low-energy spectrum of an $A_4$ UV invariant theory, but the relative weights of the amplitudes for multi-dGB emission could allow one to differentiate the two cases, see Eq.~(\ref{eq:A4:invariants_pions_6}) versus  Eq.~(\ref{I_4_A5_3}). 
\end{itemize}

In order to illustrate this last point, consider the interaction between SM fields (singlets of the discrete symmetry) and  $\I_3^{(\boldsymbol{4}, A_5)} $\,,
\begin{equation}
\mathcal{L}^{\text{int}}\propto 
\frac{1}{M^m}\,\mathcal{O}^{SM} \I_3^{(\boldsymbol{4}, A_5)}  + \cdots \,
\label{effective-SM-dGB_A5-4}
\end{equation}
It follows from Eq.~(\ref{probability}) that the $A_4$-symmetric spectrum (which has resulted from the quadruplet of a UV $A_5$ invariant theory) lead to the experimental ratios: 
\begin{align}
&&&&
\frac{\sigma(\text{SM} \rightarrow 2 \pi)}{\sigma(\text{SM} \rightarrow 3 \pi)} 
	&= \frac{15 f^2 }4 \frac{ \Pi_2 }{ \Pi_3 }\,,\qquad 
	&\frac{\sigma(\text{SM} \rightarrow 3 \pi)}{\sigma(\text{SM} \rightarrow 4 \pi)}
		&=  \frac{ 6 f^2  }{(41)^2}  \frac{ \Pi_3 }{ \Pi_4 } \,,&&&&
\label{ratios:A5-4}
\\[10pt]
&&&&
\frac{\sigma\text{SM} \rightarrow 2 \pi)}{\sigma(\text{SM} \rightarrow 3 \pi)} 
	&= 120 \pi^2 \frac{ f^2 }{ E_\text{CM}^2 }\,,\qquad 
	&\frac{\sigma(\text{SM} \rightarrow 3 \pi)}{\sigma(\text{SM} \rightarrow 4 \pi)}
		&= \left( \frac{ 24 \pi }{41} \right)^2 \frac{ f^2 }{ E_\text{CM}^2} \,, &&&&
\label{ratios:A5-4}
\end{align}
to be compared with the rations expected for a $Z_3$-symmetric spectrum (from a triplet of a UV $A_4$ invariant theory), see Eq.~(\ref{ratios:A4-3}), and with those expected for a $Z_5$-symmetric spectrum
(from a triplet of a UV $A_5$ invariant theory) in Eq.~(\ref{ratios:A5-3}).  

Analogous remarks to those in Sect.~\ref{signals-derivative-couplings} for the signals expected from interaction terms containing $\partial_\mu U$ components apply in this case.

\newpage

\section{Conclusions}
\label{sec:conclusions}

 The symmetry-protected masses of discrete  Nambu-Goldstone bosons  offer promising theoretical  avenues to soften the ultraviolet sensitivity of 
  BSM theories.   We have discussed how exact --albeit non-linearly realized-- discrete symmetries allow one to give masses to those scalars without explicitly breaking the discrete  symmetry. 
  
  Examples of non-linearly realized UV  abelian discrete symmetries  ($Z_N$) have been previously developed in depth, showing that a much-lighter than usual axion can result~\cite{Hook:2018jle,DiLuzio:2021pxd,DiLuzio:2021gos}. Here we focus instead on the much richer case of UV non-abelian symmetries. We illustrated the theoretical analysis for various simple non-abelian discrete groups with a real scalar field in an irreducible representation.

 The non-linearity condition itself, quadratic in the scalar fields, implies the absence of quadratic sensitivity to high scales for the dGB masses, because no other quadratic form is allowed by the symmetry. 
 By the same token, the UV discrete invariance has the potential to strongly separate the dGB masses from the  scale of the effective theory. A key ingredient to assess the smallness of dGB masses is the dimensionality of the first scalar operator in the dGB potential that is invariant under the UV  discrete symmetry. The higher its dimension,  the stronger the mass suppression expected.  We showed how this depends on the group and the representation, illustrating  this effect with some simple cases. The first discrete scalar invariant of the dGB potential may already appear at cubic order, or  quartic, or even appear only at the non-renormalizable level, thus guaranteeing extremely suppressed dGB masses.

  The model-independent approach of EFT and invariant analysis has been used. In all cases, we have focused on the  MaNa minima of the potential, that is,  those points that are guaranteed to be extrema  and are independent of the values of the parameters in the potential. All such extrema have been identified, together with the manifolds spanned by the invariants, for various irreducible real representations of  $A_4$, $A_5$ and $S_4$, as representative examples. 
  
  For the sake of theoretical illustration, we have discussed as well the connection with the customary analysis of pion masses in spontaneously broken global continuous theories, which always require  additional arbitrary  potentials that break explicitly the continuous symmetry.  The findings in this paper could be then simply viewed as the result of requiring  the potential for explicit breaking  to remain invariant under a discrete subgroup of the continuous symmetry.  That is, no discrete group is added { \it ad hoc}, while exact invariance under a subgroup of the global symmetry itself is required (e.g. $A_4$ or $A_5$ for $SO(3)$). The point is that exact discrete symmetries often generate approximate continuous symmetries and, when
non-linearly realized, the corresponding pGB masses are protected: not only do the quadratically divergent contributions to their masses  from putative higher scales cancel exactly but also the generated potential may acquire an exponential suppression, which is controlled by the dimensionality of the first invariant operator that breaks the continuous symmetry but preserves the discrete one.  Nevertheless, our results did not rely on any choice of continuous group embedding. In order to better illustrate the mechanism, we have also explicitly computed the cancellation of the quadratic divergent diagrams in two UV complete models of dGBs whose realization involved either fermions or scalars.

 We have also shown that the low-energy theory around the  MaNa extrema remains explicitly invariant under a discrete subgroup of the UV discrete symmetry, which is identified. That is, the spectra of dGBs are shown to exhibit an explicit --i.e. {\it \`a la Wigner}-- discrete symmetry. The latter may be abelian (e.g. $Z_3$, $Z_5$ in some examples studied), or non-abelian (e.g. $A_4$, $S_3$). This results in tantalizing tell-tale experimental signals:
 \begin{itemize}
 \item Doublets of degenerate (dGB) scalars are to be expected for  abelian symmetric low-energy theories. In contrast,  three or more degenerate dGBs are possible for low-energy non-abelian symmetric spectra.\footnote{Two degenerate pions can also stem from very small  non-abelian symmetries at low energy, such as $S_3$. The key is the dimensionality of the irreducible representations of the UV symmetry.}
 \item No event with single dGB emission is expected, as long as the SM fields are singlets under the discrete symmetries. The simultaneous emission of two, three or more dGBs are to be expected in this case from SM initiated processes.
 \item The relative weights of multi-dGB emission provide hints of the exact discrete UV symmetry: 
 the analysis of the low-energy dGB spectrum and interactions cannot identify the UV symmetry unequivocally, but it can delineate the possibilities by consistency.
 \end{itemize}
 While the first two points above are not exclusive signals of a discrete symmetry (as they could be expected for exotic particles charged under an exact and explicit new symmetry), it is the ensemble of the three points that would allow one to see the presence in Nature of a protective discrete symmetry.
 
 Note as well that the second point above is at variance with usual ALP searches, for which single ALP emission is the first signal hunted (e.g. a single missing energy track if the ALP is stable or too long-lived to decay within the detector). Would the SM fields be charged under the UV discrete symmetry, the dGB degeneracy would remain untouched, while different experimental detection patterns can follow (to be developed elsewhere).  
   In the flourishing realm of ALP searches, it is  a pertinent quest to hunt for the sets of observables indicated, as a possible trademark of an unbroken but hidden discrete symmetry, and thus of a BSM theory with enhanced UV protection. 
   
    Theoretical applications of the results in this paper could  range from the search for UV stable theories of multicomponent dark matter, to the solution or alleviation of other fine-tuning and hierarchy issues in the known physics laws.


\begin{small}

\section*{Acknowledgments}
We thank Enrique Alvarez, Daniel Alvarez-Gavela, Quentin Bonnefoy, Ferruccio Feruglio, Anson Hook, 
 Aneesh Manohar, Luca Merlo and Mathias Pierre for illuminating discussions. 
R.H.~is supported by the STFC under Grant No. ST/T001011/1.
P.Q.~acknowledges support by the Deutsche Forschungsgemeinschaft under Germany's Excellence Strategy - EXC 2121 Quantum Universe - 390833306 and by the Deutsche Forschungsgemeinschaft (DFG, German Research Foundation) - 491245950. The work of P.Q. is supported in part by the U.S. Department of Energy Grant No. DE-SC0009919.
The work of V.E. was supported by the Spanish MICIU through the National Program FPI-Severo Ochoa (grant number PRE2020-094281). 
V.E. and B.G. acknowledge partial financial support by the Spanish Research Agency (Agencia Estatal de Investigaci\'on) through the grant IFT Centro de Excelencia Severo Ochoa No CEX2020-001007-S and by the grant PID2019-108892RB-I00 funded by MCIN/AEI/ 10.13039/501100011033. All authors acknowledge support by the European Union's Horizon 2020 research and innovation programme under the Marie Sklodowska-Curie grant agreement No 860881-HIDDeN.  
B.G. and P.Q. thank respectively DESY and Madrid IFT for their warm hospitality where part of this work was carried out.


\normalsize
\appendix

\section{Brief summary of \texorpdfstring{$S_4$}{} }
\label{App:S4}

As advanced at the end of Sec.\,\ref{triplet-A4}, the scenario with  an ultraviolet $S_4$ symmetry is closely related to that of $A_4$, but it presents  some suggestive differences, discussed next for its two triplet representations, $\boldsymbol{3}$ and $\boldsymbol{3'}$.

\subsection{\texorpdfstring{$S_4$}{} in its \texorpdfstring{$\boldsymbol{3}$}{} representation}
The Molien function for the $\boldsymbol{3}$ of $S_4$,  reads
\begin{equation}
\begin{split}
\F_{S_4}(\boldsymbol{1};\boldsymbol{3};\lambda) &= \frac{1}{(1-\lambda^2)(1-\lambda^3)(1-\lambda^4)}\,,\\
\end{split}
\label{eq:S4:3:Molien_function}
\end{equation}
which indicates the same primary invariants as  for the $\boldsymbol{3}$ of $A_4$, of dimension two, three and four: $\I_2, \I_3, \I_4$ in Eqs.(\ref{eq:invariant_2})-(\ref{eq:invariant_4}). The Molien function for the $\boldsymbol{3'}$ representation is the same except that it has a factor of $(1+\lambda_6)$ in the numerator; in other words, 
the difference is that the secondary invariant $\I_6$ defined in Eq.~(\ref{eq:A4:invs:syzigy}) by a syzygy of the $A_4$ triplet, is forbidden for $S_4$. As a consequence, differences between both theories arise only at the non-renormalizable level through this six-dimensional operator. It follows in turn that  the  MaNa extrema of the potential are the same as those of $A_4$, although their little groups are extended because of the larger size of the group. That is, the manifold of invariants is that in Fig.~\ref{fig:A4:invariant_manifold} albeit with the symmetries at the  MaNa extrema changed as follows:
\begin{table}[h!]
\centering
\begin{align*}
	\begin{array}{|c||c|}
	\hline
	\text{Point} & \text{Little group}\\
	\hline\hline
	A & V_4 = Z_2\times Z_2\\
	B & V_4 = Z_3\times Z_2\\
	C & S_3 = Z_3\times Z_2\\
	\hline
	\end{array}
\end{align*}
\caption{Symmetries of the  MaNa extrema for an invariant potential for the $\boldsymbol{3}$ of $S_4$, where $A$, $B$ and $C$ are the points defined for $A_4$ in Table~\ref{tab:A4:critical_points} with the corresponding manifold as depicted in Fig~\ref{fig:A4:invariant_manifold}.}
\label{tab:S4:3:natural_singular_points}
\end{table}

In terms of the pions as defined by Eq.\,(\ref{eq:pions definition}), the absence of $\I_6$ translates into the exclusion of one of the $Z_3$-invariant combinations of order 3 introduced by Eqs.\,(\ref{eq:A4:invariants_Z3}) and\,(\ref{eq:A4:sizygy_Z3}). Even though both expressions can be reshuffled as in Eq.\,(\ref{beta-cubic-invariant}) due to pion degeneracy, this results in the presence of a single combination of both operators in the Lagrangian regardless of the basis chosen. From another point of view, this is a consequence of the extended $S_3$ invariance of the formerly $Z_3$-symmetric  MaNa extrema $B$ and $C$, which is more stringent in forbidding combinations of the pions. Interestingly then, $S_3$ is a non-abelian group ---although the smallest existing one--- and it will remain explicitly realized in the spectrum in this case.
Subtle differences at sub-leading orders in the production rates or scattering cross sections predicted by both theories might follow, but at leading order the physical predictions stemming from the $\boldsymbol{3}$ of $A_4$ and $S_4$ are identical.

\begin{figure}[h!]
    \centering
    \includegraphics[width = 0.6\textwidth]{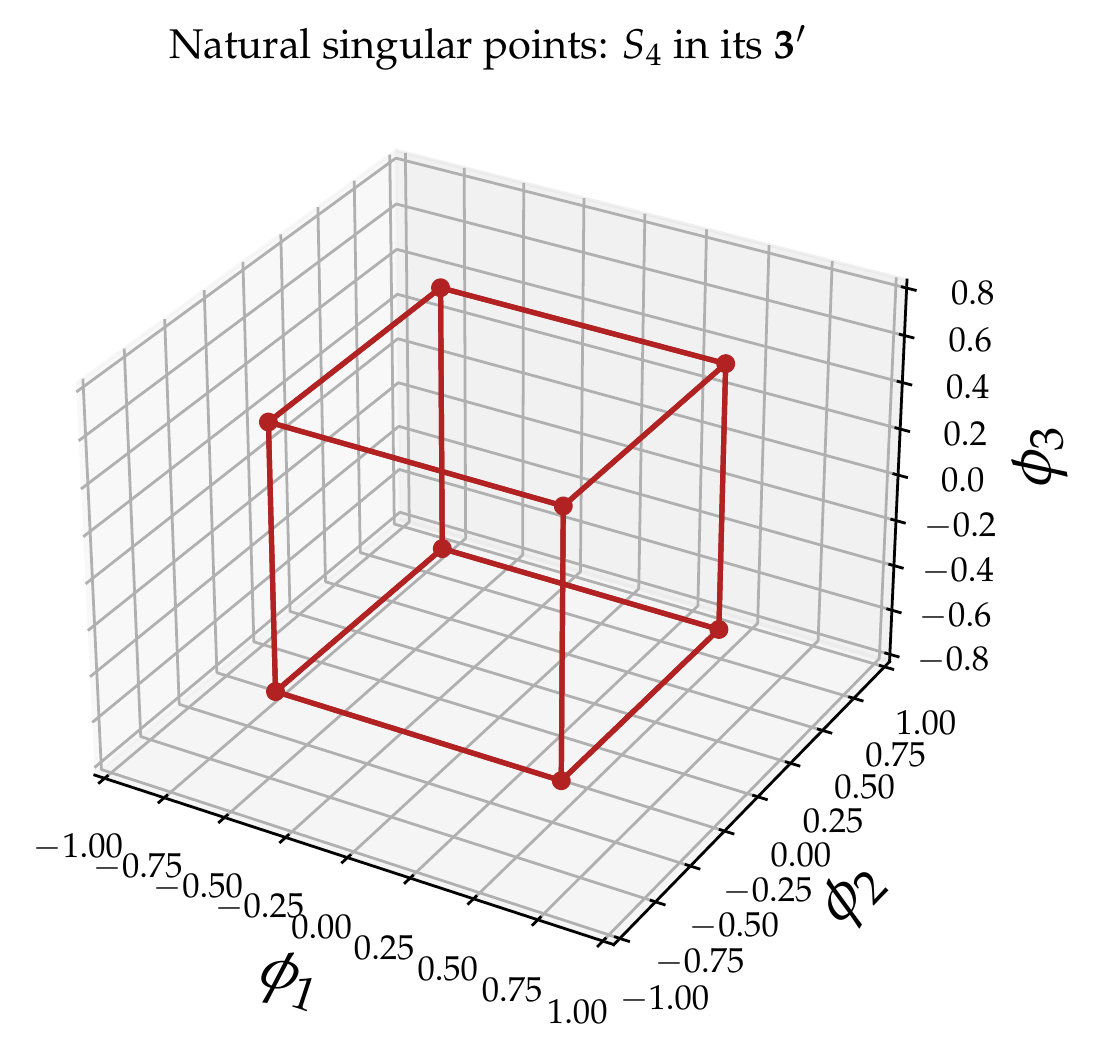}
    \caption{Geometrical distribution in field space of the $S_3$-symmetrical  MaNa extrema (B) with a $\boldsymbol{3'}$ of $S_4$ .}
    \label{fig:S4:singular_points}
 \end{figure}

\subsection{The \texorpdfstring{$\boldsymbol{3'}$}{} of \texorpdfstring{$S_4$}{} }

The $\boldsymbol{3'}$ is of considerable more interest, with more unique features that stem from its structural differences in relation to the $\boldsymbol{3}$ of $A_4$ \cite{ishimori2010non}. Looking at the Molien function,
\begin{equation}
\begin{split}
\F_{S_4}(\boldsymbol{1};\boldsymbol{3'};\lambda) &= \frac{1+\lambda^9}{(1-\lambda^2)(1-\lambda^4)(1-\lambda^6)},\\
\end{split}
\label{eq:S4:3p:Molien_function}
\end{equation}
one immediately spots the similarities and differences with the former cases. First of all, the invariant of order 3 is forbidden, so non-trivial contributions to the potential will appear only at the marginal level. Explicitly, the invariants are 
\begin{equation}
	\begin{split}
		\I_{2}^{(\boldsymbol{3'},\,S_4)}(\phi) &= \phi_1^2 +  \phi_2^2 + \phi_3^2,\\
		\I_{4}^{(\boldsymbol{3'},\,S_4)}(\phi) &= \phi_1^4 + \phi_2^4 + \phi_3^4,\\
		\I_{6}^{(\boldsymbol{3'},\,S_4)}(\phi) &= (\phi_1\phi_2\phi_3)^2,
	\end{split}
	\label{eq:S4:3p:primary_invariants}
\end{equation}
meaning that the third order invariant has been substituted by its square; instead of $\I_3$ we have $\I_6^{(\boldsymbol{3'},\, S_4)} = \I_3^2$. The non-trivial secondary invariant is of order $9$, and thus extremely suppressed. As it happened in $A_4$, it can be found to be proportional to the determinant of the Jacobian of the primary invariants. We define
\begin{align}
	\operatorname{det} \J_{(\boldsymbol{3'},\, S_4)} = -16\,\I_9^{(\boldsymbol{3'},\, S_4)} = -16\,\phi_1\phi_2\phi_3\left(\phi_1^{2}-\phi_2^{2}\right)\left(\phi_2^{2}-\phi_3^{2}\right)\left(\phi_1^{2}-\phi_3^{2}\right).
	\label{eq:S4:3p:Jacobian_det}
\end{align}
The set of  MaNa extrema is enlarged in this case; one can see from Fig.\,\ref{fig:S4:3p:invariant_manifold} that the structure of the invariant manifold has indeed changed. Now there are
\begin{itemize}
	\item 6 MaNa extrema $A$ with $V_4 = Z_2\times Z_2$ as their little group, which are stable.
	\item 8  MaNa extrema $B$ with $S_3 = Z_2\times Z_3$ as their little group, which are stable.
	\item 12  MaNa extrema $C$ with $Z_2$ as their little group, which are saddles.
\end{itemize} 

Table~\ref{eq:S4:3p:invariants:pions_S3} contains all the relevant information about all these points, including some representatives for each case. 
\begin{table}[h!]
\centering
\begin{align*}
	\begin{array}{|c||cc|ccc|c|c|}
		\multicolumn{8}{c}{\text{\textbf{ MaNa extrema for the} $\boldsymbol{3'}$ \textbf{of} $\boldsymbol{S_4}$}}\\
		\hline
		\multirow{2}{*}{\text{Point}} &\multirow{2}{*}{$\I_4^{(\boldsymbol{3'},\, S_4)}$} &\multirow{2}{*}{$\I_6^{(\boldsymbol{3'},\, S_4)}$} & \phi_{1}\, &\phi_{2}\, &\phi_{3} & \multirow{2}{*}{\text{Little group}}& \multirow{2}{*}{\text{Nature}}\\
		&	&	&	 \multicolumn{3}{c|}{\textit{(Representatives)}} & & \\[1pt]
		\hline\hline \rule{0pt}{2.3ex}
		\multirow{2}{*}{A} & \multirow{2}{*}{$\frac{1}{3}$} & \multirow{2}{*}{$\frac{1}{27}$} & \frac{1}{\sqrt{3}} & \frac{1}{\sqrt{3}} & \frac{1}{\sqrt{3}} & \multirow{2}{*}{\text{$S_3$}} & \multirow{2}{*}{\text{Minima}}\\
													  &											&	& -\frac{1}{\sqrt{3}} & -\frac{1}{\sqrt{3}} & -\frac{1}{\sqrt{3}} & & \\[0.8ex]
		\hline	\rule{0pt}{2.3ex}
        \multirow{3}{*}{B} & \multirow{3}{*}{1} & \multirow{3}{*}{0} & 0 & 0 & \pm1 & \multirow{3}{*}{\text{$V_4$}}& \multirow{3}{*}{\text{Minima}}\\
                              &                       &                       & 0 & \pm1 & 0 &                        & \\
                              &                       &                       & \pm1 & 0 & 0 &                        & \\
        \hline  \rule{0pt}{2.3ex}
		\multirow{1}{*}{C} & \multirow{1}{*}{$\frac{1}{2}$} & \multirow{1}{*}{$0$} & \pm \frac{1}{\sqrt{2}} & 0 & \pm \frac{1}{\sqrt{2}}  & \multirow{1}{*}{\text{$Z_2$}} & \multirow{1}{*}{\text{Saddles}}\\[0.8ex]
	\hline
	\end{array}
\end{align*}
\caption{ MaNa critical points of the invariants of the $\boldsymbol{3'}$ of $S_4$. The manifold of primary invariants relevant for the dGB potential is depicted in Fig.~\ref{fig:S4:3p:invariant_manifold}.}
\label{tab:S4:3p:critical_points}
\end{table}
\begin{figure}[h!]
\centering
\includegraphics[width=0.7\textwidth]{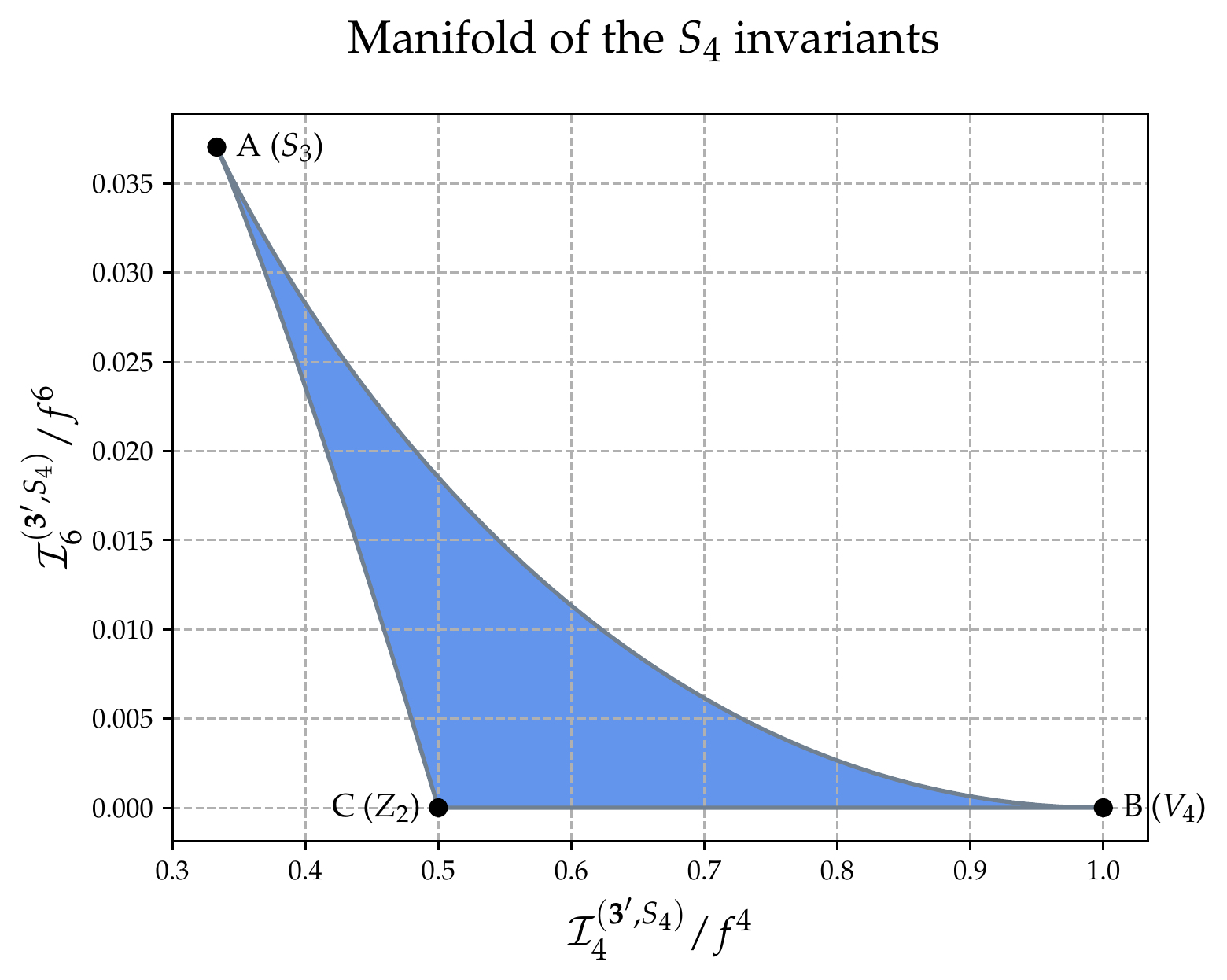}
\caption{Manifold defined by $\mathcal{I}^{(\boldsymbol{3'},\, S_4)}_4(\Phi)$ and $\mathcal{I}^{(\boldsymbol{3'},\, S_4)}_6(\Phi)$ invariants of $S_4$ when $\Phi$ is in the $\boldsymbol{3'}$ of $S_4$ and fulfills $\phi_1^2 + \phi_2^2 + \phi_3^2  = f^2$.}
\label{fig:S4:3p:invariant_manifold}
\end{figure}

\subsubsection*{Expansion in terms of  pions}
There are two sets of stable  MaNa extrema, denoted $A$ and $B$ in the manifold in Fig.~\ref{fig:S4:3p:invariant_manifold}. Therefore, two different realizations in terms of the physical dGBs are possible: either $A$ or $B$ is a minimum, depending on the sing of the $\hat c_i$ coefficients.  Assuming that the full potential can be approximated by the first invariant $V_{\rm dGB}\sim -\I_4^{(\boldsymbol{3}, S_3)}$, the $V_4$-symmetric minima in $B$ are global minima, consistent with the Michel-Radicati result~\cite{michel1971properties}.

\paragraph{Expanding around \texorpdfstring{$A$}{}.}  On defining the dGBs around the $V_4$-invariant  MaNa extremum $\Phi_{B} = (0, 0, 1)^\top$, the contributions to the potential will read
\begin{align}
\I_4^{(\boldsymbol{3'},\, S_4)} &= 
		f^4
		- 2f^2(\pi_1^2+\pi_2^2) 
		+ \frac{8}{3}(\pi_1^2+\pi_2^2)^2 - 2\pi_1^2\pi_2^2 
		+ \mathcal{O}\left(\pi_1^5\right)\,,\\
\I_6^{(\boldsymbol{3'},\, S_4)} &=
		\pi_1^2\pi_2^2 
		+ \mathcal{O}\left(\pi_i^5\right)\,,
\label{eq:S4:3p:invariants:pions_V4}
\end{align}
where the pions are arranged followed the pattern dictated by $V_4$ invariance. Using the Molien function formalism one can easily see that the two primary invariants of the doublet of $V_4$ are indeed $\I_2^{(\boldsymbol{2}, V_4)} = \pi_1^2 + \pi_2^2$ and $\I_4^{(\boldsymbol{2}, V_4)} = \pi_1^2\pi_2^2$. Consequently, this case forbids the production of an odd number of pions to all orders, distinguishing its predicted signal from all others that are discussed in this paper.

\paragraph{Expanding around \texorpdfstring{$B$}{} : } The expansion around a $S_3$-symmetric  MaNa extremum yields a quite different outcome:
\begin{align}
\I_4^{(\boldsymbol{3'},\, S_4)} &= 
		\frac{f^4}{3}
  		+ \frac{4}{3}f^2 \left(\pi 1^2+\pi 2^2\right)
  		-\frac{2}{3}f \left(\pi_1^3-3\pi_1 \pi_2^2\right)
  		-\frac{29}{18} \left(\pi _1^2+\pi _2^2\right){}^2
  		+ \mathcal{O}\left(\pi_i^5\right)\,,\\
\I_6^{(\boldsymbol{3'},\, S_4)} &= 
		\frac{f^6}{27} 
		- \frac{2}{9}f^4\left(\pi_1^2+\pi _2^2\right)
		-\frac{f^3}{27} \left(\pi_1^3-3\pi_1 \pi_2^2\right)
		+ \frac{53}{108}f^2 \left(\pi_1^2+\pi_2^2\right){}^2 + \mathcal{O}\left(\pi_i^5\right)\,,\\
\label{eq:S4:3p:invariants:pions_S3}
\end{align}
where the primary invariants of the doublet of $S_3$ are precisely $\I_2^{(\boldsymbol{2}, S_3)} = \pi_1^2 + \pi_2^2$ and $\I_3^{(\boldsymbol{2}, S_3)} = \pi_1^3-3\pi_1 \pi_2^2$. Contrarily to what happens  in the case of $Z_3$, there is no further allowed combination of order 3, so subtle variances are expected again with respect to the physical predictions coming from the $Z_3$ invariant points of $A_4$. The pions are in fact arranged in the same structures as for the $S_3$-symmetric minima from the $\boldsymbol{3}$ of $S_3$, but this case is anyhow different in that the leading contribution comes from the marginal order operator.

\subsection{Phenomenological signals from a triplet of \texorpdfstring{$S_4$}{} }
 Since the lowest order invariant relevant for the dGB potential is now $\I_4$, the leading interaction between SM fields and  pure dGB scalar invariants ---under the same assumptions as used when considering the $\boldsymbol{3}$ of $A_4$, see Eq.(\ref{effective-SM-dGB})--- reads
\begin{equation}
\mathcal{L} = \frac{1}{M^m}\mathcal{O}_{SM} \I_4\,,
\end{equation}
leading to a different prediction for the production ratios.  The results around the $S_3$-invariant minima have already been presented in Eq.\,(\ref{eq:S4:ratios34}). If Nature settles instead around the $V_4$ minima, the expected production rates are unique in that those for an odd number of pions are forbidden. The first expected production ratio is then
\begin{align}
\frac{\sigma(\text{SM}\rightarrow 2\pi)}{\sigma(\text{SM}\rightarrow 4\pi)} &= \frac{27(8\pi^2)}{19\cdot2^9}\frac{f^4}{E_\text{CM}^4}.
\end{align}

\vspace*{\fill}
\pagebreak

\newpage

\bibliographystyle{utphys.bst}
\bibliography{bibliography.bib}

\end{small}

\end{document}